\documentclass[prd,letterpaper,showpacs,groupedaddress,superscriptaddress,nofootinbib,floatfix,preprintnumbers,tightenlines,11pt]{revtex4}

\usepackage{amsmath,amssymb}
\usepackage{graphicx}
\usepackage{bm}
\usepackage{comment}
\usepackage{color}
\usepackage{subfigure}
\usepackage{array}
\usepackage{multirow}
\usepackage{natbib}
\usepackage{url}
\usepackage[]{units}
\usepackage{dcolumn}% Align table columns on decimal point
\usepackage{slashed}
\usepackage{soul}
\usepackage{mathtools}
\usepackage{float}
\usepackage{afterpage}

\usepackage[T1]{fontenc}
\usepackage[latin9]{inputenc}
\usepackage{graphicx}
\usepackage{esint}
\usepackage{hyperref}
\usepackage{atbegshi}
\usepackage{lipsum}
\usepackage{braket}
\usepackage{epsfig}

\begin{document}

\dimen\footins=5\baselineskip\relax

\preprint{\vbox{
\hbox{INT-PUB-17-017, MIT-CTP-4912, NSF-ITP-17-076}
}}

\begin{figure}[!t]
 \vskip -1.1cm \leftline{
 	\includegraphics[width=3.0 cm]{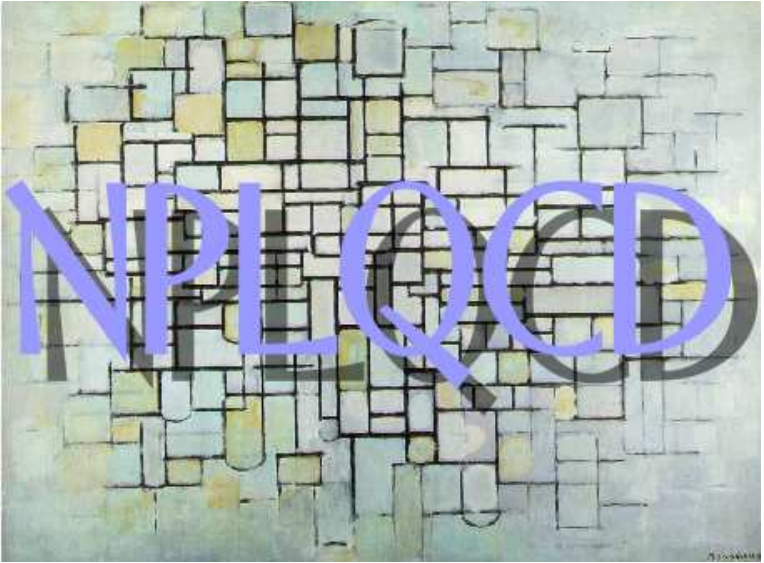}} \vskip
 -0.5cm
\end{figure}

\title{
Baryon-Baryon Interactions and Spin-Flavor Symmetry 
\\
from Lattice Quantum Chromodynamics
}

\author{Michael~L.~Wagman} 
\affiliation{Department of Physics,
	University of Washington, Box 351560, Seattle, WA 98195, USA}
\affiliation{Institute for Nuclear Theory, University of Washington, Seattle, WA 98195-1550, USA}

\author{Frank~Winter}
\affiliation{Jefferson Laboratory, 12000 Jefferson Avenue, 
	Newport News, VA 23606, USA}

\author{Emmanuel~Chang}
\noaffiliation

 \author{Zohreh~Davoudi} \affiliation{
 	Center for Theoretical Physics, 
 	Massachusetts Institute of Technology, 
 	Cambridge, MA 02139, USA}

 \author{William~Detmold} \affiliation{
 	Center for Theoretical Physics, 
 	Massachusetts Institute of Technology, 
 	Cambridge, MA 02139, USA}
 
 \author{Kostas~Orginos}
 \affiliation{Department of Physics, College of William and Mary, Williamsburg,
 	VA 23187-8795, USA}
 \affiliation{Jefferson Laboratory, 12000 Jefferson Avenue, Newport News, VA 23606, USA}
   
 \author{Martin~J.~Savage}
 \affiliation{Department of Physics,
	University of Washington, Box 351560, Seattle, WA 98195, USA}
	 \affiliation{Institute for Nuclear Theory, University of Washington, Seattle, WA 98195-1550, USA}
 
  \author{Phiala~E.~Shanahan } \affiliation{
 	Center for Theoretical Physics, 
 	Massachusetts Institute of Technology, 
 	Cambridge, MA 02139, USA}

\collaboration{NPLQCD Collaboration}

\date{\today}


\begin{abstract} 
	
Lattice quantum chromodynamics  is used to constrain the interactions of two octet baryons at the $SU(3)$ flavor-symmetric point, with quark masses that are heavier than those in nature (equal to that of the physical strange quark mass
 and corresponding to a pion mass of $\approx 806~\tt{MeV}$). Specifically, the $S$-wave scattering phase shifts of two-baryon systems at low energies are obtained with the application of L\"uscher's formalism, mapping the energy eigenvalues of two interacting baryons in a finite volume to the two-particle scattering amplitudes below the relevant inelastic thresholds. The values of the leading-order low-energy scattering parameters in the irreducible representations of $SU(3)$ are consistent with an approximate $SU(6)$ spin-flavor symmetry in the nuclear and hypernuclear forces that is predicted in the large-$N_c$ limit of QCD. The two distinct $SU(6)$-invariant interactions between two baryons are constrained at this value of the quark masses, and their values indicate an approximate accidental $SU(16)$ symmetry. The $SU(3)$ irreps containing the $NN~({^1}S_0)$, $NN~({^3}S_1)$ and $\frac{1}{\sqrt{2}}(\Xi^0n+\Xi^-p)~({^3}S_1)$ channels unambiguously exhibit a single bound state, while the irrep containing the $\Sigma^+ p~({^3}S_1)$ channel exhibits a state that is consistent with either a bound state or a scattering state close to threshold. These results are in agreement with the previous conclusions of the NPLQCD collaboration regarding the existence of two-nucleon bound states at this value of the quark masses.

\end{abstract}
\maketitle

%%%%%%%%%%%%%%%%%%%%%%%%%%%%%%%%%%%%%%%%%%%
\section{INTRODUCTION  
\label{sec:Intro} 
}
\noindent
It is speculated that hyperons, the counterparts of nucleons in which some of the valence quarks in the nucleon are replaced by strange quarks, play an important role in the composition of dense matter, such as that in the interior of neutron stars (for a comprehensive review, see Ref.~\cite{Page:2006ud}). The interactions between two nucleons are precisely constrained by experiment over a wide range of energies. However, those between a nucleon and a hyperon, or between two hyperons, are not well known~\cite{Gal:2005ni, Hashimoto:2006aw, Balewski:1998pd, Sewerin:1998ky, Kowina:2004kr, Bilger:1998jf, AbdelBary:2004ts, Gasparyan:2003cc, Batty:1997zp, Ahn:2005gb, Holzenkamp:1998tq, Reuber:1995vc, Rijken:1998yy, Haidenbauer:2005zh, Rijken:2006ep, Polinder:2006zh, Haidenbauer:2013oca, Beane:2012ey}, and are challenging to probe experimentally because of the short lifetime of hyperons and hypernuclei. Precise information on how hyperons interact, in particular in a nuclear medium, is essential to establish their effects on the equation of state of dense matter and other observables. On the theoretical side, the only reliable method with which to determine these interactions is to calculate them from the underlying strong interactions among quarks and gluons described by quantum chromodynamics (QCD). This can be achieved using the non-perturbative method of  lattice QCD (LQCD), which involves numerically evaluating path integrals representing Euclidean correlation functions using Monte Carlo sampling methods. This approach is taken in this work to constrain the scattering amplitudes of several classes of nucleon-nucleon, hyperon-nucleon and hyperon-hyperon systems, albeit in world that exhibits an exact $SU(3)$ flavor symmetry, with degenerate light and strange quark masses tuned to produce pions and kaons with masses of $\approx 806~\tt{MeV}$. The calculations are performed in the absence of quantum electrodynamics (QED). This work extends our previous studies of such systems using the same ensembles of gauge-field configurations~\cite{Beane:2012vq, Beane:2013br}, and complements previous and ongoing studies of hyperon interactions using LQCD, see for example Refs.~\cite{Beane:2006gf, Nemura:2008sp, Beane:2009py, Beane:2010hg, Beane:2011zpa, Beane:2011iw, Inoue:2010es, Inoue:2011ai, Beane:2012ey, Etminan:2014tya, Yamada:2015cra, Nemura:2017bbw, Doi:2017cfx, Ishii:2017xud, Sasaki:2017ysy}. 

L\"uscher's finite-volume (FV) methodology~\cite{Luscher:1986pf, Luscher:1990ux, Rummukainen:1995vs, Beane:2003da, Feng:2004ua, Kim:2005gf, Davoudi:2011md, Leskovec:2012gb, Gockeler:2012yj, Briceno:2013lba, He:2005ey, Hansen:2012tf, Briceno:2012yi, Li:2012bi, Briceno:2013hya, Briceno:2014oea} is used to constrain the scattering amplitudes of two-baryon systems below the relevant inelastic thresholds from the corresponding energy eigenvalues of two interacting baryons in a finite cubic volume with periodic boundary conditions. The extraction of energies relies on the identification of low-lying states in Euclidean correlation functions. Discussions of the methods of energy determination used in this work are presented in Sec.~\ref{subsec:EMP}, along with careful analyses of the scattering amplitudes that result from the extracted energies in Sec.~\ref{subsec:results}. 
In particular, it is shown that, in agreement with our previous conclusions in Refs.~\cite{Beane:2012vq, Beane:2013br}, there is clear evidence for the existence of a bound state in each of the $SU(3)$-symmetric two-baryon channels containing the $NN~({^1}S_0)$, $NN~({^3}S_1)$ and $\frac{1}{\sqrt{2}}(\Xi^0n+\Xi^-p)~({^3}S_1)$ systems. The phase shifts in these channels, and in the channels containing the $\Sigma^+ p~({^3}S_1)$ system, pass all of the so-called ``sanity checks'' introduced in Ref.~\cite{Iritani:2017rlk}, contradicting the claims in that reference. 

$SU(3)$ flavor symmetry has important consequences for the interactions of two octet baryons. Despite there being 64 flavor states that can be constructed from two octet baryons, $SU(3)$ symmetry dictates that there are only six independent interactions between two octet baryons, namely those in the $27$, $\overline{10}$, $10$, $8_A$, $8_S$ and $1$ irreducible representations (irreps). $SU(3)$ flavor symmetry is only an approximate symmetry in nature, given the different masses of the light quarks and the strange quark, but is exact within the present numerical study, enabling a simple classification of these interactions. In particular, at leading order (LO) in an effective field theory (EFT) expansion~\cite{Savage:1995kv}, only six coefficients, corresponding to the six irreps, need be determined. Because of the structure of the interpolating fields implemented in this work, scattering information has been obtained only for channels belonging to the first four irreps listed above. 

The results of our calculations allow an exploration of the spin-flavor symmetries of nuclear and hypernuclear interactions that are predicted from QCD (with degenerate flavors) in the limit of a large number of colors, $N_c\rightarrow\infty$~\cite{Kaplan:1995yg}. In this limit, the six LO interactions of the $SU(3)$-symmetric low-energy theory are defined by only two independent constants, reflecting a manifest $SU(6)$ spin-flavor symmetry. Corrections to the constraints imposed by the $SU(6)$ symmetry scale as $1/N_c$. The calculations performed in this work provide an opportunity to examine the large-$N_c$ relations without contamination from $SU(3)$ breaking effects that are present in nature. Sec.~\ref{subsec:results} includes the results of this investigation, which demonstrate for the first time the $SU(6)$-symmetric nature of interactions in the two-baryon channels (even at $N_c=3$), and further point to an accidental $SU(16)$ symmetry. Assuming the $SU(6)$ spin-flavor symmetry, predictions are made for the leading LECs of the effective $SU(3)$-symmetric baryon-baryon interactions. Future studies of the two-baryon channels belonging to the  $8_S$ and $1$ irreps are needed to confirm these conclusions.

As the values of the parameters of QCD in this study differ from those in nature, there is no direct connection between the present results and phenomenology. However, our work presents an exploration of a non-Abelian gauge theory that is continuously connected to the strong-interaction sector of nature through the variation of the masses of the light quarks. These calculations establish and verify formal, numerical and algorithmic technologies that are needed for future explorations of multi-baryon systems at the physical values of the quark masses. Additionally, the possibility of changing the parameters of QCD in LQCD studies is itself a unique feature that has been shown to reveal insights into the structure of QCD that would be impossible to discover experimentally. Perhaps most importantly, refinements of the chiral nuclear forces requires calculations over a range of quark masses. Nucleon-nucleon interactions are speculated to be finely tuned in nature, and the calculations in this work test how robust this fine tuning is with regard to changes in the quark masses~\cite{Beane:2002vs, Bedaque:2010hr, Chen:2010yt, Soto:2011tb}, investigations that are only possible with LQCD. The first study of the unnaturalness of nucleon-nucleon interactions using the same ensembles of gauge-field configurations as in this work has already been conducted in Ref.~\cite{Beane:2013br}. Here, this study is extended to scattering channels involving hyperons. In addition, important progress has been made recently in applying EFTs and nuclear many-body techniques to extend the range of predictions of QCD with $m_{\pi} \approx 806~\tt{MeV}$ to heavier nuclei~\cite{Barnea:2013uqa, Kirscher:2015yda, Kirscher:2017fqc}, pointing to the ground state of $^{16}$O being likely unbound~\cite{Contessi:2017rww}. Further investigations are needed to confirm this result and study its implications on the periodic table of nuclide at heavy quark masses. Such studies can be extended to hyperon systems with the aid of the LQCD results presented in this work, and will provide more insight into the robustness of the properties of nuclear and hypernuclear systems with respect to  variations in the parameters of QCD.

The rest of this paper is organized as follows. The formalism required to analyze and interpret the numerical results of this study is presented in Sec.~\ref{sec:formalism}. In particular, Sec.~\ref{subsec:FV} contains a summary of the method used to extract the scattering amplitudes below the relevant inelastic thresholds from LQCD energy eigenvalues, along with discussions of the volume dependence of bound-state energies. Sec.~\ref{subsec:SU(3)} summarizes the expectations of $SU(3)$ flavor symmetry for two-baryon channels, and subsequent predictions for an extended $SU(6)$ symmetry present in the limit of large $N_c$. 
Details of the numerical study and the results are presented in Sec.~\ref{sec:numerical}. A summary and conclusion follow in Sec.~\ref{sec:conclusion}. The paper includes four appendices: Appendix~\ref{app:SU(3)} presents more detail on the $SU(3)$ structure of baryon-baryon systems. Appendix~\ref{app:LO-amplitude} tabulates the values of LO scattering amplitudes in mixed flavor channels. Appendix~\ref{app:tables} contains the full tables of the results for energies and phase shifts. Finally, Appendix~\ref{app:checks} is devoted to examining the so-called ``sanity checks'' of Ref.~\cite{Iritani:2017rlk}, demonstrating the definitive presence of physical bound states in the two-nucleon channels at this value of the quark masses, and the source-independence of the results presented, contrary to the claims presented in Ref.~\cite{Iritani:2017rlk}.

%%%%%%%%%%%%%%%%%%%%%%%%%%%%%%%%%%%%%%%%%%%
\section{Formalism
\label{sec:formalism} 
}
\noindent
The goal of the numerical calculations presented here is to constrain scattering amplitudes in various baryon-baryon channels. Scattering information is obtained from the energy spectra of two baryons in a finite volume, and a summary of L\"uscher's methodology for mapping finite-volume energy eigenvalues to scattering amplitudes is presented in this section. Signatures of bound states in LQCD calculations of two-baryon spectra are further discussed. This section also contains  theoretical background relevant for scattering processes with $SU(3)$ symmetry 
and the predictions of the large-$N_c$ limit of QCD.

%%%%%%%%%%%%
\subsection{Two-baryon systems in a finite volume and L\"uscher's methodology
\label{subsec:FV}
}
Below all relevant inelastic thresholds, the interacting energies of two particles in a finite volume determine the scattering amplitudes through a direct mapping given by L\"uscher's quantization condition (QC)~\cite{Luscher:1986pf, Luscher:1990ux}. 
This mapping is valid as long as the interactions have a finite range that is contained inside the lattice volume. For typical hadronic systems, the range of interactions is set by the Compton wavelength of the pion. This gives rise to corrections to the QC that are suppressed as $e^{-m_{\pi}L}$, where $L$ denotes the spatial extent of a cubic volume~\cite{Luscher:1985dn}.

Two octet baryons can be in either a spin-singlet (${^1}S_0$) or a spin-triplet (coupled ${^3}S_1-{^3}D_1$) state. However, states in a finite cubic volume with periodic boundary conditions can not be characterized with well-defined angular-momentum quantum numbers. As a consequence, the FV QC mixes scattering amplitudes in all partial waves, preventing the extraction of scattering parameters. At low energies, however, only the lowest partial waves are expected to be significant, and the QC can be truncated to a finite space. Therefore, for two-baryon systems in a spin-singlet state, a simple algebraic relation enables the $S$-wave scattering phase shift, $\delta_S$, to be accessed from the FV energy eigenvalues at low energies,\footnote{The $S$-wave scattering amplitude in the spin-singlet two-baryon channel is
\begin{eqnarray}
\mathcal{M}_s = \frac{4\pi}{M_Bk^*}\frac{i}{\cot \delta_S -i}.
\label{eq:M-singlet}
\end{eqnarray}
}
\begin{eqnarray}
k^*\cot\delta_S=4 \pi
c_{00}^{\mathbf{d}}(k^{*2};  L).
\label{eq:QC}
\end{eqnarray}
Here, $k^*$ is the relative momentum of each baryon in the center-of-mass (CM) frame and $\mathbf{d}$ denotes the total CM momentum of the system in units of $2\pi/L$. $c_{lm}^{\mathbf{d}}(k^{*2};  L)$ is a kinematic function related to the three-dimensional zeta function, 
$\mathcal{Z}^\mathbf{d}_{lm}$,
\begin{eqnarray}
\hspace{1cm} 
c^\mathbf{d}_{lm}(k^{*2};  L)
& = & \frac{\sqrt{4\pi}}{\gamma L^3}\left(\frac{2\pi}{L}\right)^{l-2}\mathcal{Z}^\mathbf{d}_{lm}[1;(k^*  L/2\pi)^2],
\label{eq:clm}
\end{eqnarray}
where $\gamma = E/E^*$ is the relativistic gamma factor, with $E$ and $E^*$ denoting the total energy of two baryons in the lab and CM frames, respectively~\cite{Luscher:1986pf, Luscher:1990ux, Rummukainen:1995vs, Christ:2005gi,Kim:2005gf, Briceno:2013lba, Briceno:2013bda}. Further,
\begin{eqnarray}
\mathcal{Z}^\mathbf{d}_{lm}[s;x^2]
& = & \sum_{\mathbf{n}}\frac{|\mathbf{r}|^lY_{l,m}(\mathbf{r})}{(|\mathbf{r}|^2-x^2)^s},
\label{eq:Zlm}
\end{eqnarray}
where, for two baryons with equal masses, $\mathbf{r} = \frac{2\pi}{L} \hat{\gamma}^{-1} (\mathbf{n}-\frac{1}{2}\mathbf{d})$. $\mathbf{n}$ denotes a triplet of integers and $\hat{\gamma}^{-1}$ acting on a vector rescales the component of the vector parallel to the boost vector by $1/\gamma$ while leaving the perpendicular component intact. The zeta function in Eq.~(\ref{eq:Zlm}) can be numerically evaluated most efficiently using an equivalent exponential form~\cite{Luscher:1990ux, Beane:2011sc, Leskovec:2012gb}. For two-baryon systems at the energies considered below, the relativistic corrections due to the deviation of the $\gamma$ factor from unity are at the sub-percent level. As a result, for boost vectors whose components (in units of $2\pi/L$) are equal to each other modulo a factor of 2, the corresponding QCs in Eq.~(\ref{eq:QC}) are approximately the same. The other consequence of the proximity to the non-relativistic (NR) limit is that for boost vectors of the form $\mathbf{d}=(2n_1,2n_2,2n_3)$, with each $n_i$ being an integer, the leading contamination to the $S$-wave QC arises from nonvanishing $G$-wave interactions, which are expected to be suppressed relative to the $S$-wave interactions. With boost vectors that do not take the above form, the leading contamination arises from $D$-wave interactions~\cite{Briceno:2013lba}.

For two-baryon systems in a spin-triplet state at low energies, the physical mixing between $S$ and $D$ partial waves must be taken into account. A low-energy EFT of two-baryon interactions suggests that the $S$-$D$ mixing parameter, $\epsilon$, contributes to the low-energy expansion of the scattering amplitude at the same order as the effective range parameter, and may not be ignored~\cite{Briceno:2013bda}. In the \emph{Blatt-Biedenharn} parametrization of a coupled-channel scattering amplitude~\cite{Blatt:1952zza}, the mixing parameter has an analytic expansion in energy near the bound-state pole, and the scattering amplitude  exhibits a simple condition for the location of such a pole, $\cot \delta_{\alpha} = i$. Here, $\delta_{\alpha}$ is the counterpart of the $S$-wave phase shift of the \emph{barred} parametrization~\cite{Stapp:1956mz}, which has a small  $D$-wave admixture as well. The Blatt-Biedenharn parametrization will be adopted in this work for scattering in the spin-triplet channels.

The mixing parameter, $\epsilon$, adds an extra unknown to the QC in the spin-triplet channels. Constraining this parameter, as discussed in Ref.~\cite{Briceno:2013bda}, requires knowledge of the spectra of two-baryon systems with the total spin aligned both parallel and perpendicular to the boost vector, and with boost momenta that have at least one component equal to unity modulo $2$ (in units of $2\pi/L$). As not all distinct orientations of total spin with respect to the boost momenta are constructed in forming the correlation functions of spin-triplet systems in this work, the $\epsilon$ parameter cannot be constrained here for the spin-triplet channels. This also implies that for boost vector $\mathbf{d}=(0,0,1)$, the corrections to the QC from the s-d mixing might be significant at the order of low-energy EFT considered, and constraints on the $\alpha$-wave phase shift arising from a simple $\alpha$-wave QC may be contaminated. On the other hand, for boost vectors of the form $\mathbf{d}=(2n_1,2n_2,2n_3)$, with each $n_i$ being an integer, in particular for the $(0,0,0)$ and $(0,0,2)$ boost vectors that are considered in this work, the $\alpha$-wave QC,
\begin{eqnarray}
k^*\cot\delta_{\alpha}=4 \pi
c_{00}^{\mathbf{d}}(k^{*2};  L),
\label{eq:QC-alpha}
\end{eqnarray}
is exact up to corrections from $\beta$-wave interactions.\footnote{In the Blatt-Biedenharn parametrization, the spin-triplet coupled-channel scattering amplitude is
%s
\begin{eqnarray}
\mathcal{M}_{\alpha-\beta}=\frac{4\pi}{M_Bk^*}\left(
\begin{array}{cc}
 \cot \delta_{\alpha} \cos ^2\epsilon+\cot \delta_{\beta} \sin ^2\epsilon-i & \sin \epsilon \cos \epsilon ~ (\cot \delta_{\alpha}-\cot \delta_{\beta}) \\
 \sin \epsilon \cos \epsilon ~(\cot \delta_{\alpha}-\cot \delta_{\beta}) & \cot \delta_{\beta} \cos ^2\epsilon+\cot \delta_{\alpha} \sin ^2\epsilon-i \\
\end{array}
\right)^{-1}.
\label{eq:M-triplet}
\end{eqnarray}
} These corrections are subleading at the order in the EFT considered below and will therefore be neglected. Given that this QC is identical to the $S$-wave QC in the spin-singlet channels, the $s$ and $\alpha$ subscripts on the phase shifts will be suppressed in the rest of this paper, as their assignment should be clear from the channels under consideration.

Once the phase shifts are determined at several CM energies, a low-energy parametrization of the scattering amplitude as a function of energy, with only a few unknown parameters, can be constrained over a given range of energies. In the baryon-baryon channels well below the t-channel cut, the most common parametrization is the effective range expansion (ERE). For $S$-wave ($\alpha$-wave) interactions, the ERE is an expansion of the $k^*\cot \delta$ function in ${k^*}^2$,
\begin{eqnarray}
k^*\cot\delta=-\frac{1}{a}+\frac{1}{2}r{k^*}^2+P{k^*}^4+\dots,
\label{eq:ERE}
\end{eqnarray}
where $a$, $r$ and $P$ are the scattering length, effective range and the leading shape parameter, respectively. The ellipsis denotes terms that are higher order in the momentum expansion. L\"uscher's QC condition provides (up to exponentially small volume corrections and discretization effects) an exact constraint on the amplitude at corresponding energies regardless of the complexities present in the analytic structure of the amplitude below the inelastic thresholds. It is the output of the QC that allows the efficacy of given parametrizations of the amplitude to be assessed. For example, although the ERE is guaranteed to have a nonzero radius of convergence around $k^{*2}=0$, the convergence rate is not known a priori, and fits with higher order terms in the ERE may be needed. With numerical calculations for a range of momenta, the appropriateness of a given truncation of the ERE must be carefully tested.

L\"uscher's QC contains information about possible bound states in the system through an analytic continuation of the condition to negative energies. In particular, it is straightforward to show that for ${k^*}^2<0$, and for boost vectors of the type $\mathbf{d}=(2n_1,2n_2,2n_3)$,
\begin{eqnarray}
\left|k^*\right|&=&\kappa^{(\infty)}+\frac{Z^2}{L}\left[6e^{-\kappa^{(\infty)} L}+\frac{12}{\sqrt{2}}e^{-\sqrt{2}\kappa^{(\infty)} L}+\frac{8}{\sqrt{3}}e^{-\sqrt{3}\kappa^{(\infty)} L} \right]+\mathcal{O}\left(\frac{e^{-2\kappa^{(\infty)} L}}{L}\right),
\label{eq:extrapolation}
\end{eqnarray}
 in the NR limit~\cite{Konig:2011nz,Bour:2011ef,Davoudi:2011md,Briceno:2013bda}. Here, $\kappa^{(\infty)}$ is the infinite-volume binding momentum of the state and $Z^2$ is the residue of the scattering amplitude at the bound-state pole. 
 Note that the occurrence of negative ${k^*}^2$ values in a system in a finite volume  is not necessarily an indication of a bound state, and the movement of the state on the real energy axis must be examined as function of volume, according to the above form, to ascertain that the energy (shift) remains in the negative region towards infinite volume. Here, this will be referred to as a \emph{direct} method to obtain the binding energy. A crucial feature of calculations performed in this work is that two-baryon systems are studied at multiple volumes in order to provide unambiguous signatures for the existence of bound states once negative-valued energy shifts are observed. In particular, for the largest volume used, with a spatial extent of $\approx 6.7~{\tt{fm}}$, the FV corrections to the infinite-volume binding momenta are very small for the bound states in the $27$, $\overline{10}$ and $8_A$ irreps, see Sec.~\ref{subsec:results}. Since the closed form of the FV corrections to the binding momenta are known~\cite{Konig:2011nz,Bour:2011ef,Davoudi:2011md,Briceno:2013bda}, the significance of the terms that are dropped from the expansion in Eq.~(\ref{eq:extrapolation}) can be evaluated order by order.

Another method of obtaining information about a bound state is to first constrain the scattering amplitude and its parametrization in terms of energy using L\"uscher's methodology. An analytic continuation to negative energies then allows the bound state energy to be obtained from the pole location(s) of the scattering amplitude,
\begin{eqnarray}
\left. k^*\cot \delta\right|_{k^*=i\kappa^{(\infty)}}+\kappa^{(\infty)}=0.
\label{eq:pole}
\end{eqnarray}
Since this method involves an intermediate step to obtain the binding energies, it is referred to here as an \emph{indirect} method. The advantages of this method are that it makes no assumption about the suppression of higher-order exponentials in the extrapolation form as in Eq.~(\ref{eq:extrapolation}), and that it provides information about the existence or absence of a bound state even near threshold. The disadvantage of this method is that it relies on a parametrization of the scattering amplitude. Often, including additional parameters to improve the goodness of the fit increases the uncertainty of constraints on the location of the pole. Bound state(s) extracted this way must be shown to be robust against changes in the parameterization, and the scattering amplitude at the bound state energy must be shown to satisfy certain physical conditions. These features will become more apparent in Sec.~\ref{subsec:results}, where the determinations of the binding energies in the various baryon-baryon channels are discussed.

%%%%%%%%%%%%
\subsection{Two-baryon scattering with $SU(3)$ flavor symmetry and large-$N_c$ predictions
\label{subsec:SU(3)}
}
The number of distinct FV spectra in the baryon-baryon systems is dictated by the $SU(3)$ flavor symmetry of the present calculations. The flavor representation of two octet baryons, each transforming in the $8$ irrep of $SU(3)$, has a decomposition of the form:
\begin{eqnarray}
8 \otimes 8 = 27 \oplus 10 \oplus \overline{10} \oplus 8_S \oplus 8_A \oplus 1.
\label{eq:8times8}
\end{eqnarray}
Flavor channels belonging to the totally symmetric irreps $27$, $8_S$ and $1$ have a total spin equal to zero, while those belonging to the totally antisymmetric irreps $10$, $\overline{10}$ and $8_A$ have a total spin equal to one. The $SU(3)$ classification of the  flavor channels is summarized in Appendix~\ref{app:SU(3)} for reference. The use of interpolating operators that transform under irreps of the $SU(3)$ decomposition of the product of two octet baryons allows for these distinct spectra to be determined in a LQCD calculation. The two-baryon interpolating operators used in this study, however, transform under the isospin subgroup of $SU(3)$, with strangeness treated as a quantum number. As a result, the excited spectra corresponding to the $8_S$ and $1$ irreps cannot be rigorously determined unless multiple interpolating operators in flavor space are used to isolate the lowest-lying states of the systems. For example, to obtain the energy eigenvalues beyond the ground state in the $1$ irrep, a matrix of correlation functions in flavor space must be formed from interpolating operators corresponding to spin-singlet $\Lambda \Lambda$, $\frac{1}{\sqrt{2}}(\Xi^0n+\Xi^-p)$ and $\frac{1}{\sqrt{3}}(\Sigma^+\Sigma^-+\Sigma^0\Sigma^0+\Sigma^-\Sigma^+)$ states, see Fig.~\ref{fig:J1-irreps}. Since such a complete basis of operators was not used to form the correlation functions~\cite{Beane:2012vq}, direct constraints on scattering amplitudes in these two irreps could not be obtained.\footnote{The channel with the quantum numbers of $\Lambda \Lambda~({^1}S_0)$ in $S$-wave exhibits a somewhat deep bound state~\cite{Beane:2012vq}. As a result, there is a sufficiently large gap to the second-lowest energy level that even a single interpolating operator should obtain the ground-state energy correctly. This becomes more challenging for closely-spaced excited states that can only be constrained with multiple interpolating operators. A very deeply bound H-dibaryon in nature is conjectured to have significant cosmological consequences~\cite{Farrar:2002ic}.} On the other hand, the $27$, $\overline{10}$, $10$ and $8_A$ irreps each contain at least one flavor channel that does not suffer from mixing into other flavor channels. For example, $NN~({^1}S_0)$, $NN~({^3}S_1)$, $\Sigma^+ p~({^3}S_1)$ and $\frac{1}{\sqrt{2}}(\Xi^0n+\Xi^-p)~({^3}S_1)$ can be used as the interpolating operators to constrain the lowest-lying spectra of the $27$, $\overline{10}$, $10$ and $8_A$ irreps, respectively, as is evident from Figs.~\ref{fig:J0-irreps}-\ref{fig:J1-irreps}.

At low energies, the leading $S$-wave interactions of two octet baryons can be described by a Lagrange density\cite{Savage:1995kv} in a pionless EFT~\cite{Chen:1999tn} of the form,
\begin{eqnarray}
\mathcal{L}_{BB}^{(0)}&=&
-c_1 \text{Tr}(B_i^{\dagger}B_iB_j^{\dagger}B_j)
-c_2 \text{Tr}(B_i^{\dagger}B_jB_j^{\dagger}B_i)
-c_3 \text{Tr}(B_i^{\dagger}B_j^{\dagger}B_iB_j)
\nonumber
\\
&& -c_4 \text{Tr}(B_i^{\dagger}B_j^{\dagger}B_jB_i)
-c_5 \text{Tr}(B_i^{\dagger}B_i)\text{Tr}(B_j^{\dagger}B_j)
-c_6 \text{Tr}(B_i^{\dagger}B_j)\text{Tr}(B_j^{\dagger}B_i).
\label{eq:BB-Lagrangian}
\end{eqnarray}
Here, $B$ is the octet baryon matrix,
\begin{eqnarray}
B=\begin{bmatrix}
       \frac{\Sigma^0}{\sqrt{2}}+\frac{\Lambda}{\sqrt{6}} & \Sigma^+ & p           \\[0.3em]
       \Sigma^- & -\frac{\Sigma^0}{\sqrt{2}}+\frac{\Lambda}{\sqrt{6}}         & n \\[0.3em]
       \Xi^-           & \Xi^0 & -\sqrt{\frac{2}{3}}\Lambda
    \end{bmatrix},
\label{eq:octet-matrix}
\end{eqnarray}
where Roman indices on the B fields denote spin components. The Savage-Wise (SW) coefficients $c_1,\dots,c_6$ can be matched to scattering amplitudes at LO in a momentum expansion. For natural interactions, i.e., when the scattering length is comparable to the range of interactions, the relationships between the scattering lengths and the SW coefficients are presented in Ref.~\cite{Savage:1995kv} for various baryon-baryon channels. However, as is known in nature, and was deduced previously for the heavy quark masses of this work~\cite{Beane:2013br}, the $S$-wave interactions in both two-nucleon channels appear to be unnatural. The present investigation reconfirms the unnatural nature of interactions in the two-nucleon channels (belonging to the $27$ and $\overline{10}$ irreps) and further points to the similar feature in channels belonging to the $10$ and $8_A$ irreps. For unnatural $S$-wave interactions, the required power counting of the amplitude is produced in the Kaplan, Savage and Wise~\cite{Kaplan:1998tg,Kaplan:1998we} and van Kolck~\cite{vanKolck:1998bw} (KSW-vK) schemes. The relations of Ref.~\cite{Savage:1995kv} for unnatural scattering lengths in terms of $SU(3)$ coefficients $c_1,\dots,c_6$ become
\begin{eqnarray}
&&
\left[ -\frac{1}{a^{(27)}}+\mu \right]^{-1} = \frac{ M_B }{ 2\pi } \left(c_1-c_2+c_5-c_6\right),
\nonumber\\
&& \left[ -\frac{1}{a^{(\overline{10})}}+\mu \right]^{-1}  = \frac{ M_B }{ 2\pi }  \left(c_1+c_2+c_5+c_6\right),
\nonumber\\
&&\left[ -\frac{1}{a^{(10)}}+\mu \right]^{-1} = \frac{ M_B }{ 2\pi }  \left(-c_1-c_2+c_5+c_6\right),
\nonumber\\
&& \left[ -\frac{1}{a^{(8_A)}}+\mu \right]^{-1} = \frac{ M_B }{ 2\pi }  \left(\frac{3c_3}{2}+\frac{3c_4}{2}+c_5+c_6\right),
\nonumber\\
&& 
\left[ -\frac{1}{a^{(8_S)}}+\mu \right]^{-1} = \frac{ M_B }{ 2\pi }  \left(-\frac{2 c_1}{3} + \frac{2 c_2}{3} - \frac{5 c_3}{6} + \frac{5 c_4}{6} + c_5 - c_6\right),
\nonumber\\
\label{eq:SW-coeffs}
&& \left[ -\frac{1}{a^{(1)}}+\mu \right]^{-1} = \frac{ M_B }{ 2\pi }  \left(-\frac{c_1}3 + \frac{c_2}{3} - \frac{8 c_3}{3} + \frac{8 c_4}{3} + c_5 - c_6\right),
\end{eqnarray}
where $M_B$ denotes the baryon mass, and the $c_i$ coefficients on the right-hand side are evaluated at the renormalization scale $\mu$. For natural interactions, the renormalization scale $\mu$ is set equal to zero in the left-hand side of these equations, corresponding to a tree-level expansion of the scattering amplitude in these couplings.

The large-$N_c$ limit has interesting consequences and gives rise to further simplification of the interactions of two baryons~\cite{Kaplan:1995yg}. As argued in Ref.~\cite{Kaplan:1995yg}, in the limit of $SU(2)$ flavor symmetry, the interactions among two nucleons are invariant under a spin-flavor $SU(4)$ symmetry up to corrections that scale as $1/N_c^2$. Including the strange quarks and in the limit of $SU(3)$ flavor symmetry, interactions are invariant under an $SU(6)$ symmetry up to corrections that scale as $1/N_c$. Focussing on the latter case (which contains the former case as a subgroup), it can be shown that there are only two independent dimension-six $SU(6)$-symmetric interactions of two octet baryons, with coefficients $a$ and $b$.\footnote{The  $SU(6)$ coefficient ``$a$'' should not be confused with the scattering length. In the following sections, the scattering length carries a superscript denoting the irrep it corresponds to, while the $SU(6)$ coefficient $a$ is left as is. In a few cases where the subscripts on scattering lengths are omitted, these two letters can be distinguished from the context. Similarly, the $SU(6)$ coefficient ``$b$'' should not be confused with the lattice spacing.} These are expressed in terms of a baryon field that transforms as a three-index symmetric tensor under $SU(6)$~\cite{Kaplan:1995yg}. The corresponding coefficients $a$ and $b$ can, once again, be matched to the scattering amplitudes at LO in a momentum expansion. For unnaturally large scattering lengths, the $SU(3)$ relations in Eqs.~(\ref{eq:SW-coeffs}) become
%\footnote{The binding energy }
%
\begin{eqnarray}
\left[ -\frac{1}{a^{(27)}}+\mu \right]^{-1} = \frac{ M_B }{ 2\pi } (a-\frac{ b }{ 27 })+\mathcal{O}\left( \frac{1}{N_c^2} \right),~~~~\left[ -\frac{1}{a^{(\overline{10})}}+\mu \right]^{-1}  = \frac{ M_B }{ 2\pi } (a-\frac{ b }{ 27 })+\mathcal{O}\left( \frac{1}{N_c^2} \right),~~
\nonumber
\\
\left[ -\frac{1}{a^{(10)}}+\mu \right]^{-1} = \frac{ M_B }{ 2\pi } (a+\frac{ 7b }{ 27 })+\mathcal{O}\left( \frac{1}{N_c} \right),~~~~\left[ -\frac{1}{a^{(8_A)}}+\mu \right]^{-1} = \frac{ M_B }{ 2\pi } (a+\frac{ b }{ 27 })+\mathcal{O}\left( \frac{1}{N_c} \right),~~
\nonumber
\\
\label{eq:ab-coeffs}
\left[ -\frac{1}{a^{(8_S)}}+\mu \right]^{-1} = \frac{ M_B }{ 2\pi } (a+\frac{ b }{ 3 })+\mathcal{O}\left( \frac{1}{N_c} \right),~~~~\left[ -\frac{1}{a^{(1)}}+\mu \right]^{-1} = \frac{ M_B }{ 2\pi } (a-\frac{ b }{ 3 })+\mathcal{O}\left( \frac{1}{N_c} \right),~~~~~~
\end{eqnarray}
where the coefficients on the right-hand side are evaluated at the renormalization scale $\mu$. For natural interactions, $\mu$ is set equal to zero in the left-hand side of these equations, corresponding to a tree-level expansion of the amplitudes in these couplings. Note that the scattering lengths in channels belonging to the $27$ and $\overline{10}$ are the same up to $1/N_c^2$ corrections. Recalling that the $NN~({^1}S_0)$ and $NN~({^3}S_1)$ states belong to the $27$ and $\overline{10}$ irreps, respectively, this equality is a manifestation of the accidental $SU(4)$ Wigner symmetry \cite{Wigner:1936dx}, indicating that the spin-dependent $S$-wave $NN$ interaction vanishes in the large-$N_c$ limit~\cite{Kaplan:1995yg,Kaplan:1996rk}. Additionally, a larger accidental symmetry of two-baryon interactions can be realized in the limit where the $a$ coefficient is of $\mathcal{O}(1)$ or larger while the $b$ coefficient is of $\mathcal{O}(1)$ or smaller. In this case, the contributions from the $b$ coefficient to the amplitudes is suppressed relative to those of the $a$ coefficient through a numerical suppression observed in the $b$ terms in Eqs.~(\ref{eq:ab-coeffs}). This results in an $SU(16)$ symmetry of LO interactions, with only one coefficient, $a$, to be constrained.

Observation of $SU(6)$ spin-flavor symmetry and an accidental $SU(16)$ symmetry of nuclear and hypernuclear forces, although at an unphysical value of the quark masses, will be the first confirmation of the large-$N_c$ QCD predictions in hyperon-nucleon and hyperon-hyperon systems. Such investigation are presented in Sec.~\ref{subsec:results}. Identifying these spin-flavor symmetries, along with the theoretical estimate of their violation, provides important constraints on the hyperon-nucleon interactions that can be included in calculations of finite density systems. It is important to note that the calculations presented in this work exhibit an exact $SU(3)$ flavor symmetry, making the large-$N_c$ predictions above free of the $SU(3)$ breaking contaminations that are present in nature.

%%%%%%%%%%%%%%%%%%%%%%%%%%%%%%%%%%%%%%%%%%%
\section{Numerical calculations and results
\label{sec:numerical} 
}
\noindent
This section contains the main results of this paper. These include the scattering phase shifts and the constraints on the ERE parametrization of two-baryon channels belonging to the $27$, $\overline{10}$, $10$ and $8_A$ irreps of the $SU(3)$ decomposition of the product of two octet baryons, as well as an investigation of spin-flavor symmetries of interactions and a subsequent accidental symmetry predicted at large $N_c$. The main inputs to the scattering amplitude determinations are energy eigenvalues obtained from LQCD calculations of correlation functions. Some of these energies have been previously presented in Refs.~\cite{Beane:2012vq, Beane:2013br}. Here, multiple analyses are performed to determine the ground states and first excited states of two-baryon channels. The extracted energies are found to be consistent with our previous determinations.

%%%%%%%%%%%%
\subsection{Details of LQCD computations
\label{subsec:LQCD}
}
The ensembles of gauge-field configurations and the two-baryon correlation functions used in this work have previously been analyzed to obtain binding energies in two, three and four-baryon systems~\cite{Beane:2012vq}, as well as low-energy scattering phase shifts in two-nucleon systems~\cite{Beane:2013br}. Additionally, the same gauge-field configurations have been used to study the magnetic structure of light nuclei~\cite{Beane:2014ora, Chang:2015qxa, Detmold:2015daa} and some of the simplest reactions in the few-nucleon systems, such as the radiative capture process $np \to d \gamma$~\cite{Beane:2015yha} and single and double-$\beta$ decays~\cite{Savage:2016kon, Shanahan:2017bgi, Tiburzi:2017iux}. Details of the ensemble generation, as well as of the construction of the nuclear correlation functions, have been presented in those works, see for example Ref.~\cite{Beane:2012vq}. Here, some of the technical details are reviewed for completeness.

The gauge-field configurations were generated using a tadpole-improved L\"uscher-Weisz  gauge action \cite{Luscher:1984xn} and a clover action for fermions \cite{Sheikholeslami:1985ij}. The choice of stout smearing and the tadpole-improved clover coefficient used in generating the gauge configurations alleviate discretization effects to $\mathcal{O}(b^2)$, where $b$ denotes the lattice spacing. This spacing is determined, from $\Upsilon$ spectroscopy on these ensembles, to be $b=0.145(2)~{\tt{fm}}$, see Ref.~\cite{Beane:2012vq} and references therein. The physical spatial extents of these ensembles are approximately $3.4~{\tt{fm}}$, $4.5~{\tt{fm}}$ and $6.7~{\tt{fm}}$. Throughout this paper, these ensembles will be referred to as: $24^3 \times 48$, $32^3 \times 48$ and $48^3 \times 64$, respectively. The first three dimensions refer to the spatial extent of the hypercubic volume, $L$, while the last dimension refers to the temporal extent, $T$, both in lattice units (l.u.). The configurations are separated by ten Hybrid Monte Carlo evolution trajectories to reduce autocorrelations, with the total number of configurations used for each ensemble, $N_{\text{cfg}}$ given in Table~\ref{fig:ensembles}. An average of $N_{\text{src}}$ measurements are performed on each configuration. Various other properties of the ensembles are listed in Table~\ref{fig:ensembles}. Given the large values of $T$ and $L$ relative to the inverse pion mass, both the thermal contamination and the exponential  finite-volume contamination of single-hadron masses and two-baryon energies from pion propagation through the boundaries are strongly suppressed.

%
%%%%%%%%%
\begin{table}
\includegraphics[scale=1]{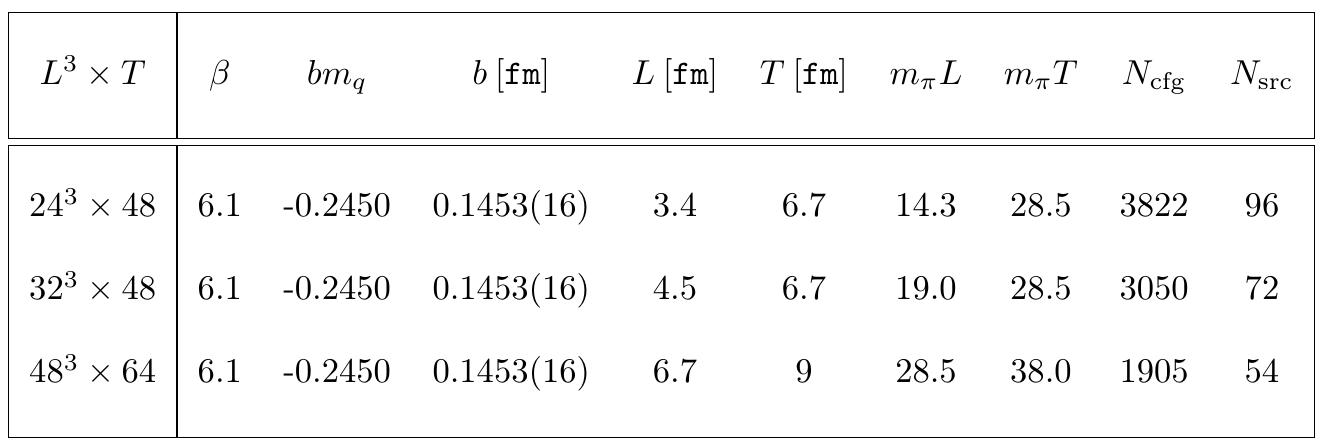}
\caption[.]{The parameters of the gauge-field ensembles used in this work. See Ref.~\cite{Beane:2012vq} for more details.}
\label{fig:ensembles}
\end{table}

Sources are smeared with a gauge-invariant Gaussian profile with stout-smeared gauge links. The quark propagators at the sink are either not smeared (smeared-point combination, SP) or are smeared with the same smearing profile as that of the source operators (smeared-smeared combination, SS). The plateau regions of the effective mass plots (EMPs) formed out of the SP and SS correlation functions are found consistent in every case. Propagators are contracted at the sink in blocks of three quarks to assemble a baryon field with given quantum numbers at the sink. In particular, the baryon blocks are projected to a fixed three-momentum, enabling the two-baryon interpolators at the sink to have either zero or non-zero CM momentum, with various possibilities for the momentum of each baryon. As the next step, a fully-antisymmetrized quark-level wavefunction with overall quantum numbers of the two-baryon system of interest is formed at the location of the source. The contraction step is defined by the selection of the appropriate indices from the baryon blocks at the source, in a way that is dictated by the quark-level wavefunction. More details regarding the contraction algorithm for a general $A$-nucleon system are presented in Ref.~\cite{Detmold:2012eu} (with a similar approach proposed in Refs.~\cite{Doi:2012xd,Gunther:2013xj}). The final products of the contraction step are two-baryon correlation functions as a function of Euclidean time. These correspond to a definite total momentum resulting from several (nearly orthogonal) choices of baryon momentum at the sink.

%%%%%%%%%%%%
\subsection{Analysis of correlation functions
\label{subsec:EMP}
}
To maximize confidence in the energy determinations and their uncertainties, five different analysis procedures were used, and the results obtained from each method were found to be consistent. The statistical and fitting systematic uncertainties on the final results are taken from one analysis, with an additional systematic uncertainty added to account for the small variations between the five analyses.

The correlation function of a single or two-baryon system, projected to the total momentum $2\pi\mathbf{d}/L$, can be written as
\begin{eqnarray}
C_{\hat{\mathcal{O}},\hat{\mathcal{O}}'}(\tau;\mathbf{d})&=&\sum_{\mathbf{x}} e^{2\pi i\mathbf{d} \cdot \mathbf{x} /L} \langle 0 | \hat{\mathcal{O}}'(\mathbf{x},\tau) \hat{\mathcal{O}}^{\dagger}(\mathbf{0},0) | 0\rangle=\mathcal{Z}'_0\mathcal{Z}_0^{\dagger}e^{-E^{(0)}\tau}+\mathcal{Z}'_1\mathcal{Z}_1^{\dagger}e^{-E^{(1)}\tau}+\dots,
\label{eq:Corr-funct}
\end{eqnarray}
where $\mathcal{Z}$ ($\mathcal{Z}'$) denotes the overlap of the interpolating operator $\hat{\mathcal{O}}$ ($\hat{\mathcal{O}}'$) onto the corresponding eigenstates of the system, with subscripts ``$0$'' and ``$1$'' referring to the ground state and the first excited state, respectively. $E^{(0)}$ and $E^{(1)}$ denote the ground and excited-state energies, respectively, and the ellipsis denotes contributions from additional higher-energy states. In principle, these correlation functions can be used to obtain the tower of energy eigenvalues of the system in a finite volume. In practice, a reliable determination of even the few lowest-lying energies is challenging. Since only a single source operator and two different sink operators were used for any given momentum configuration, it is not possible to use a Hermitian variational approach here. Nonetheless, given the exponential form in Eq.~(\ref{eq:Corr-funct}), it is clear that a linear combination of the two correlation functions can be used to remove the excited-state contamination of the lowest lying state at earlier times. Various realizations of this approach are the Matrix Prony~\cite{Beane:2009kya, Beane:2009gs} and the GPoF~\cite{Hua:1989, Sarkar:1995, Aubin:2010jc} methods. Alternatively, a correlated $\chi$-squared function can be formed to fit directly to single or two-exponential forms, with  the correlations both in time and between the different source and sink structures accounted for. As another alternative, the effective energy function defined as
\begin{eqnarray}
\mathcal{C}_{\hat{\mathcal{O}},\hat{\mathcal{O}}'}(\tau;\mathbf{d},\tau_J)\ = \ \frac{1}{\tau_J}\log\left[ \frac{C_{\hat{\mathcal{O}},\hat{\mathcal{O}}'}(\tau;\mathbf{d})}{C_{\hat{\mathcal{O}},\hat{\mathcal{O}}'}(\tau+\tau_J;\mathbf{d})} \right] ~ \stackrel{\tau \to \infty }{\longrightarrow} ~ E^{(0)},
\label{eq:EM}
\end{eqnarray}
can be fit to a constant at late times to obtain the ground-state energy, $E^{(0)}$. $\tau_J$ in Eq.~(\ref{eq:EM}) is a non-zero integer.
A detailed account of various analysis techniques employed herein has been presented in Ref.~\cite{Beane:2009kya}. While the results and the plots corresponding to a single analysis are presented below, a systematic uncertainty accounting for the small variation among the energies from different analyses is incorporated in the numbers that are reported.

\begin{figure}
\includegraphics[scale=0.6475]{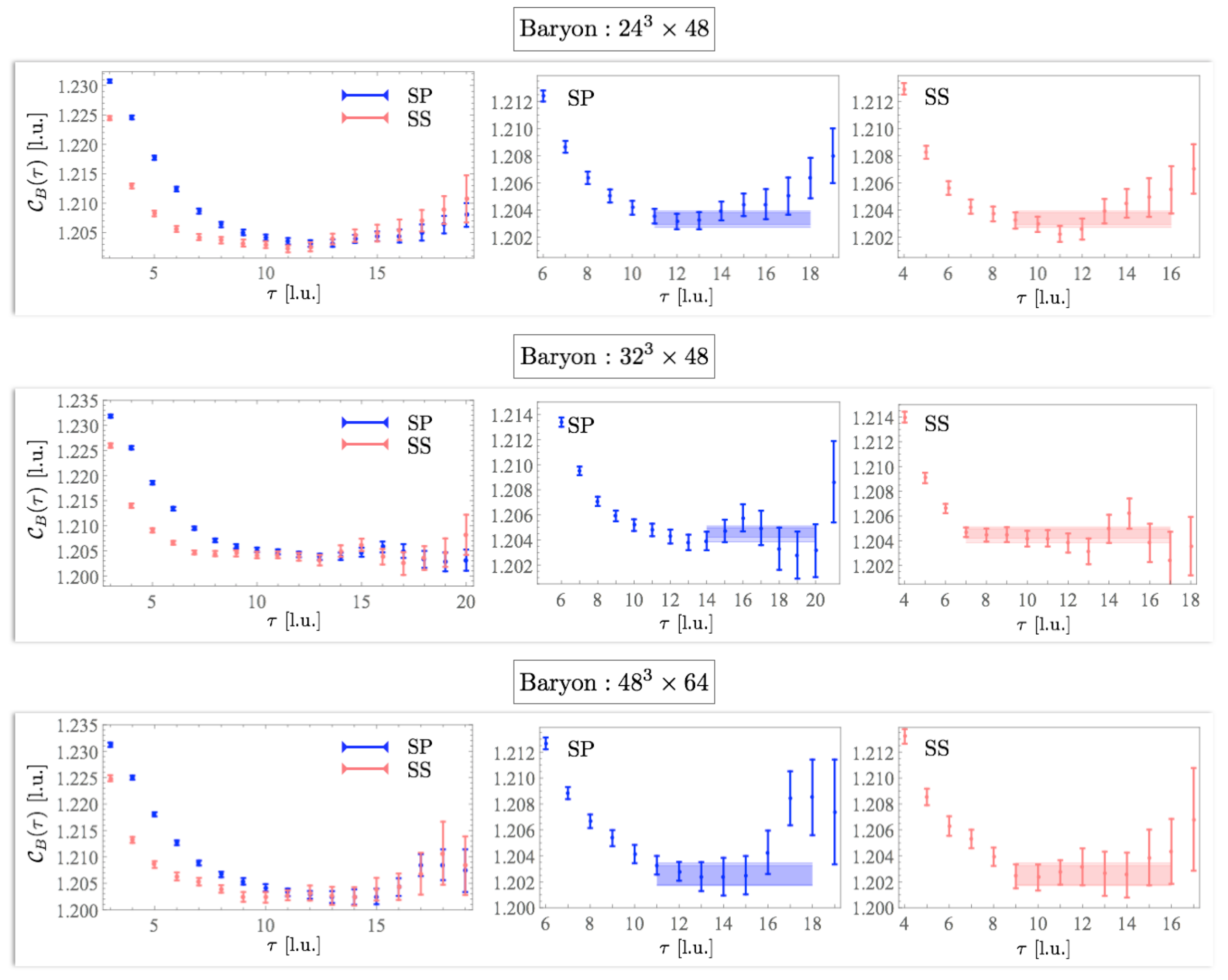}
\caption[.]{The single-baryon EMPs for the SP (blue) and SS (pink) source-sink combinations. The center and right panels present the same EMPs as in the left panel, rescaled to focus on the plateau region. The bands correspond to a correlated single-exponential fit to the SP and SS correlation functions, and obtain the mass of the baryon, $M_B$. The inner bands represent the statistical uncertainty of the fits, while the outer bands correspond to the statistical and systematic uncertainties combined in quadrature. The systematic uncertainty encompasses the variation of the fit window, as described in the text, with the longest time interval considered shown in the plots. The additional systematic resulting from multiple analyses is included in the bands. All quantities are expressed in lattice units (l.u.).}
\label{fig:EMP-baryon}
\end{figure}

\begin{figure}
\includegraphics[scale=0.745]{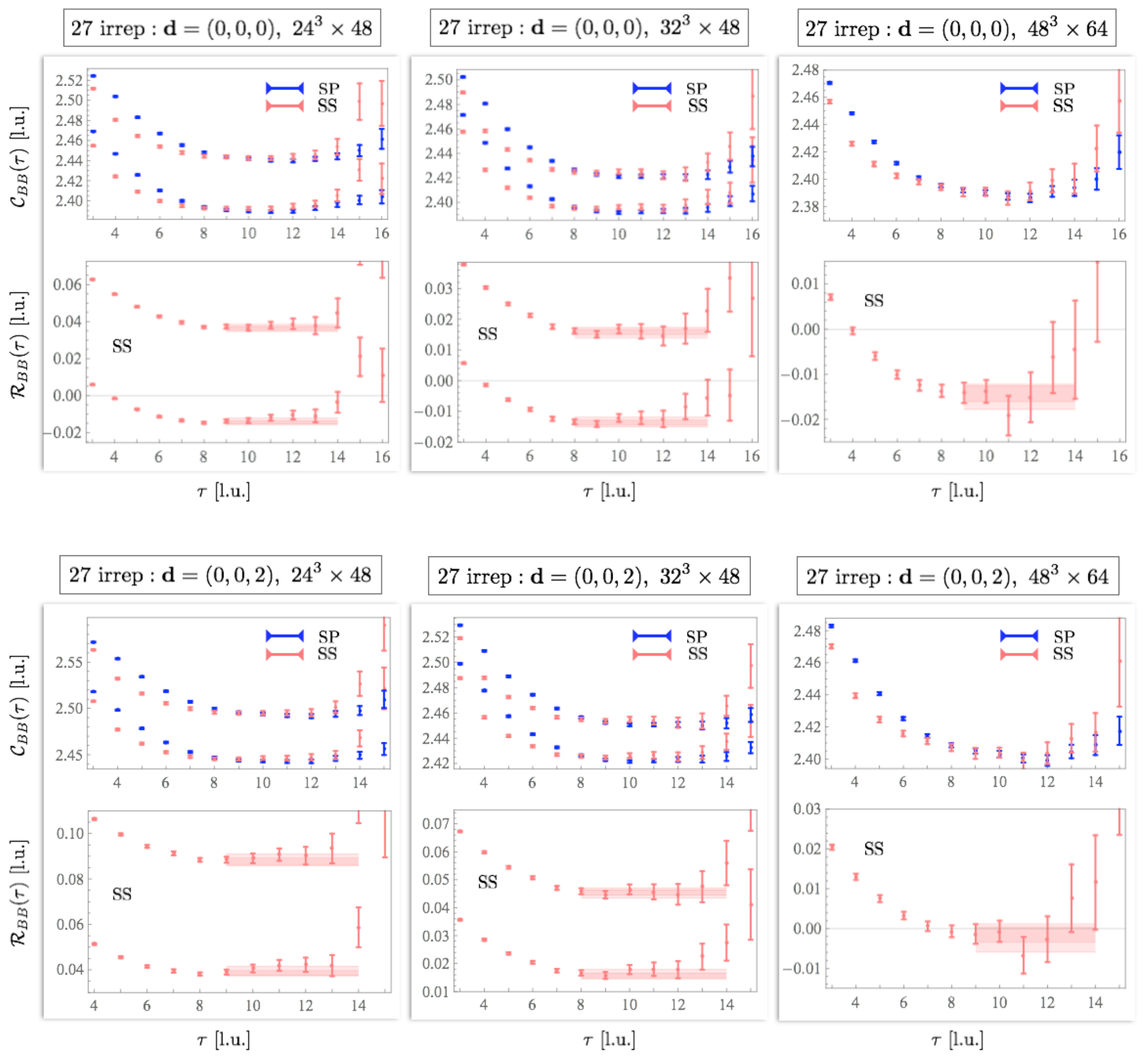}
\caption[.]{The EMPs of two baryons at rest (upper panel) and with $\mathbf{d}=(0,0,2)$ (lower panel) in the $27$ irrep for the SP (blue) and SS (pink) source-sink combinations (the upper panel of each segment), as well as the EMP (the lower panel of each segment) corresponding to the ratio of the SS two-baryon correlation function and the square of the SS single-baryon correlation function. The bands correspond to one-exponential fits to the SS/SS  correlation function ratios and obtain the energy shift $\overline{\Delta E}=E_{BB}-2M_B$. The inner bands represent the statistical uncertainty of the fits, while the outer bands correspond to the statistical and systematic uncertainties combined in quadrature. The systematic uncertainty encompasses the variation of the fit window, as described in the text, with the longest time interval considered shown in the plots. The additional systematic resulting from multiple analyses is included in the bands. All quantities are expressed in lattice units (l.u.).}
\label{fig:EMP-27}
\end{figure}

\begin{figure}[h!]
	\includegraphics[scale=0.745]{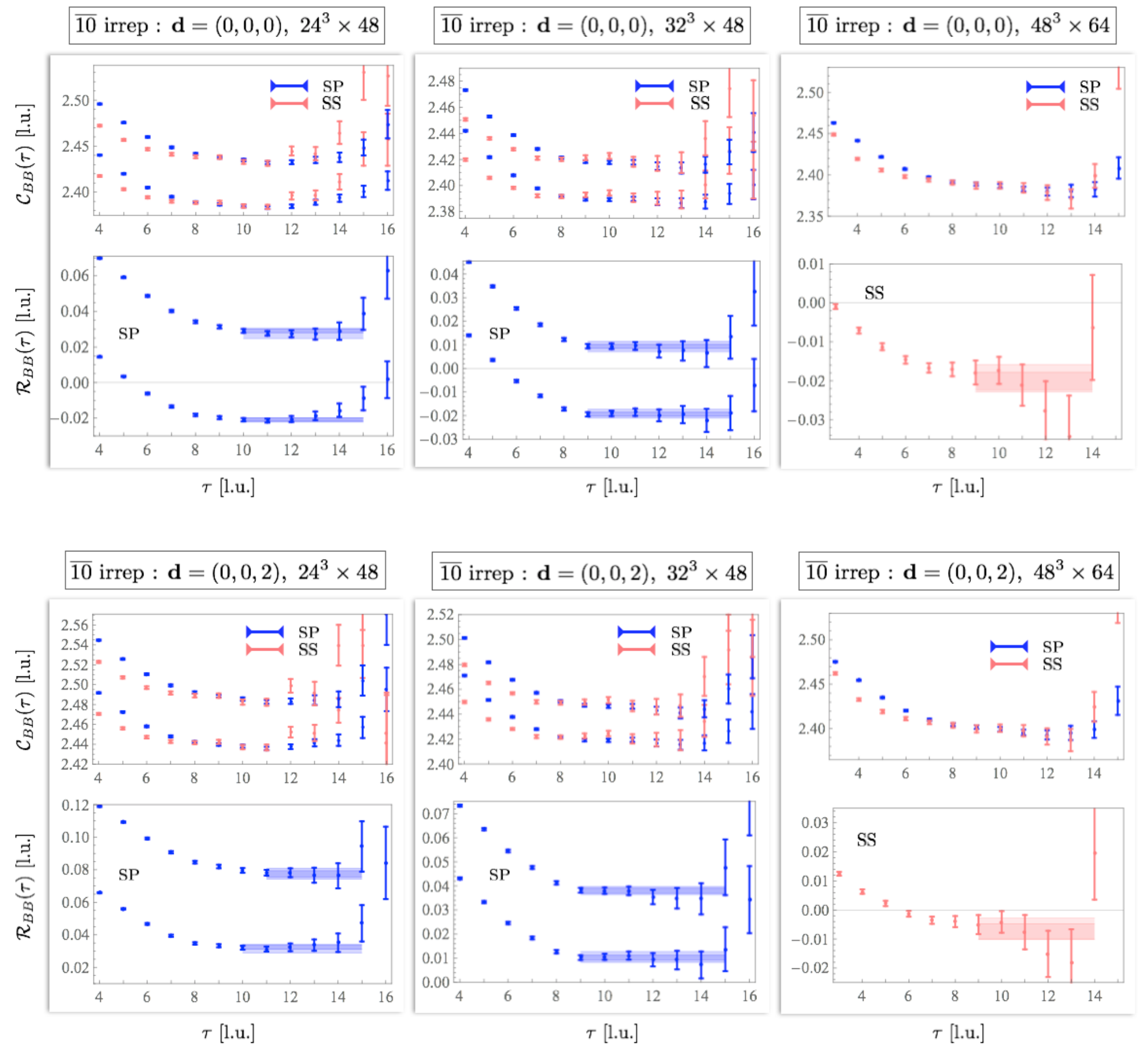}
	\caption[.]{The EMPs of two baryons at rest (upper panel) and with $\mathbf{d}=(0,0,2)$ (lower panel) in the $\overline{10}$ irrep for the SP (blue) and SS (pink) source-sink combinations (the upper panel of each segment), as well as the EMP (the lower panel of each segment) corresponding to the ratio of the two-baryon correlation function and the square of the single-baryon correlation function, the former with the SP (or SS as indicated) and the latter with the SS source-sink combinations. The bands correspond to one-exponential fits to the SP/SS (or SS/SS as indicated) ratios of correlation functions and obtain the energy shifts $\overline{\Delta E}=E_{BB}-2M_B$. See the caption of Fig.~\ref{fig:EMP-27} for more details.}
	\label{fig:EMP-10-bar}
\end{figure}

\begin{figure}[h!]
	\includegraphics[scale=0.745]{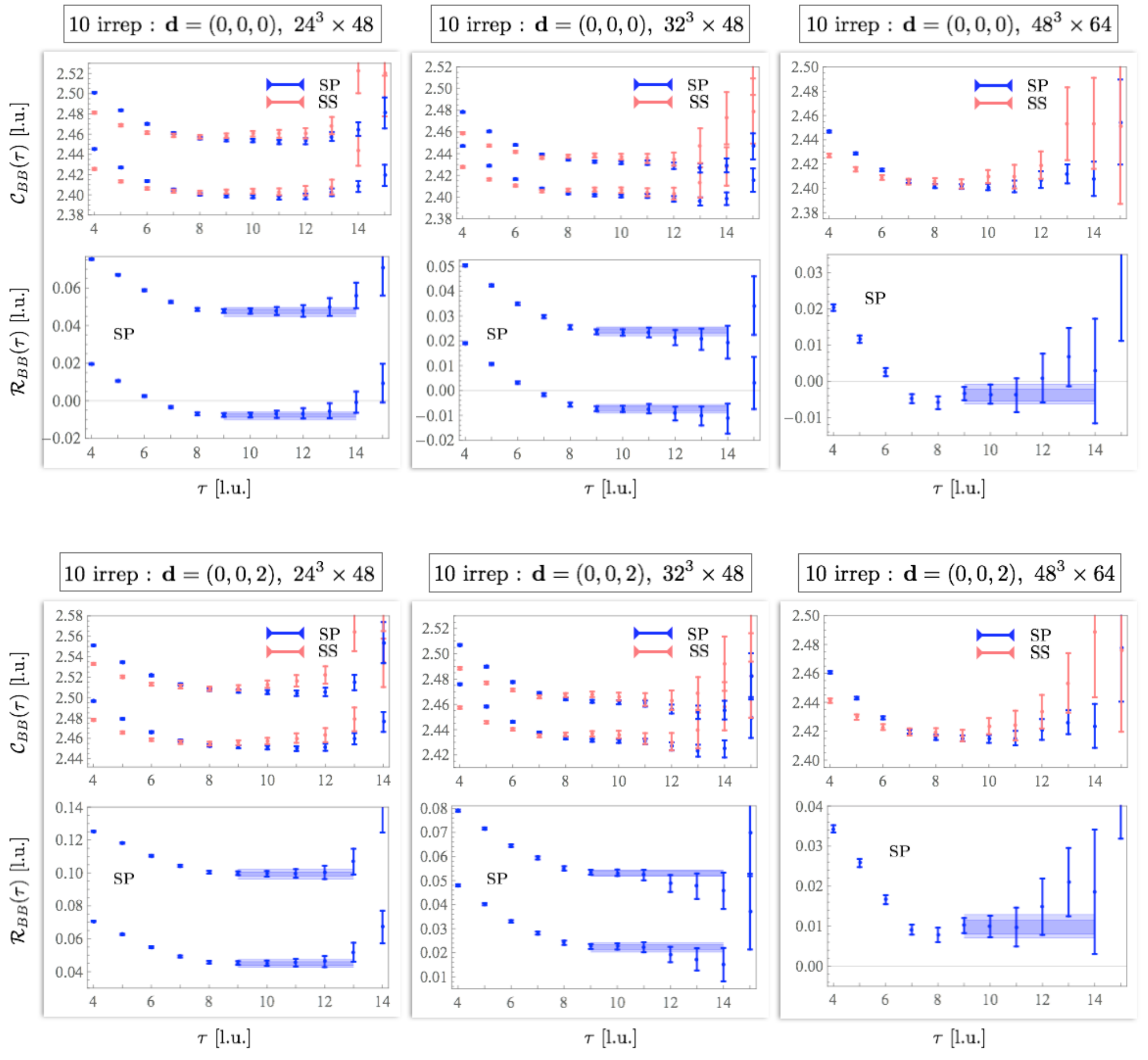}
	\caption[.]{The EMPs of two baryons at rest (upper panel) and with $\mathbf{d}=(0,0,2)$ (lower panel) in the $10$ irrep for the SP (blue) and SS (pink) source-sink combinations (the upper panel of each segment), as well as the EMP (the lower panel of each segment) corresponding to the ratio of the two-baryon correlation function and the square of the single-baryon correlation function, the former with the SP and the latter with the SS source-sink combinations. The bands correspond to one-exponential fits to the SP/SS  ratio of correlation functions and obtain the energy shifts $\overline{\Delta E}=E_{BB}-2M_B$. See the caption of Fig.~\ref{fig:EMP-27} for more details.}
	\label{fig:EMP-10}
\end{figure}

\begin{figure}[h!]
	\includegraphics[scale=0.745]{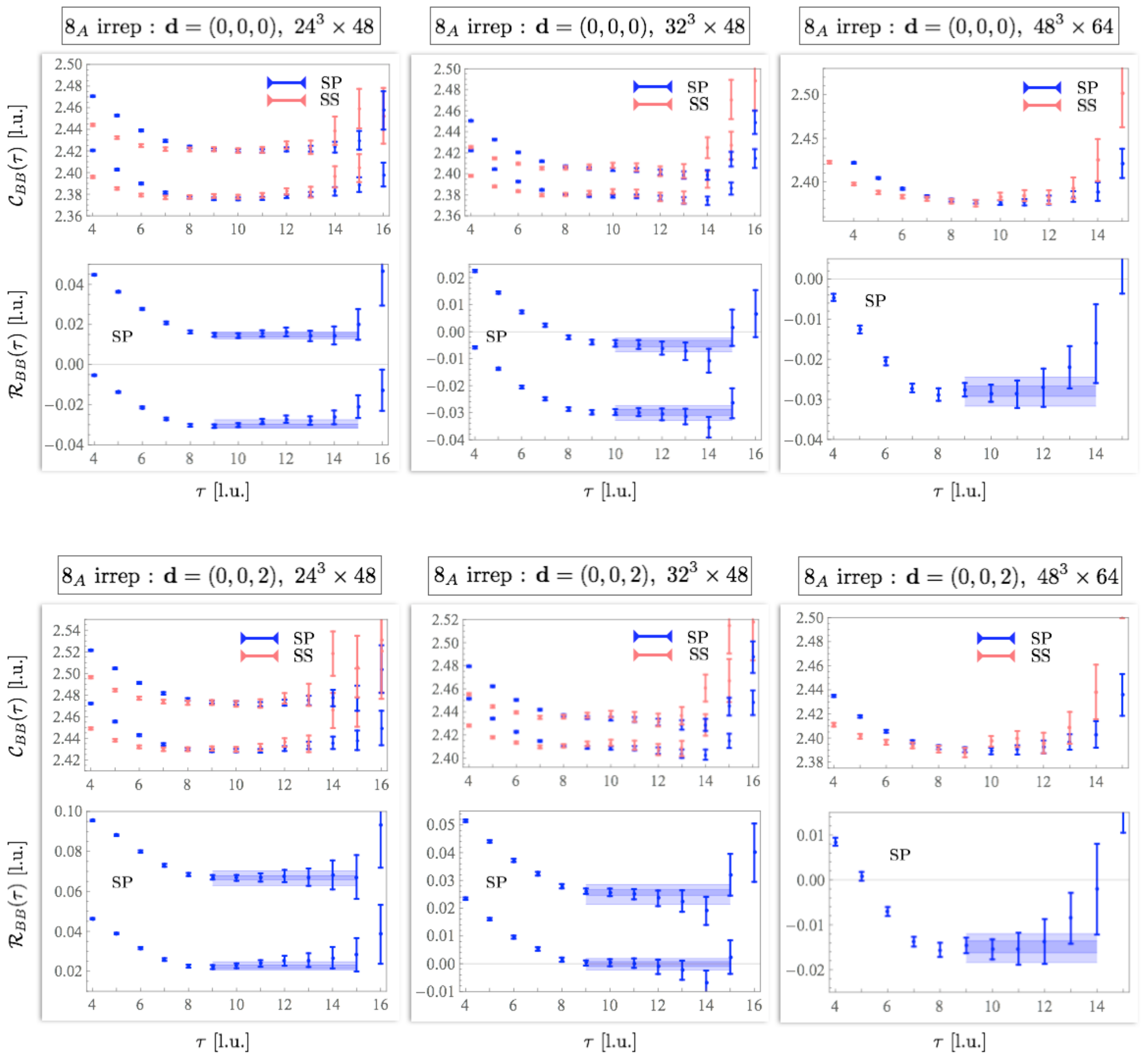}
	\caption[.]{The EMPs of two baryons at rest (upper panel) and with $\mathbf{d}=(0,0,2)$ (lower panel) in the $8_A$ irrep for the SP (blue) and SS (pink) source-sink combinations (the upper panel of each segment), as well as the EMP (the lower panel of each segment) corresponding to the ratio of the two-baryon correlation function and the square of the single-baryon correlation function, the former with the SP and the latter with the SS source-sink combinations. The bands correspond to one-exponential fits to the SP/SS ratios of correlation functions and obtain the energy shifts $\overline{\Delta E}=E_{BB}-2M_B$. See the caption of Fig.~\ref{fig:EMP-27} for more details.}
	\label{fig:EMP-8A}
\end{figure}

Statistical uncertainties were obtained for each analysis technique using bootstrap or jackknife procedures. The systematic uncertainty in each analysis includes a fitting uncertainty obtained by allowing the fit region to vary within an acceptable window. Representative fits obtained from the primary analysis are shown in the EMPs in Figs.~\ref{fig:EMP-baryon}-\ref{fig:EMP-8A}. These correspond to the largest fit intervals with a $\chi^2/\text{d.o.f} \sim 1$ in each case. 

%
%%%%%%%%%

The central values and uncertainties in the masses of the octet baryon obtained in this manner for all ensembles are overlaid with the SP (SS) EMPs in the center (right) panels of Fig.~\ref{fig:EMP-baryon}. Both the SP and SS EMPs are additionally shown at a larger scale in the left panels. The extracted values of the baryon mass at each volume are given in Table~\ref{tab:MB} in Appendix~\ref{app:tables}. The masses extracted for the three different ensembles agree within uncertainties, as expected from the large values of $m_\pi L$ in this calculation, and the infinite-volume value of the mass is taken to be the mass extracted for the largest ensemble, $M_B=1.2025(8)(3)$ in lattice units (l.u.).

The upper panels of each segment in Figs.~\ref{fig:EMP-27}--\ref{fig:EMP-8A} show the EMPs of the two-baryon systems for both the SP and SS correlation functions. The ground-state energy associated with each correlation function is determined as described above. However, the quantity that is of most interest in two-baryon channels is the shift in the energy of the system resulting from two-body interactions. The energy of two free baryons at rest, $2M_B$, can be subtracted from the two-baryon energies in a correlated manner to extract this small energy shift. Another approach that retains the correlations between the single and two-baryon correlations functions, thus reducing the statistical noise, is to form the ratio
\begin{eqnarray}
R(\tau;\mathbf{d})\ = \ \frac{C_{\hat{\mathcal{O}}_{BB},\hat{\mathcal{O}}'_{BB}}(\tau;\mathbf{d})}{\left[C_{\hat{\mathcal{O}}_B,\hat{\mathcal{O}}_B'}(\tau;\mathbf{0})\right]^2} 
=
\mathcal{A}_1e^{-(E^{(0)}_{BB}-2M_B)\tau}
\times \frac{ 1+\mathcal{A}_2e^{-(E^{(1)}_{BB}-E^{(0)}_{BB})\tau}+\dots}{ \left[1+\mathcal{A}_3e^{-(E^{(1)}_{B}-M_B)\tau}+\dots \right]^2}.
\label{eq:R}
\end{eqnarray}
Here, $\hat{\mathcal{O}}_B$ and $\hat{\mathcal{O}}_B'$ ($\hat{\mathcal{O}}_{BB}$ and $\hat{\mathcal{O}}_{BB}'$) are interpolating operators for the single(two)-baryon system and $\mathcal{A}_1$, $\mathcal{A}_2$ and $\mathcal{A}_3$ are known ratios of overlap factors of given states. At late times when the exponential factors in both the numerator and the denominator of the ratio on the right-hand side of Eq.~(\ref{eq:R}) are negligible compared with unity, a fit to a single exponential can be performed at large times, following the analysis steps described above, to obtain the energy shift $\overline{\Delta E} \equiv E^{(0)}_{BB}-2M_B$.  The effective energy-shift function associated with the ratio in Eq.~(\ref{eq:R}) can be defined as
\begin{eqnarray}
\mathcal{R}(\tau;\mathbf{d},\tau_J)\ = \ \frac{1}{\tau_J}\log\left[ \frac{R(\tau;\mathbf{d})}{R(\tau+\tau_J;\mathbf{d})} \right] ~ \stackrel{\tau \to \infty }{\longrightarrow} ~ \overline{\Delta E}.
\label{eq:calR}
\end{eqnarray}
Given the form of $R(\tau;\mathbf{d})$, flat behavior of $\mathcal{R}(\tau;\mathbf{d},\tau_J)$ in time is not a sufficient indicator that the function $R(\tau;\mathbf{d})$ is a single exponential. The values of overlap ratios in the numerator and the denominator in Eq.~(\ref{eq:R}) may conspire to give rise to flat behavior, despite neither the single-baryon nor the two-baryon systems being in their respective ground states. As a result, in fitting the quantity $R(\tau;\mathbf{d})$, none of the fit intervals must begin earlier than the beginning of the single-exponential regions in the single-baryon and two-baryon EMPs.
%
%%%%%%%%%
%

%
%%%%%%%%%
%
%
%%%%%%%%%
%
%
%%%%%%%%%

In principle, two-baryon correlation functions contain spectral information beyond ground-state energies.  Although this study did not use a large basis of operators, physical intuition regarding the differing nature of bound and scattering states of a two-baryon system suggested constructing not only the two-baryon operators that interpolate to two baryons at rest or in motion with equal velocity, but also those that interpolate to two baryons with relative back-to-back momenta. While the former can have significant overlap onto a compact state in a finite volume (corresponding to a bound state in infinite volume), they are not optimal interpolators for states corresponding to the scattering states of infinite volume. This results in correlation functions that are dominated by the ground state after a short time interval. On the other hand, interpolators with back-to-back momenta appear to predominantly overlap with states with positive energy shifts in the volume, and are almost orthogonal to the operators of the first type. The quality of plateaus in the EMPs with both types of interpolators was found to be comparable, suggesting that each set primarily overlaps onto one state and not the other. This allows the first excited states of the two-baryon systems to be extracted using the simplest back-to-back momentum configurations for baryons. The only exception is for the  $48^3 \times 64$ ensemble, where the splitting between the energy levels of the systems is small (being comparable to the uncertainties in the energies) and it can not be established that the first excited state is only minimally mixed into the nearby ground state. As a result, while for the smaller volumes two energy levels are extracted, for the $48^3 \times 64$ ensemble only the ground-state energies are reported.

The upper panels in each segment in Figs.~\ref{fig:EMP-27}-\ref{fig:EMP-8A} include not only the lowest-lying state, but also the second lowest-lying state of the two-baryon systems obtained from the correlation functions with back-to-back momenta. The lower panels of each segment show EMPs corresponding to the quantity $\mathcal{R}$ for the SS or SP (depending on the channel) correlation functions, as defined in Eq.~(\ref{eq:calR}). The same quantity can be constructed for the correlation functions that project to the first excited state, with $\overline{\Delta E}=E^{(1)}_{BB}-2M_B$. In the $27$ irrep, both the fit to the SP correlation function and a correlated fit to both the SP and SS correlation functions exhibited consistent plateaus, but the fit to the SS/SS correlation function ratio was found most precise. Similarly, in other irreps, fits to the SP/SS and SS/SS correlation function ratios, as well as a correlated fit to both of these ratios, were performed and the fit corresponding to the least uncertainty was selected, as is indicated in Figs.~\ref{fig:EMP-10-bar}--\ref{fig:EMP-8A}. The energy shifts and their uncertainties are denoted as horizontal bands in the $\mathcal{R}$ plots, and are compiled for all two-baryon channels studied in this work in Fig.~\ref{fig:energy-shifts}. The corresponding values are tabulated in Tables~\ref{tab:27}-\ref{tab:8A} of Appendix~\ref{app:tables} for reference.

%
%%%%%%%%%
\begin{figure}
\includegraphics[scale=0.705]{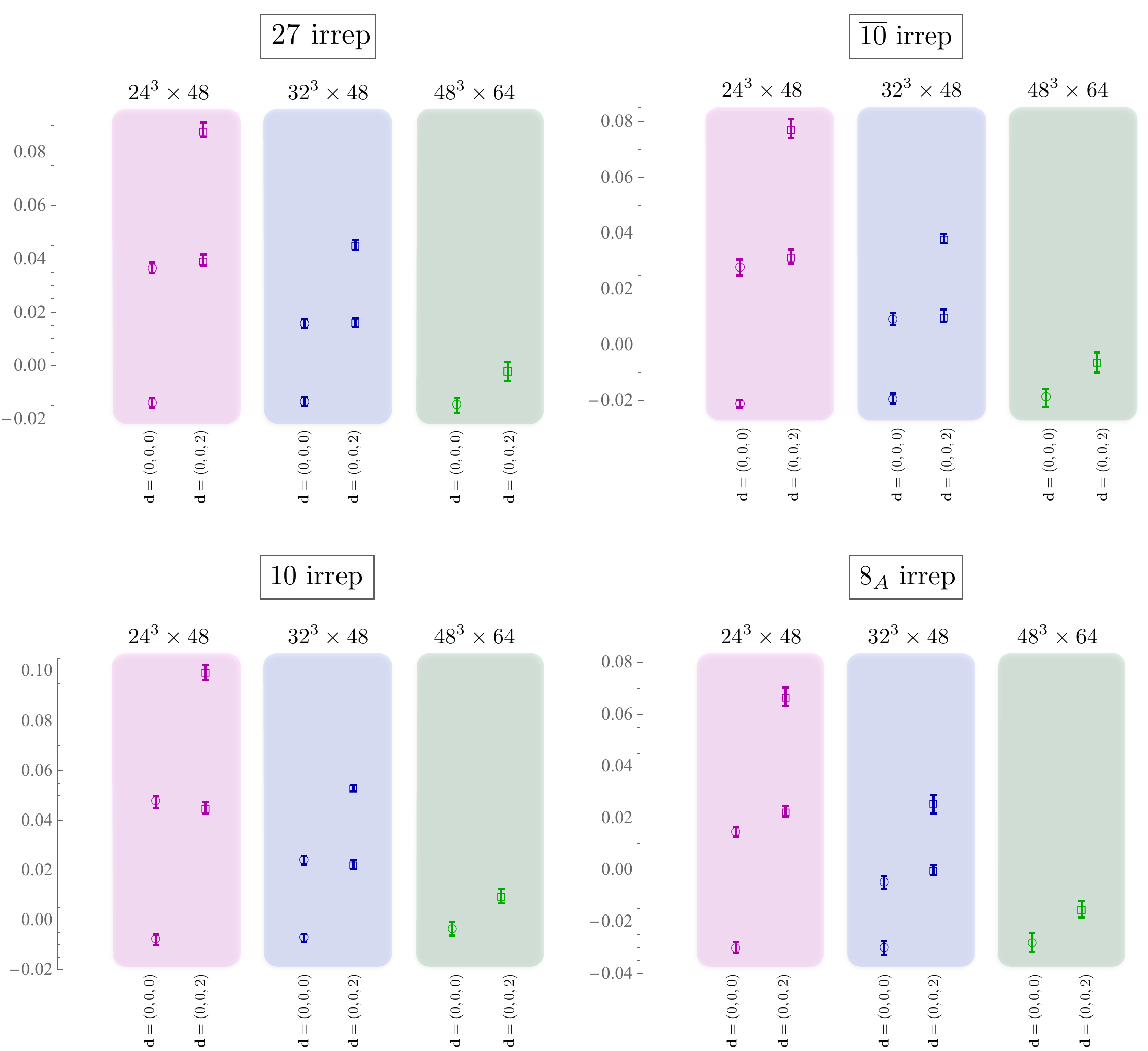}
\caption[.]{The shifts in the energy of the two-baryon systems in the $27$, $\overline{10}$, $10$ and $8_A$ irreps from that of two non-interacting baryons at rest in the three lattice volumes, i.e., $\overline{\Delta E}=E_{\text{BB}}-2M_{\text{B}}$. Energies are expressed in lattice units (l.u.). Different columns correspond to different volumes and boosts, as indicated.
}
\label{fig:energy-shifts}
\end{figure}

Recently, there have been comments by Iritani, et~al.~\cite{Iritani:2015dhu,Iritani:2016jie,Iritani:2016xmx} questioning the extraction of energy eigenvalues from the late-time behavior of correlation functions, and methods for identification of energies such as those used here. These authors present an example of two-nucleon correlation functions that exhibit a considerable mismatch in the location of the naive plateaus in the EMPs when different source and sink operators are used (namely locally-smeared and wall sources). However, as is shown by the PACS-CS collaboration~\cite{Yamazaki:2017euu}, such a mismatch disappears once both the single-nucleon and the two-nucleon systems are required to be in their ground states. The failure of wall sources to overlap well onto the ground state at early times is a well-known problem, and has no bearing on the results reported by other groups using more optimal sources, such as those used in this work. Indeed, the quality of plateaus in the two-baryon systems are comparable to those of the single-nucleon system in the present study, demonstrating that the ground state (and the first excited state) of these systems can be obtained efficiently, with the results from two different source-sink combinations being fully consistent.

\begin{figure}[t!]
\includegraphics[width=0.89\textwidth]{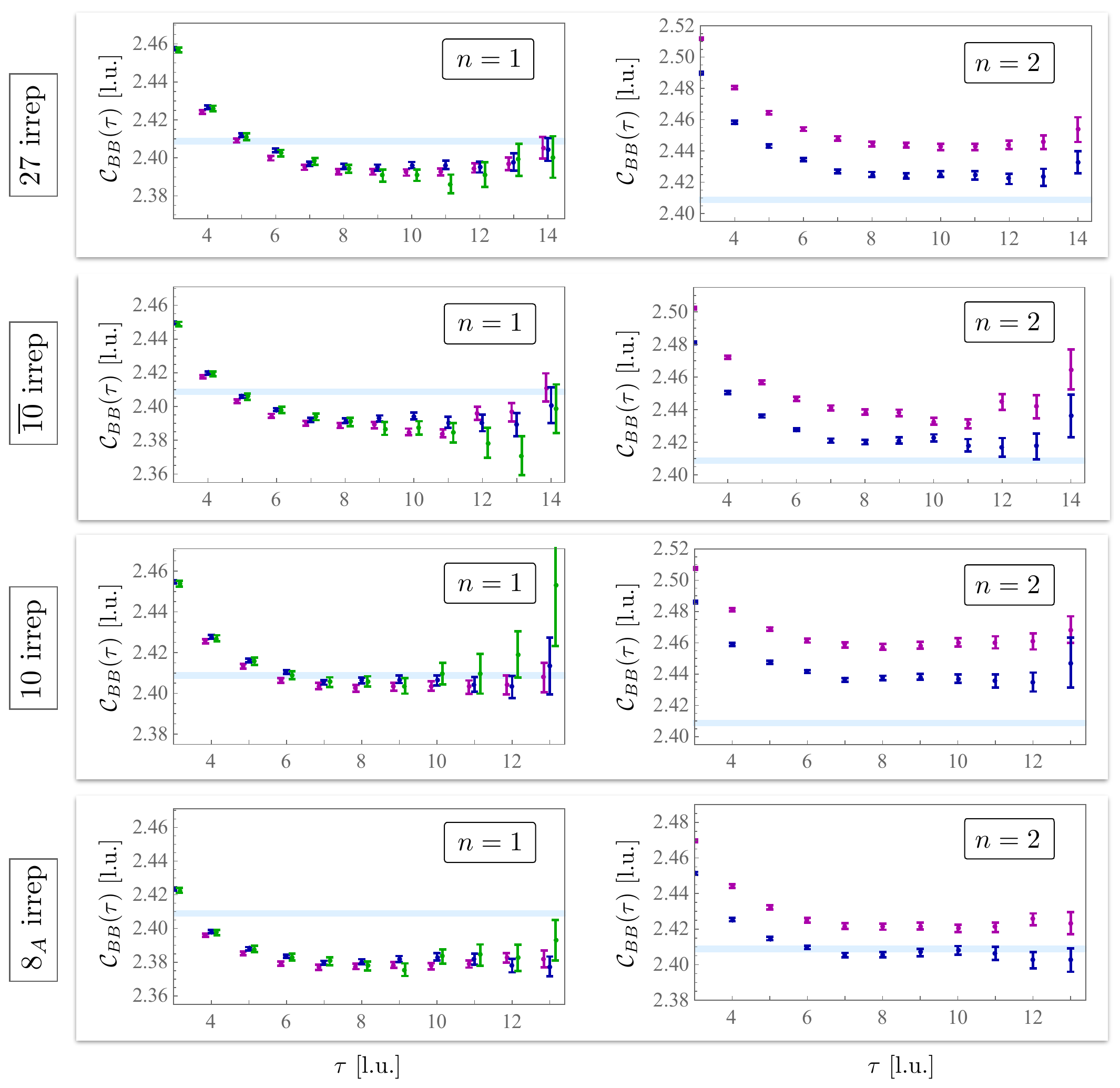}
\caption{A comparison of the SS EMPs of two-baryon channels at rest belonging to the four irreps, $27$, $\overline{10}$, $10$ and $8_A$, for the lowest-lying states ($n=1$) in the lattice volumes $L=24~\text{l.u.}$ (dark magenta), $L=32~\text{l.u.}$ (dark blue) and $L=48~\text{l.u.}$ (green) in the left panels, and for the second lowest-lying states ($n=2$) in the lattice volumes $L=24~\text{l.u.}$ (dark magenta) and $L=32~\text{l.u.}$ (dark blue) in the right panels. The points from different volumes in the panels on the left have been slightly shifted in the time direction for display purposes. The light-blue band corresponds to twice the mass of the baryon and its uncertainty in the $32^3 \times 48$ ensemble. Quantities are expressed in lattice units (l.u.).}
\label{fig:volumecomp}
\end{figure}
Another argument to consider when assessing the claims by Iritani {\it et al.} regarding the occurrence of so-called ``mirage plateaus'' in two-baryon systems  follows from observations of the volume dependence of the correlation functions. Fig.~\ref{fig:volumecomp} shows the EMPs in each of the two-baryon channels studied in this work in the three different lattice volumes. Volume dependence is clearly visible in states identified as scattering states. No significant volume dependence is observed for the lowest-lying state, strongly supporting the hypothesis that the ground state in these channels is a bound state. If the plateaus observed for the lowest-lying state are to be identified as ``mirages'' (that is, if the systems do not exhibit bound ground states), such fake plateaus could only result from cancellations between the FV states above the two-baryon threshold that contribute to the correlation function with opposites signs, and whose contributions depend upon the source structures. The spectrum of these states changes rapidly with power-law scaling as the volume is increased, in contrast with an exponential scaling for a compact state. In order for the ``mirage plateaus'' to be nearly coincident over the large range of volumes considered here, $V=39$~--~300~{\tt fm}$^3$, the linear combination of states would also have to change very rapidly, and in a finely-tuned manner, in order to keep the plateau regions approximately volume independent. As the employed sources are volume independent and compact on the scale of all the spatial volumes, such behavior is exceedingly unlikely. There is no indication that the values of the ground-state energies in the two-baryon correlation functions scale as a power-law with the volume, and any multi-level model of these correlation functions with the exclusion of one or more possible bound state(s) fails to reproduce the behavior shown in Fig.~\ref{fig:volumecomp}.

In summary, the ``mirage plateau'' issue posed by Iritani~{\it et al.}~\cite{Iritani:2017rlk} appears to be irrelevant to the calculations presented here. The results of the present work are consistent with the correlation functions in each of two-baryon channels relaxing into a bound state at late times, with their binding energies determined in the next section. Iritani~{\it et al.} additionally question the validity of the scattering amplitudes arising from these spectral studies, but again these claims  have no bearing on the current results as is shown in Appendix~\ref{app:checks} (see also Ref.~\cite{Beane:2017edf}, where a coherent rebuttal of Ref.~\cite{Iritani:2017rlk} is presented).

%%%%%%%%%%%%
\subsection{Results and discussions
\label{subsec:results}
}
In this section, the results for the LQCD spectra will be used to: \emph{1)} obtain the $S$-wave\footnote{The term $S$-wave is collectively used to refer to $S$-wave in spin-singlet channels and $\alpha$-wave in spin-triplet channels.} scattering amplitudes, explicitly the $k^*\cot \delta$ function, at low energies, \emph{2)} constrain the ERE parametrization of the scattering amplitudes, \emph{3)} determine bound states and their binding energies, \emph{4)} examine the naturalness of $S$-wave baryon-baryon interactions, and \emph{5)} provide constraints on the leading $SU(3)$-symmetric interactions and well as the leading $SU(6)$-symmetric interactions in the limit of large $N_c$.

%%%%%%%%%%%%
\subsubsection{$k^* \cot \delta$ function
\label{subsubsec:kcotd}
}
Given the ten FV energy eigenvalues determined in the previous section for each two-baryon channel, each scattering amplitude can be constrained at ten kinematic points via L\"uscher's QCs, Eqs.~(\ref{eq:QC}) and~(\ref{eq:QC-alpha}). In the NR limit, the CM energy eigenvalues corresponding to a two-baryon system at rest must be identical to that of the system in motion with two units of momentum in one Cartesian direction (the direction of total spin in a spin-triplet system)~\cite{Briceno:2013lba}. Therefore, two sets of energy eigenvalues obtained from $\mathbf{d}=(0,0,0)$ and $\mathbf{d}=(0,0,2)$ measurements on the same ensemble do not provide constraints on scattering amplitude at distinct kinematic points. Nonetheless, given that these are obtained from separate sets of measurements (they are different Fourier projections of correlation functions with the same interpolating operators), including both sets in the analysis leads to better constraints on  the scattering parameters and the binding energies.

%
%%%%%%%%%
\begin{figure}
\includegraphics[scale=0.2975]{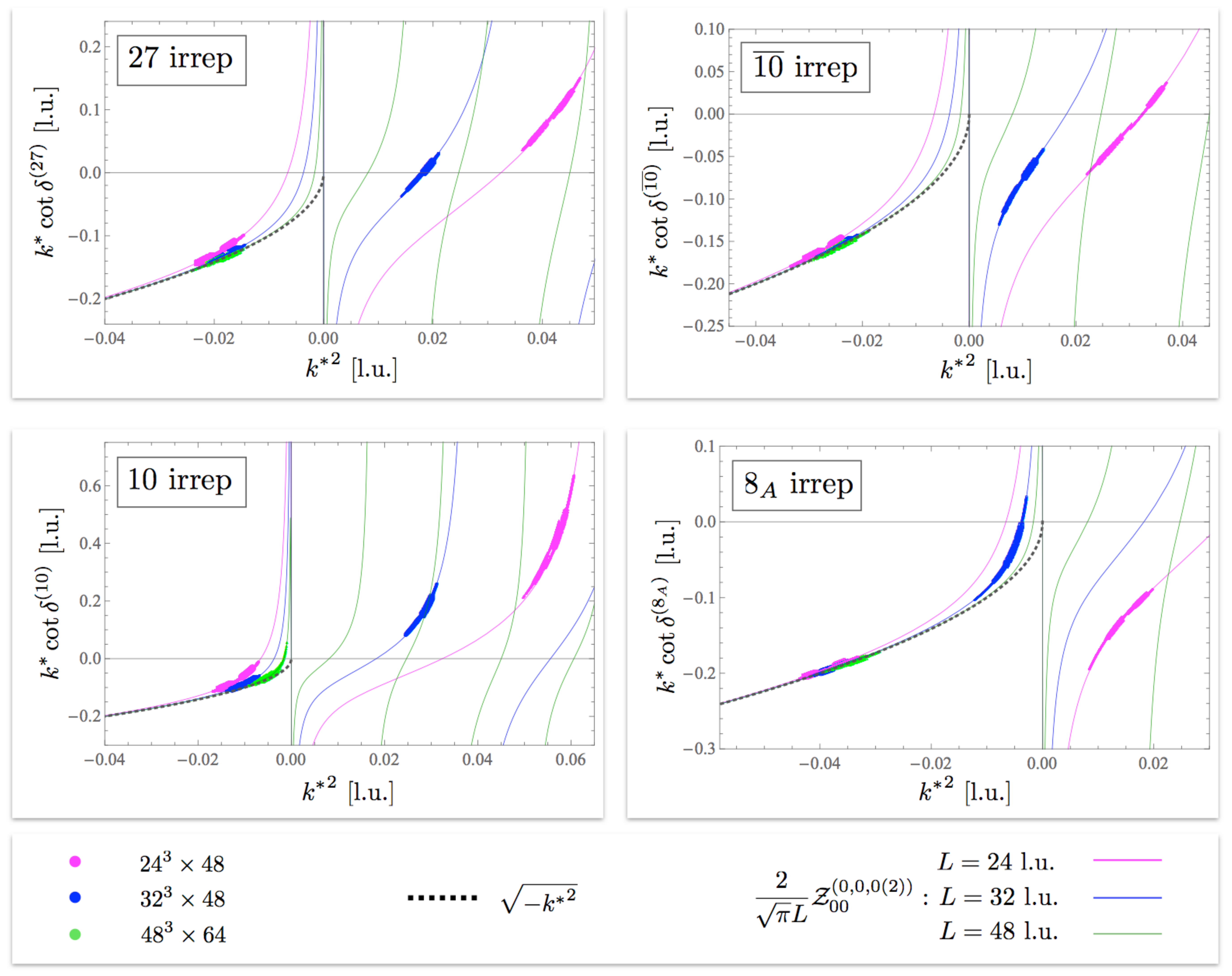}
\caption[.]{$k^*\cot \delta$ values in the two-baryon channels belonging to the four irreps $27$, $\overline{10}$, $10$ and $8_A$, obtained by solving Eq.~(\ref{eq:QC}) at the corresponding values of the square of the CM momentum of the two baryons, ${k^*}^2$.  The functions on the left-hand side of this equation for $l=m=0$, i.e., $\frac{2}{\sqrt{\pi}L}\mathcal{Z}^\mathbf{d}_{00}[1;(k^* L/2\pi)^2]$, are also shown at the corresponding volumes and CM boost momenta. The thick points cover the statistical uncertainty in the results, while the thin points cover the statistical and systematic uncertainties combined in quadrature. Quantities are expressed in lattice units (l.u.).}
\label{fig:Z-functions}
\end{figure}

The $S$-wave scattering amplitude of the two-baryon channels with $\mathbf{d}=(0,0,0)$ and $\mathbf{d}=(0,0,2)$ belonging to the $27$ irrep, e.g. $NN~({^1}S_0)$, is parametrized by a single phase shift, whose value can be constrained at a given CM momentum using the QC in Eq.~(\ref{eq:QC}), up to contaminations from $G$-wave interactions that are neglected. The resulting $k^* \cot \delta$ function is shown in Fig.~\ref{fig:Z-functions} for the ten energy eigenvalues obtained in the previous section. The figure includes the corresponding $\frac{2}{\sqrt{\pi}L}\mathcal{Z}^\mathbf{d}_{00}[1;(k^* L/2\pi)^2]$ functions from which $k^* \cot \delta$ is obtained, see Eqs.~(\ref{eq:QC}) and~(\ref{eq:clm}) with $l=m=0$. The $(-\sqrt{-{k^*}^2})$ function, whose intersection with $k^* \cot \delta$ determines the location of the bound state pole in the amplitude (see Eq.~(\ref{eq:M-singlet})), is also shown in Fig.~\ref{fig:Z-functions}.

For the spin-triplet channels $NN~({^3}S_1)$, $N\Sigma~({^3}S_1)$ and $\frac{1}{\sqrt{2}}(\Xi^0n+\Xi^-p)~({^3}S_1)$ associated with the $\overline{10}$, $10$ and $8_A$ irreps, respectively, additional mixing into the $D$-wave in anticipated.
 As discussed in Sec.~\ref{subsec:FV}, in the Blatt-Biedenharn parametrization, and with the boost vectors $\mathbf{d}=(0,0,0)$ and $\mathbf{d}=(0,0,2)$, the $\alpha$-wave phase shift can be constrained at a given CM momentum using the QC in Eq.~(\ref{eq:QC-alpha}), up to negligible contaminations from $\beta$-wave interactions. The resulting $k^* \cot \delta$ functions are plotted in Fig.~\ref{fig:Z-functions} for the ten energy eigenvalues obtained in the previous section in each of these channels.

%%%%%%%%%%%%
\subsubsection{Effective range expansion parameters 
\label{subsubsec:ERE}
}
\begin{figure}[b!]
\includegraphics[scale=0.615]{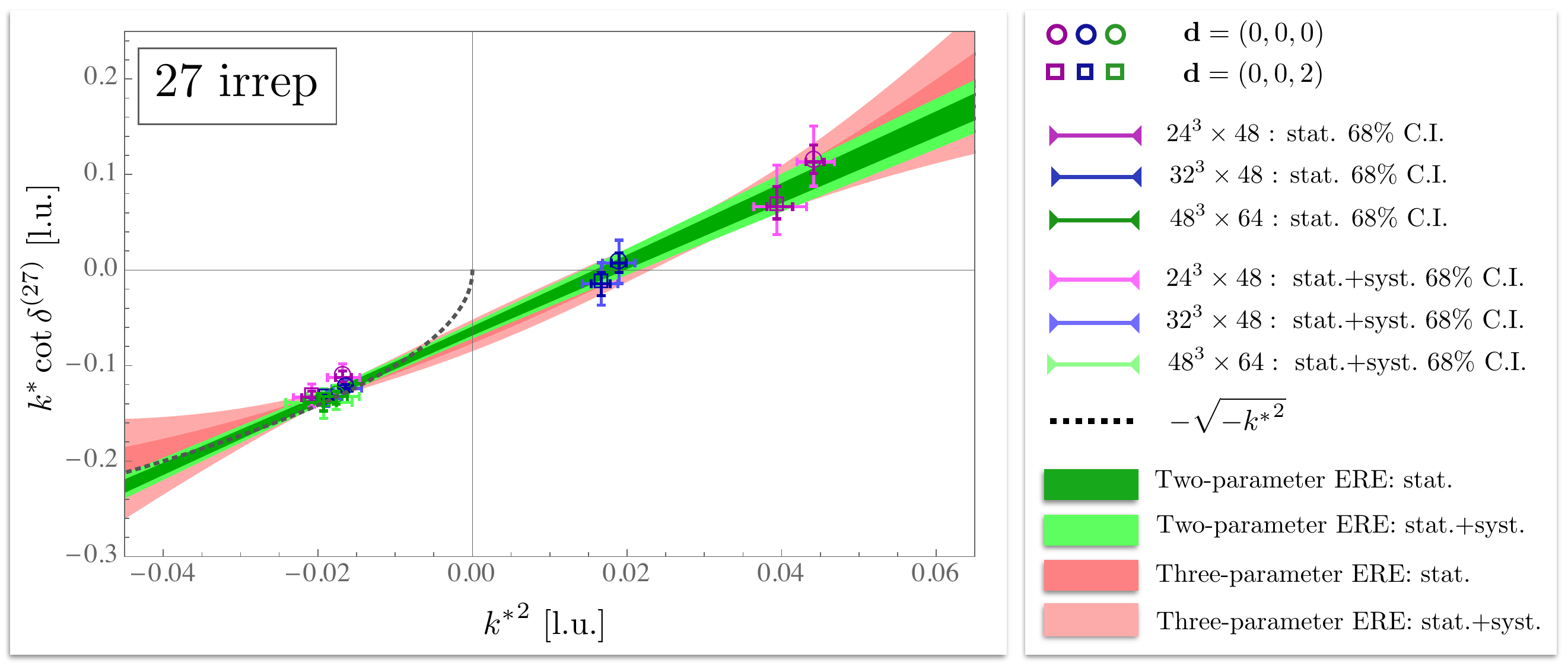}
\caption[.]{$k^*\cot \delta$ versus the square of the CM momentum of the two baryons, ${k^*}^2$, in the $27$ irrep. The bands represent fits to the two and three-parameter EREs. Quantities are expressed in lattice units (l.u.).}
\label{fig:27-phase-shifts}
\end{figure}

\begin{figure}[t!]
\includegraphics[scale=0.615]{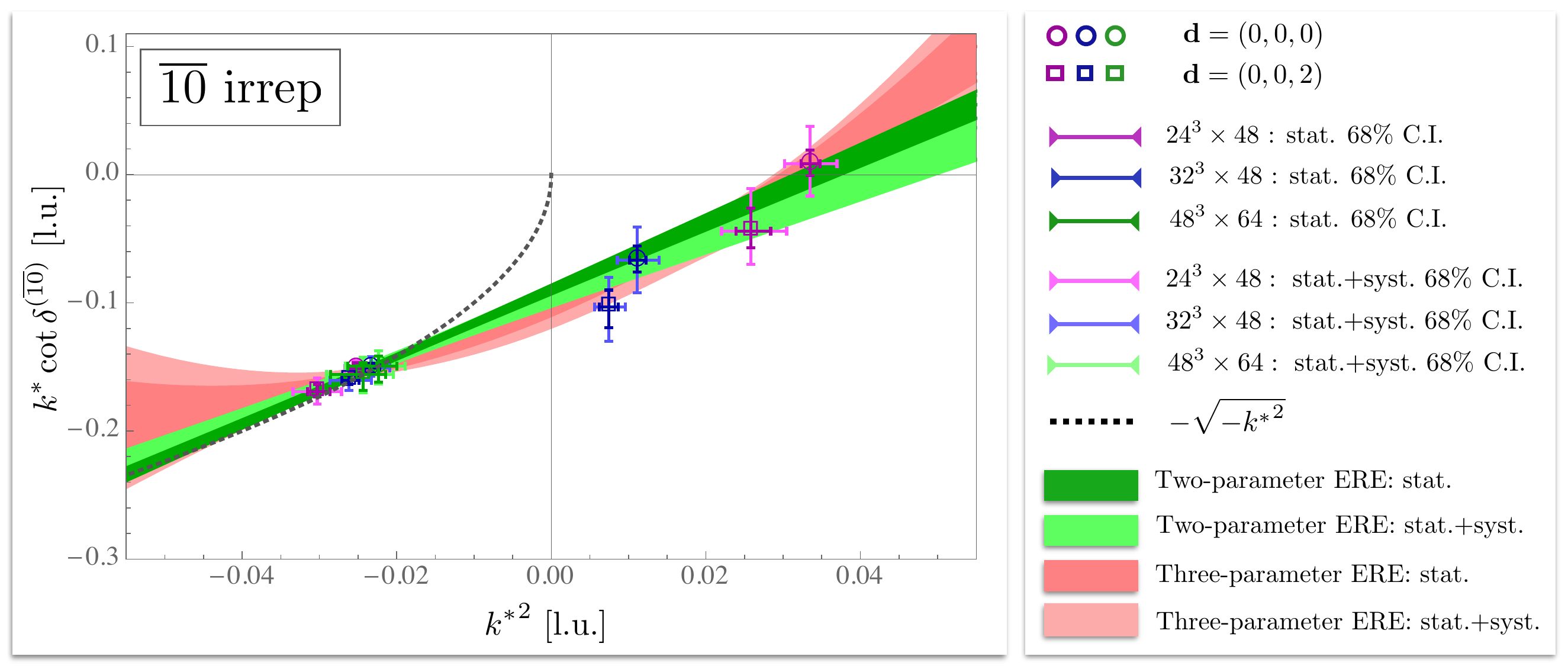}
\caption[.]{$k^*\cot \delta$ versus the square of the CM momentum of the two baryons, ${k^*}^2$, in the $\overline{10}$ irrep. The bands represent fits to the two and three-parameter EREs. Quantities are expressed in lattice units (l.u.).}
\label{fig:10-bar-phase-shifts}
\end{figure}

\begin{figure}[t!]
\includegraphics[scale=0.615]{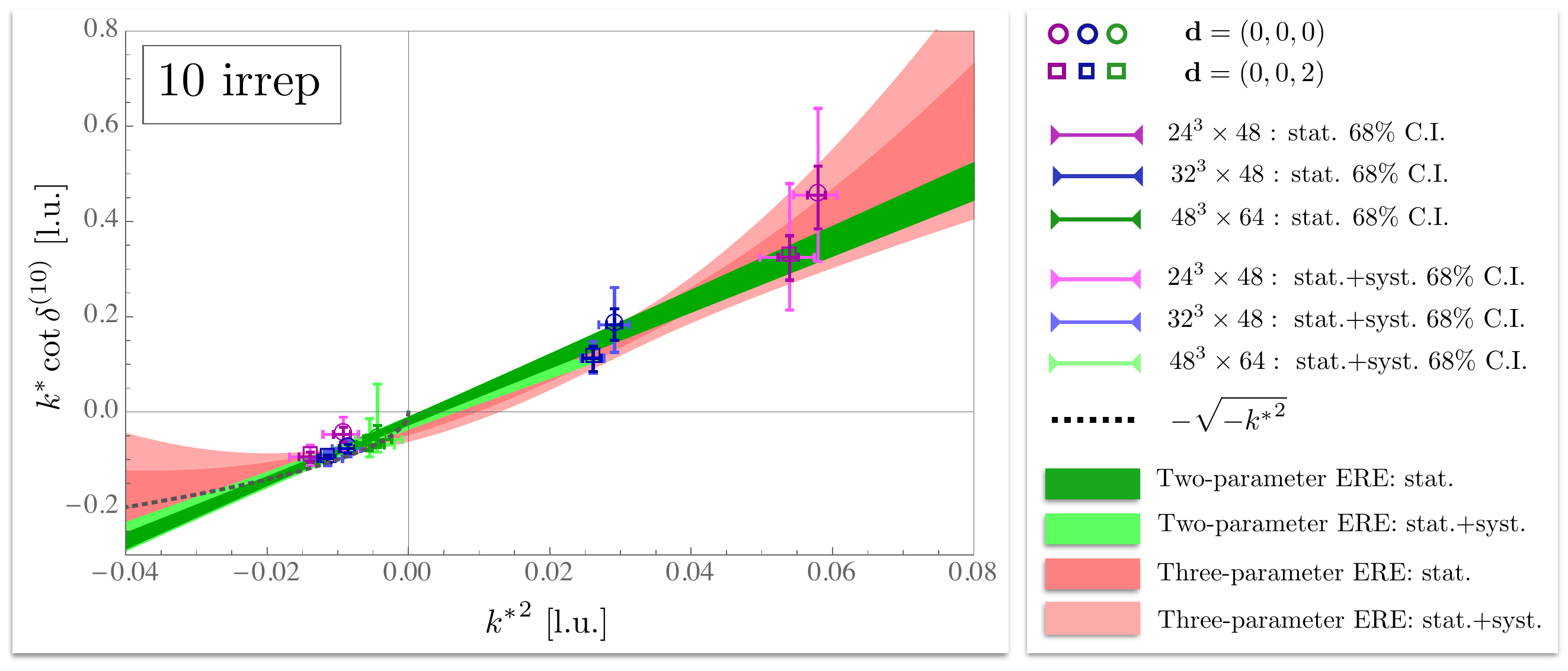}
\caption[.]{$k^*\cot \delta$ versus the square of the CM momentum of the two baryons, ${k^*}^2$, in the $10$ irrep. The bands represent fits to the two and three-parameter EREs. Quantities are expressed in lattice units (l.u.).}
\label{fig:10-phase-shifts}
\end{figure}

\begin{figure}[h!]
\includegraphics[scale=0.615]{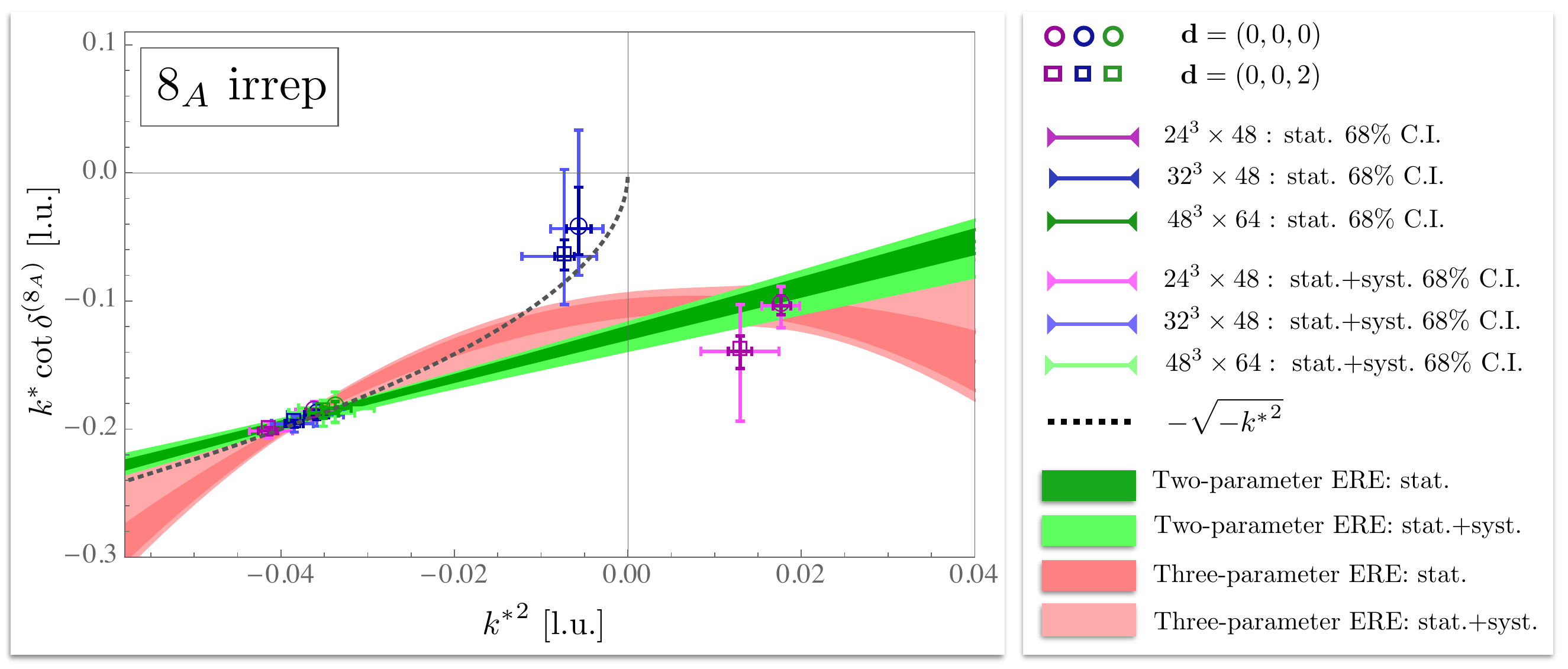}
\caption[.]{$k^*\cot \delta$ versus the square of the CM momentum of the two baryons, ${k^*}^2$, in the $8_A$ irrep. The bands represent fits to the two and three-parameter EREs. Quantities are expressed in lattice units (l.u.).}
\label{fig:8A-phase-shifts}
\end{figure}

\begin{figure}[h!]
\includegraphics[scale=0.6375]{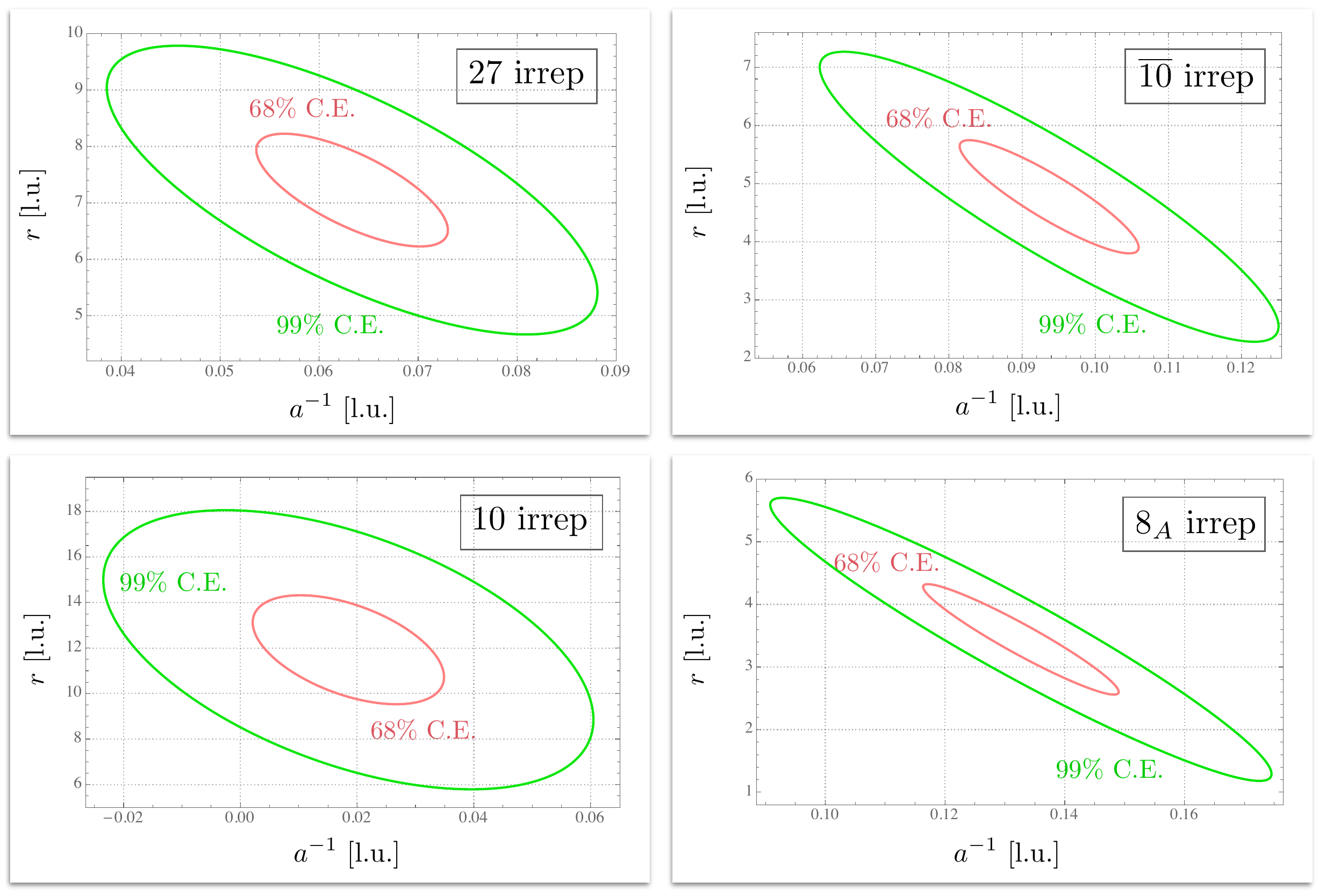}
\caption[.]{The $68\%$ (red) and $99\%$ (green) confidence ellipses (C.E.) corresponding to the full uncertainty (statistical and systematic combined in quadrature) of the inverse scattering length and effective range in channels belonging to each of the four irreps as denoted in the label of the plots.}
\label{fig:ar-confint}
\end{figure}

Below the start of the t-channel cut, the $k^* \cot \delta$ function for the $S$-wave ($\alpha$-wave) amplitude is anticipated to be well described by an ERE, see Eq.~(\ref{eq:ERE}). Assuming that the pion is the lightest hadron exchanged between the baryons at this value of the quark masses, the t-channel cut starts at $\left|{k^*}^2\right|=m_\pi^2/4\approx 0.088~\text{l.u.}$, considerably higher than the $\left|{k^*}^2\right|$ values obtained from the FV spectra in all channels. The constrained values of $k^* \cot \delta$ as a function of ${k^*}^2$ can thus be fit by  two and three-parameter forms in each of the two-baryon channels, and the resulting fit bands are shown in Figs.~\ref{fig:27-phase-shifts}-\ref{fig:8A-phase-shifts}. The $k^* \cot \delta$ values at the ten kinematic points considered here are also shown. Note that the vertical and horizontal error bars are displayed for simplicity and do not reflect the strongly correlated distributions of the $k^* \cot \delta$ and ${k^*}^2$ results. The precise form of the uncertainties are those shown in Fig.~\ref{fig:Z-functions}. In all channels, a two-parameter ERE describes the data well. The three-parameter ERE fits provide only small improvements in the values of $\chi^2/\text{d.o.f}$ of the fits, with the resulting scattering lengths and effective ranges being consistent with those of the two-parameter fit but with larger uncertainties. The values of the inverse scattering lengths and effective ranges from the two and three-parameter fits, as well as the shape parameters from the three-parameter fits, are listed in Table~\ref{fig:scatt-param-table}. The fit parameters are correlated, with their best values described by a multi-dimensional confidence ellipsoid. The $68\%$ and $98\%$ confidence ellipses from the two-parameter ERE are shown in Fig.~\ref{fig:ar-confint}, with the values of the center of the ellipses, their semi-minor and semi-major axes, as well as the slope of the semi-major axis of each ellipse listed in Table~\ref{tab:ellipse} of Appendix~\ref{app:tables}.

%
%%%%%%%%%
%
%
%%%%%%%%%
The values of the inverse scattering lengths and effective ranges of the two-parameter EREs that are tabulated in Table~\ref{fig:scatt-param-table} in lattice units can be expressed in physical units:
\begin{eqnarray}
&&27 \text{ irrep:} ~~~~~~a^{-1}=0.44^{(+4)(+8)}_{(-5)(-8)}~{\tt{fm}}^{-1},~~~r=1.04^{(+10)(+18)}_{(-10)(-18)}~{\tt{fm}},
\label{eq:aInv-r-phys-27}
\\
\nonumber\\
&&\overline{10} \text{ irrep:} ~~~~~~a^{-1}=0.63^{(+6)(+10)}_{(-5)(-11)}~{\tt{fm}}^{-1},~~~r=0.70^{(+16)(+12)}_{(-2)(-20)}~{\tt{fm}},
\label{eq:aInv-r-phys-10-bar}
\\
\nonumber\\
&&10 \text{ irrep:} ~~~~~~a^{-1}=0.16^{(+15)(+6)}_{(-13)(-6)}~{\tt{fm}}^{-1},~~~r=1.74^{(+36)(+34)}_{(-16)(-48)}~{\tt{fm}},
\label{eq:aInv-r-phys-10}
\\
\nonumber\\
&&8_A \text{ irrep:} ~~~~~~a^{-1}=0.88^{(+8)(+14)}_{(-7)(-14)}~{\tt{fm}}^{-1},~~~r=0.50^{(+10)(+14)}_{(-6)(-14)}~{\tt{fm}}.
\label{eq:aInv-r-phys-8A}
\end{eqnarray}
The numbers in the first and second parentheses denote, respectively, the statistical uncertainty, and the systematic uncertainty propagated from the corresponding uncertainties in the energies. The uncertainty in the lattice spacing is small compared with other uncertainties. Although these calculations have been performed for heavy quark masses at the flavor-symmetric point and without QED interactions, it is still interesting to compare these parameters with those in nature. While constraints on hyperon-nucleon and hyperon-hyperon scattering are not precise enough for a useful comparison, there exist precise determinations of nucleon-nucleon scattering parameters at low energies. In particular, the experimental values of the $nn$ and $np~({^3}S_1)$ $S$-wave scattering lengths and effective ranges are
\begin{eqnarray}
nn ~~~~~~&&a_{\text{phys.}}^{-1} \approx -0.05~{\tt{fm}}^{-1},~~~r_{\text{phys.}} \approx 2.75~{\tt{fm}},
\label{eq:aInv-r-phys-27-phys}
\\
\nonumber\\
np~({^3}S_1) ~~~~~~&&a_{\text{phys.}}^{-1} \approx 0.18~{\tt{fm}}^{-1},~~~r_{\text{phys.}} \approx 1.75~{\tt{fm}},
\label{eq:aInv-r-phys-10-bar-phys}
\end{eqnarray}
which should be compared with the scattering parameters in the $27$ and $\overline{10}$ irreps above, respectively. It is observed that the ranges of interactions in both channels, as characterized by the effective range parameters, are larger in nature than they are in the present work with heavier quark masses. Furthermore, the scattering lengths are smaller in this calculation than they are in nature. A more in-depth discussion of these parameters, in particular with regard to the unnaturalness of interactions and their spin-flavor symmetries, will be presented in Secs.~\ref{subsubsec:unnaturalness} and~\ref{subsubsec:large-Nc}. 
%
%%%%%%%%%
%
%
%%%%%%%%%
%
%
%%%%%%%%%
\begin{table}[t!]
\includegraphics[scale=1.0]{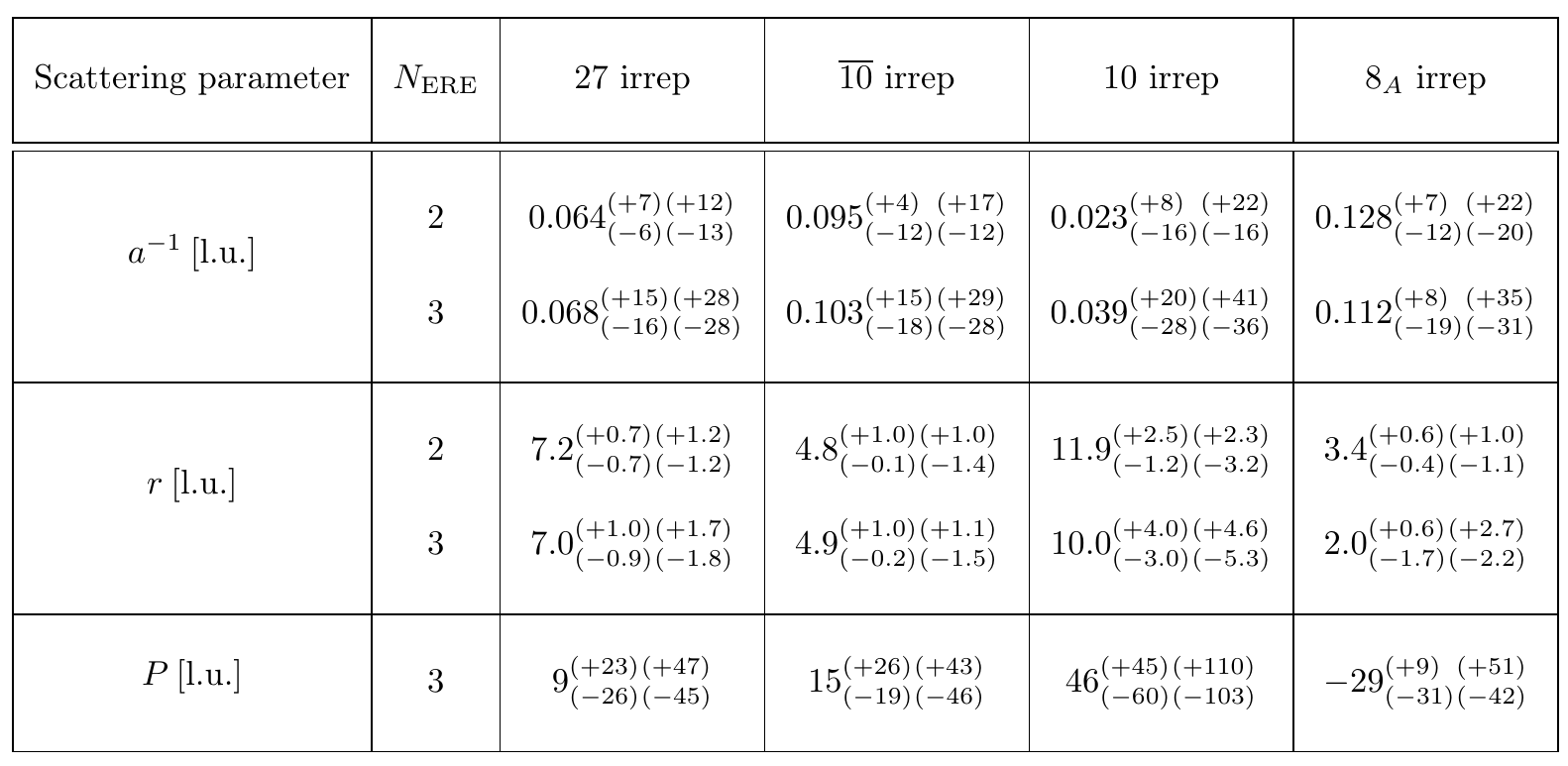}
\caption[.]{The values of the inverse scattering length, $a^{-1}$, effective range, $r$, and the first shape parameter, $P$, from fits to the two-parameter ($N_{\text{ERE}}=2$) and three-parameter ($N_{\text{ERE}}=3$) EREs, in channels belonging to the four different irreps.}
\label{fig:scatt-param-table}
\end{table}
%
%
%%%%%%%%%
%

%%%%%%%%%%%%
\subsubsection{Bound states and binding energies
\label{subsubsec:bound-state}
}
There is clear evidence for the existence of a bound state in each of the channels belonging to the $27$, $\overline{10}$ and $8_A$ irreps of the $SU(3)$ decomposition of the product of two octet baryons. First, as the volume is increased, the ground-state energies of two-baryon systems converge to a negatively-shifted energy far away from the two-particle threshold. Second, the analytic continuation of the amplitudes in these channels is consistent with the presence of a pole in the amplitude for ${k^*}^2<0$, as is evident from the intersection of the ERE bands with the $(-\sqrt{-{k^*}^2})$ function in Figs.~\ref{fig:27-phase-shifts},~\ref{fig:10-bar-phase-shifts} and~\ref{fig:8A-phase-shifts}. The location of this intersection determines the square of the binding momentum in these channels. However, given the uncertainties in the ERE fits, the infinite-volume extrapolation of negatively-shifted energies leads to a more precise determination of the binding energies. With energies determined in three volumes, a controlled extrapolation to infinite volume is possible in the present work. Fitting to the truncated form of the FV QC for negative ${k^*}^2$ values, Eq.~(\ref{eq:extrapolation}), the infinite-volume binding momenta, $\kappa^{(\infty)}$, can be obtained in each channel. These results are presented in Table~\ref{tab:binding-momenta} for measurements with $\mathbf{d}=(0,0,0)$ and $(0,0,2)$, with complete agreement seen between the two determinations. The bootstrap samples of extracted $\kappa^{(\infty)}$ values from each case can be combined to obtain a conservative estimate of the binding momenta and their uncertainties, given in the last row of Table~\ref{tab:binding-momenta}. The omitted terms in the truncated form in Eq.~(\ref{eq:extrapolation}) are negligible as $e^{-\sqrt{3}\kappa^{(\infty)}L}$ is at most $\sim 10^{-3}$ for the channels belonging to the $27$, $\overline{10}$ and $8_A$ irreps. The stability of the extracted binding momenta has been verified by excluding lower-order terms and by adding higher-order terms to the fits. 

Table~\ref{tab:binding-momenta} also includes the $\kappa^{(\infty)}$ values for the channels belonging to the $10$ irrep. As is seen from Fig.~\ref{fig:10-phase-shifts}, the ground-state energy in the largest volume is close to threshold. Nonetheless, assuming that there is a bound state in this channel, a determination of $\kappa^{(\infty)}$ based on the fit to Eq.~(\ref{eq:extrapolation}) is fully consistent with the ground-state energies at the largest volume, as well as with the location of the pole in the scattering amplitude.  From these results, the existence of a bound state in the $10$ irrep cannot be confirmed or excluded with statistical significance. Future calculations with higher statistics are needed in order to draw robust conclusions about the nature of the ground state in the $10$ irrep.

In physical units, the binding energies of these states are:
\begin{eqnarray}
27 \text{ irrep:}&&~~~~~~B=20.6{}_{(-2.4)}^{(+1.8)}{}_{(-1.6)}^{(+2.8)}~\tt{MeV},
\label{eq:binding-phys-27}
\\
\nonumber\\
\overline{10} \text{ irrep:}&&~~~~~~B=27.9{}_{(-2.3)}^{(+3.1)}{}_{(-1.4)}^{(+2.2)}
 ~\tt{MeV},
\label{eq:binding-phys-10-bar}
\\
\nonumber\\
10 \text{ irrep:}&&~~~~~~B=6.7{}_{(-1.9)}^{(+3.3)}{}_{(-6.2)}^{(+1.8)}
 ~\tt{MeV},
\label{eq:binding-phys-10}
\\
\nonumber\\
8_A \text{ irrep:}&&~~~~~~B=40.7{}_{(-3.2)}^{(+2.1)}{}_{(-1.4)}^{(+2.4)}
 ~\tt{MeV},
\label{eq:binding-phys-8A}
\end{eqnarray}
where $B=-2\sqrt{-{\kappa^{(\infty)}}^2+M_B^2}+2M_B$. Again, the first uncertainty is statistical and the second uncertainty encompasses both a fitting uncertainty and an uncertainty encoding variation among multiple analyses. The uncertainty in the lattice spacing is small compared with other uncertainties. These binding energies are consistent with our previous determination in Ref.~\cite{Beane:2012vq, Beane:2013br}, and with the binding energies obtained on the same ensembles of gauge-field configurations in Ref.~\cite{Berkowitz:2015eaa} for the ground states of the two-nucleon channels in the $27$ and $\overline{10}$ irreps.
%
%%%%%%%%%
\begin{table}
\includegraphics[scale=1]{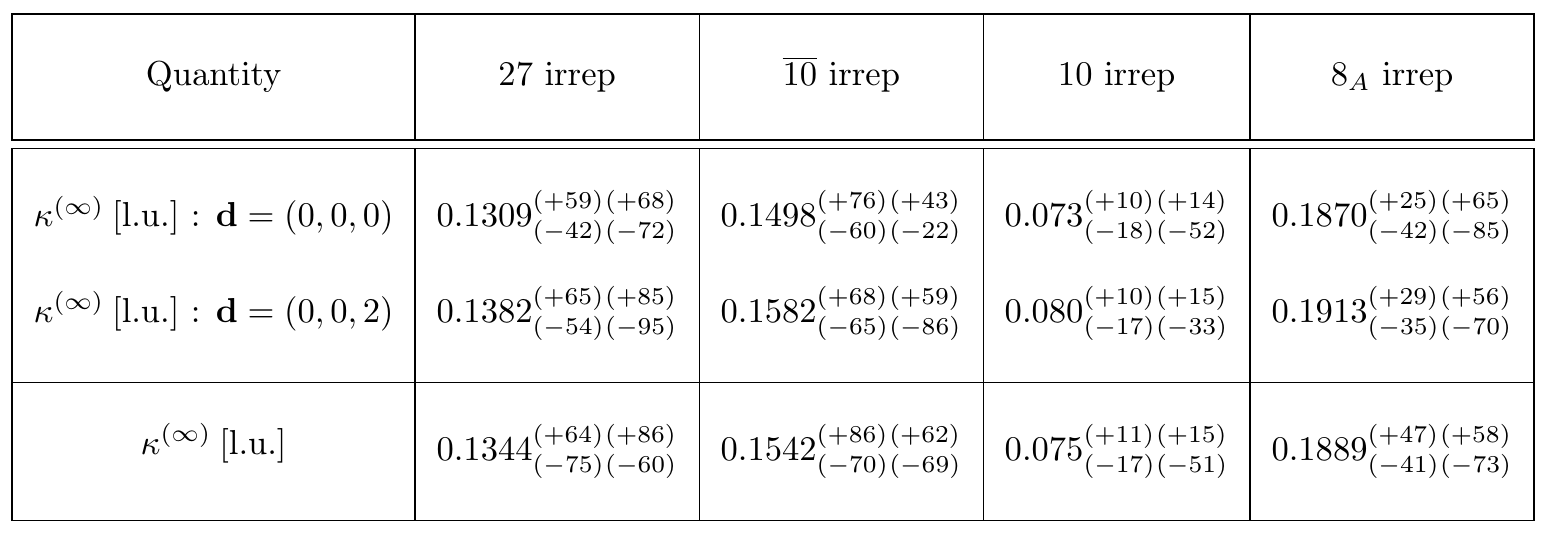}
\caption[.]{The infinite-volume binding momenta for bound states in channels belonging to different irreps at the $SU(3)$ flavor-symmetric point. These results are obtained by fitting negative ${k^*}^2$ values in Tables~\ref{tab:27}-\ref{tab:8A} for $\mathbf{d}=(0,0,0)$ and $\mathbf{d}=(0,0,2)$ with the extrapolation formula in Eq.~(\ref{eq:extrapolation}), as well as using a combined fit to both sets of values.}
\label{tab:binding-momenta}
\end{table}
%

%%%%%%%%%%%%
\subsubsection{$S$-wave baryon-baryon interactions and naturalness
\label{subsubsec:unnaturalness}
}
Interactions are considered unnatural if they give rise to some characteristic length scale of the system that is much larger than their range. There are at least two measures to assess naturalness in a two-particle system. For scattering states at low energies, scattering length defines a characteristic length scale, and the range of interactions can be approximated by the effective range. As an example, $S$-wave interactions in the spin-singlet and spin-triplet two-nucleon channels in nature produce effective range to scattering length ratios, $r/a$, that are $\approx -0.14$ and $ \approx 0.32$, respectively. This indicates that both channels are unnatural, particularly the spin-singlet channel. When interactions support a bound state, another characteristic length scale of the two-particle system is the inverse of the binding momentum, which defines an intrinsic size for the bound state. Considering the exchange of the pion to be the dominant contribution to the long-range part of effective interactions among two nucleons, the ratio of the binding momentum to the pion mass provides another measure of unnaturalness of interactions. In nature, $|\kappa|/m_{\pi} \approx 0.07$ and $0.33$ for the di-neutron and deuteron, respectively, again indication that both channels are unnatural.\footnote{The scattering length and binding momentum are not, however, independent quantities.} One may ask whether this is a generic property of QCD with any value of the quark masses or if naturalness is strongly sensitive to the input parameters of QCD. High sensitivity would suggest that the properties of two-nucleon interactions in nature require fine tuning of the quark masses. It is interesting to ask if a similar feature is observed for interactions involving hyperons. Such questions can be partially addressed using the results obtained in the previous sections for the scattering parameters and binding momenta of two-baryon systems at the heavy quark masses used in this calculation.

The ratios of the scattering lengths to effective ranges obtained from both the two and three-parameter ERE fits are shown in Table~\ref{fig:r-over-a} for the $27$, $\overline{10}$, $10$ and $8_A$ two-baryon channels. Interestingly, these ratios are universally consistent with $\sim 0.5$ within uncertainties, a feature that points to a spin-flavor symmetry of interactions as will be discussed in Sec.~\ref{subsubsec:large-Nc} (see also Ref.~\cite{Beane:2013br}). This value indicates that all channels are governed by $S$-wave interactions that are only slightly less unnatural than those in the spin-triplet two-nucleon system in nature. This also implies that the $S$-wave interactions in a spin-singlet two-nucleon state undergo a more dramatic change as a function of the quark masses and appear more finely tuned in nature, a feature that was also pointed to in Ref.~\cite{Beane:2013br}.

The ratios of the binding momenta to the pion mass of this calculation ($m_{\pi}=0.59426(12)(11)~\text{l.u.}$) for the two-baryon channels in the $27$, $\overline{10}$, $10$ and $8_A$ irreps are generally close to $\sim 0.2-0.3$, similar to the value of $|\kappa|/m_{\pi}$ for the deuteron in nature. However, a comparison between the effective range in each channel and the inverse pion mass of this calculation suggest that pion exchange may not be the dominant contribution to the long-range forces between the baryons~\cite{Beane:2013br}. In particular, the intrinsic size of the bound state in each channel (set by the inverse binding momenta), is comparable to the corresponding effective range in each channel (with larger uncertainties in the $10$ irrep), a feature that is consistent with the results obtained for the ratio of the effective ranges to the scattering lengths, $r/a \approx1/2$. Nonetheless, while the bound states in the $27$, $\overline{10}$ and $8_A$ irreps appear to have a natural intrinsic size, the scattering lengths in all channels are still large compared with the binding momenta of the bound states. Consequently, an EFT treatment of these channels at low energies with the KSW-vK power-counting scheme is justified.
%
%%%%%%%%%
\begin{table}
\includegraphics[scale=1]{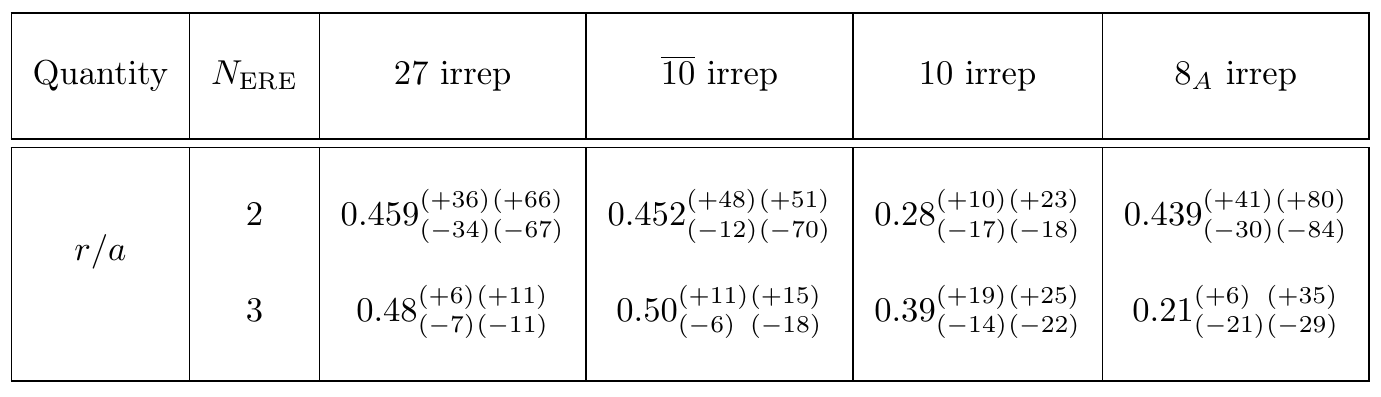}
\caption[.]{The ratio of effective range to scattering length in each channel, determined from the two-parameter ($N_{\text{ERE}}=2$) and three-parameter ($N_{\text{ERE}}=3$) ERE fits to the $k^*\cot \delta$ functions. This ratio provides a measure of the unnaturalness of interactions, as discussed in the text.}
\label{fig:r-over-a}
\end{table}
%
%
%%%%%%%%%%%%
\subsubsection{Large-$N_c$ limit and leading $SU(6)$ interactions in effective field theory
\label{subsubsec:large-Nc}
}

\begin{figure}
\includegraphics[scale=0.65]{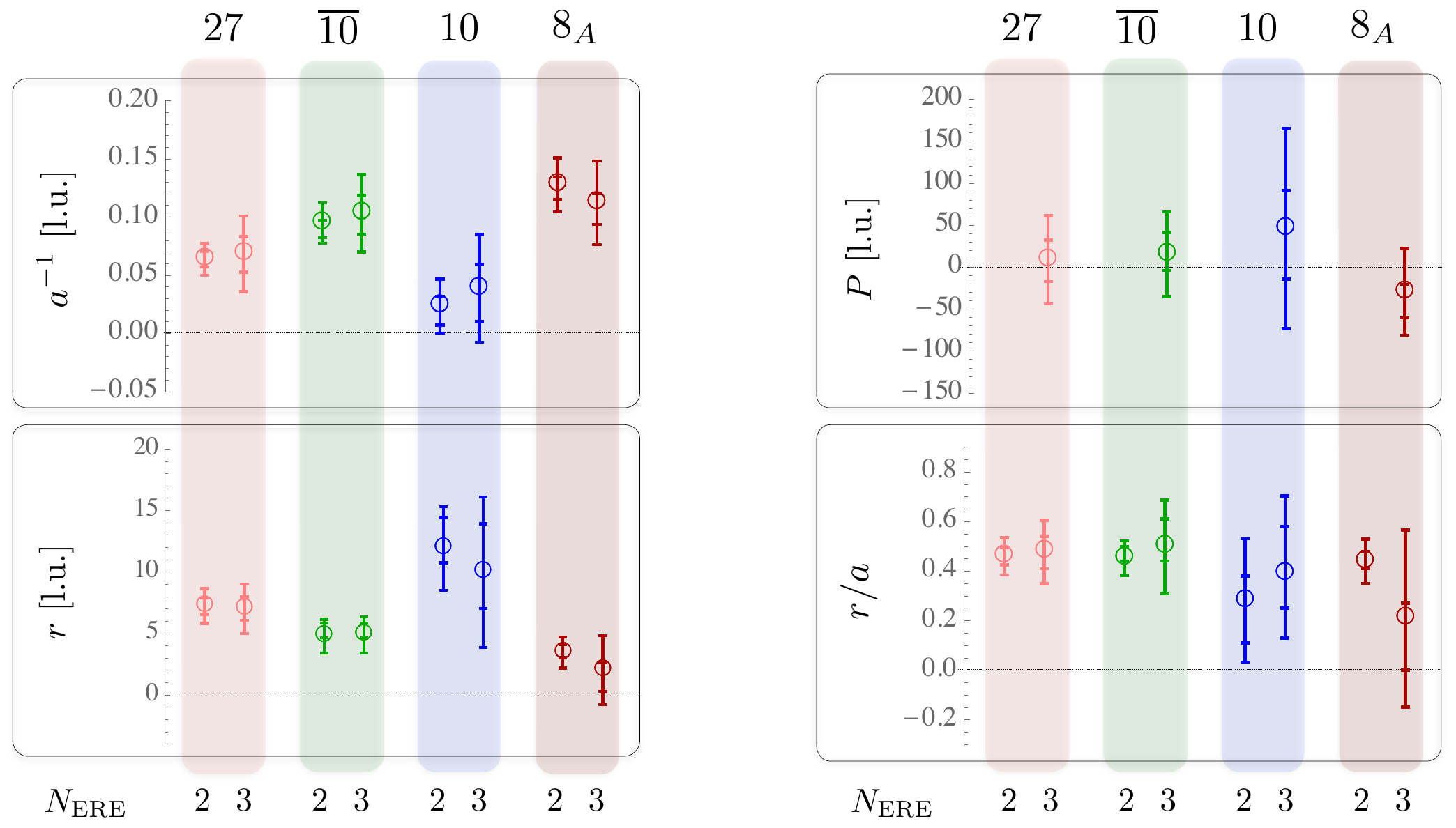}
\caption[.]{
A comparison of the values of the inverse scattering length, $a^{-1}$, effective range, $r$, the first shape parameter, $P$, and the ratio $r/a$, obtained from fits to the two-parameter ($N_{\text{ERE}}=2$) and three-parameter ($N_{\text{ERE}}=3$) EREs, in channels belonging to the four different irreps.}
\label{fig:a-inv-r-rho-compared}
\end{figure}

The scattering parameters in two-baryon channels belonging to different $SU(3)$ irreps have similar values in the present calculations, 
differing by at most $2\sigma$ from an average value. This feature is illustrated in Fig.~\ref{fig:a-inv-r-rho-compared}, in which the scattering lengths and effective ranges from the two and three-parameter ERE fits, and the shape parameters from the three-parameter ERE fits, are compared. Additionally, the correlated $r/a$ ratios from both ERE fits are shown for the four irreps. 
All of these quantities are broadly consistent between different channels, in particular between channels belonging to the $27$ and $\overline{10}$ irreps. This is a manifestation of an approximate spin-flavor symmetry of nuclear and hypernuclear forces as predicted to exist in the large-$N_c$ limit of QCD~\cite{Kaplan:1995yg}, as introduced in Sec.~\ref{subsec:SU(3)}. Deviations from  $SU(4)$ symmetry (involving  the $27$ and $\overline{10}$ irreps) are expected to scale as $1/N_c^2 \sim 10 \%$ while the deviations from $SU(6)$ symmetry are expected to scale as $1/N_c \sim 30\%$. This is consistent with the almost identical nature of the channels belonging to the $27$ and $\overline{10}$ irreps in the results presented here. Additional deviations from the spin-flavor symmetry that occur as a result of the $SU(3)$ flavor-symmetry breaking in nature are absent in the present 
calculations, making them an ideal testing ground for the large-$N_c$ relations.

\begin{figure}[h!]
\includegraphics[scale=0.545]{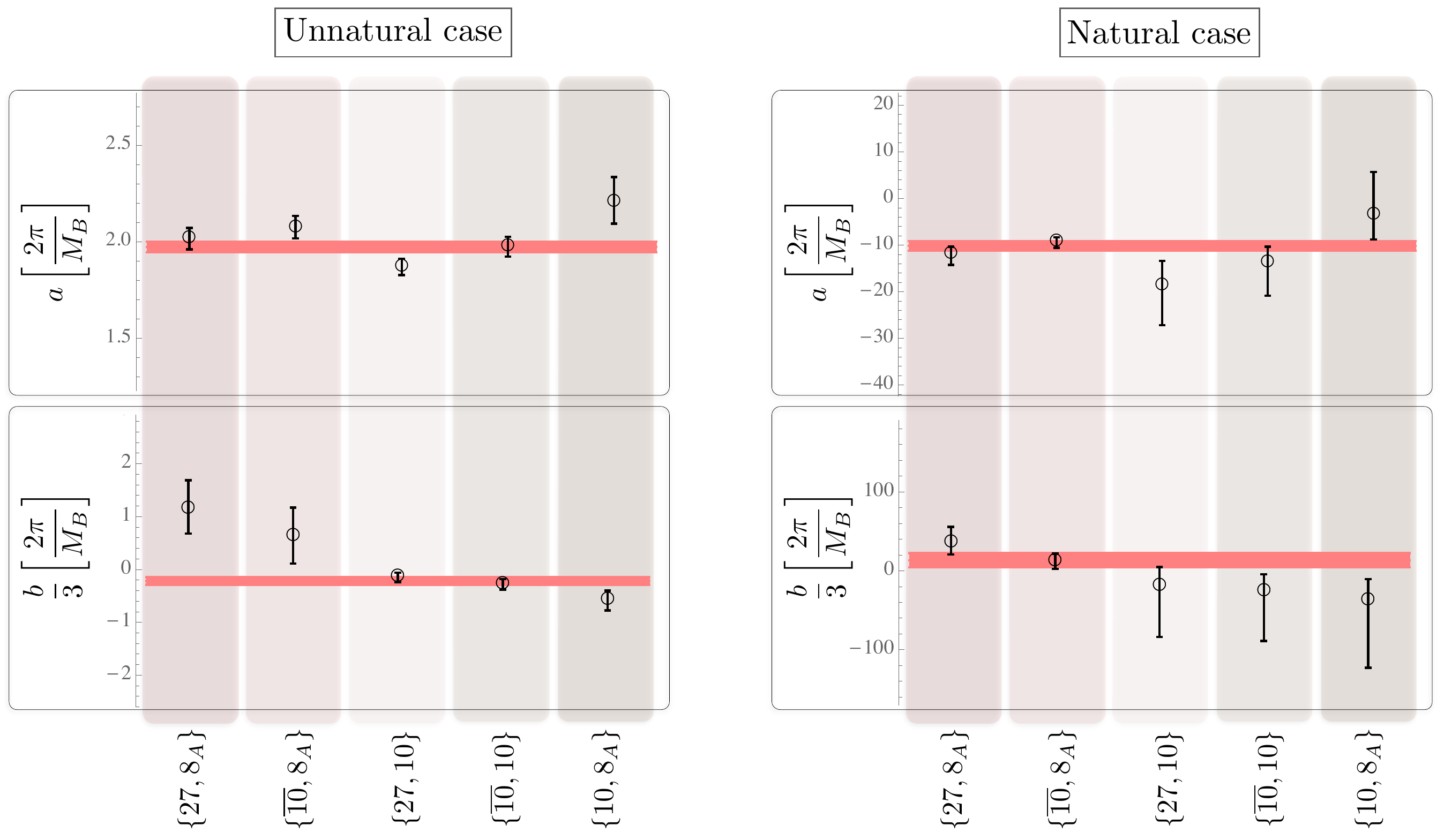}
\caption[.]{
The coefficients, $a$ and $b/3$, of the leading $SU(6)$ effective interactions obtained by fitting pairs of inverse scattering lengths in the channels belonging to each of the four $SU(3)$ irreps considered in this work. The left panel uses a renormalization scheme relevant for unnatural interactions given in Eq.~(\ref{eq:ab-coeffs}) with $\mu=m_\pi$. The right panel corresponds to a tree-level expansion of the scattering amplitudes with natural interactions giving rise to Eq.~(\ref{eq:ab-coeffs}) with $\mu=0$. Note the different  plot ranges in the two panels. The pink bands represent a combined constant fit to all five different determinations of $a$ and $b/3$ in each case. The couplings are expressed in units of $[\frac{2\pi}{M_B}]$, with $M_B$ being the baryon mass in this calculation, expressed in lattice units.}
\label{fig:ab-coeffs}
\end{figure}

Given an approximate $SU(6)$ symmetry of $S$-wave interactions, constraints can be obtained on the $SU(6)$ coefficients, $a$ and $b$, using Eqs.~(\ref{eq:ab-coeffs}). Two cases are considered here: unnatural interactions and natural interactions. In the unnatural case,  the leading $S$-wave interactions are summed to all orders in perturbation theory (implementing KSW-vK power counting) introducing a UV-scale dependence in the coefficients. A convenient choice of renormalization scale is $\mu=m_{\pi}$. However, any scale much above the largest inverse scattering length, but below the cut off of the theory,  would lead to manifest power counting and RG-scale independence. The values of $a$ and $b/3$ obtained from each pair of equations in~(\ref{eq:ab-coeffs}) are tabulated in Table~\ref{tab:ab-coeffs}, and are shown in Fig.~\ref{fig:ab-coeffs}. Within the uncertainty of each determination, these values are in agreement with each other. Note that from Eq.~(\ref{eq:ab-coeffs}), contributions from the $b$ coefficient are suppressed by at least a factor of 3 compared with those from the $a$ coefficient, thus the rescaled coefficient $b/3$ is considered. A combined fit of a constant to the five determinations results in the values for $a$ and $b/3$ that are listed in Table~\ref{tab:ab-coeffs} and shown as pink bands in Fig.~\ref{fig:ab-coeffs}.

Assuming the systems to be natural in the EFT analysis results in large uncertainties in the $b$ coefficient. This precludes conclusions to be drawn regarding its significance compared to the $a$ coefficient. Additionally, determinations that involve the $10$ irrep yield large uncertainties in the coefficients, signaling the inappropriate assumption of naturalness for interactions in a channel that is almost at unitarity within uncertainties. As the observations in Sec.~\ref{subsubsec:unnaturalness} point to primarily an unnatural scenario for all interactions considered, the values of the $a$ and $b$ coefficients in the unnatural scenario, defined with KSW-vK power counting, are found to be more relevant. In particular, it is observed that the value of $b/3$ is approximately an order of magnitude smaller than the value of $a$. This is a signature of an accidental $SU(16)$ symmetry of nuclear and hypernuclear forces which was first predicted in Ref.~\cite{Kaplan:1995yg}, and is studied here directly with QCD for the first time.

Without constraints on the scattering parameters belonging to the $8_S$ and $1$ irreps, a conclusive statement regarding the $SU(6)$ spin-flavor symmetry in the interactions is not possible. However, with the observations in other irreps pointing to  such a symmetry, a prediction can be made for the scattering amplitudes in the $8_S$ and $1$ irreps, giving $a^{-1}_{(8_S)}=a^{-1}_{(1)}=0.08(3)(2)~\text{l.u.}$, where the second uncertainty accounts for $\mathcal{O}\left( 1/N_c \right)$ corrections to the prediction of $SU(6)$ symmetry. This result enables an extraction of all SW coefficients of the LO $SU(3)$-symmetric interactions. 

%
%%%%%%%%%
%
%
%%%%%%%%%
\begin{table}[h!]
\includegraphics[scale=0.975]{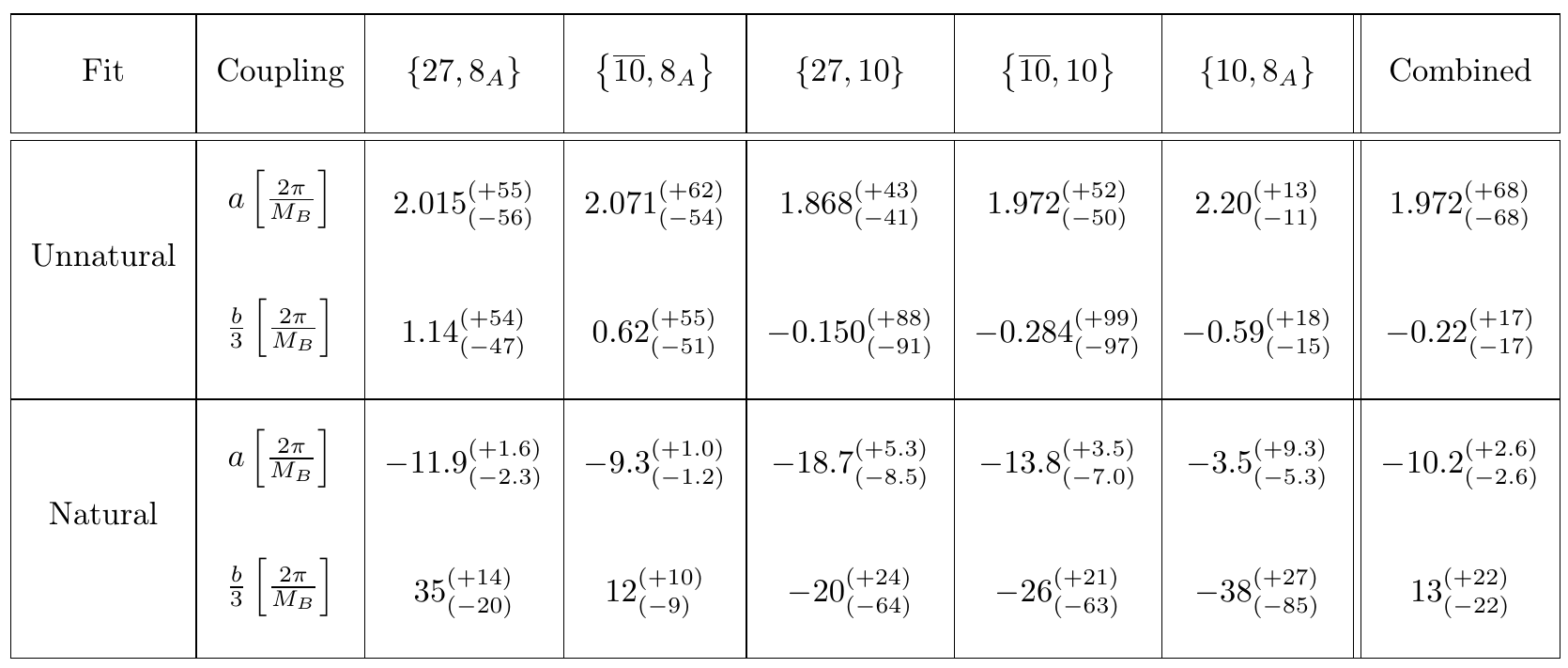}
\caption[.]{
The coefficients, $a$ and $b/3$, of the leading $SU(6)$ effective interactions 
obtained by solving the pairs of equations in Eq.~(\ref{eq:ab-coeffs}) with $\mu=m_\pi$ for the unnatural case and $\mu = 0$ for the natural case (corresponding to a tree-level expansion of the scattering amplitudes). The last column shows the result of a constant fit to all five determinations. The coefficients $a$ and $b/3$ are given in units of $[\frac{2\pi}{M_B}]$, with $M_B$ being the baryon mass in this calculation, expressed in lattice units.
}
\label{tab:ab-coeffs}
\end{table}
%
%
%%%%%%%%%
%

%%%%%%%%%%%%
\subsubsection{Leading $SU(3)$ interactions in effective field theory
\label{subsubsec:large-Nc}
}

The scattering lengths in the channels belonging to the $27$, $\overline{10}$, $10$ and $8_A$ irreps can be used to constrain 
various linear combinations of the coefficients of the LO $SU(3)$-symmetric Lagrange density, i.e., SW coefficients, at  $m_{\pi}\approx806~\tt{MeV}$. These constraints arise from Eq.~(\ref{eq:SW-coeffs}), which for unnatural interactions read
\begin{eqnarray}
\label{eq:SW-combo-U-I}
\left(c_1-c_2+c_5-c_6\right)^{-1}-\mu = -0.06(1)~\text{l.u.},
\\
\label{eq:SW-combo-U-II}
\nonumber\\
\left(c_1+c_2+c_5+c_6\right)^{-1}-\mu = -0.09(2)~\text{l.u.},
\\
\label{eq:SW-combo-U-III}
\nonumber\\
\left(-c_1-c_2+c_5+c_6\right)^{-1}-\mu = -0.02(2)~\text{l.u.},
\\
\label{eq:SW-combo-U-IV}
\nonumber\\
\left(\frac{3c_3}{2}+\frac{3c_4}{2}+c_5+c_6\right)^{-1}-\mu = -0.13(3)~\text{l.u.}
\end{eqnarray}
The coefficients $c_i$ depend on the scale $\mu$ and are expressed in units of $[\frac{2\pi}{M_B}]$, where $M_B$ is the mass of the baryon in this calculation in lattice units. The statistical and systematic uncertainties are combined in quadrature. Imposing $SU(6)$ spin-flavor symmetry, constrains the scattering lengths in the $8_S$ and $1$ irreps, and provides further constraints on the SW coefficients, 
\begin{eqnarray}
\label{eq:SW-combo-U-V}
(-\frac{2 c_1}{3} + \frac{2 c_2}{3} - \frac{5 c_3}{6} + \frac{5 c_4}{6} + c_5 - c_6)^{-1}-\mu = -0.08(4)~\text{l.u.}
\\
\label{eq:SW-combo-U-VI}
\nonumber\\
(-\frac{c_1}3 + \frac{c_2}{3} - \frac{8 c_3}{3} + \frac{8 c_4}{3} + c_5 - c_6)^{-1}-\mu = -0.08(4)~\text{l.u.}
\end{eqnarray}
Setting $\mu=0$ recovers the results for natural systems.
%
%%%%%%%%%
\begin{table}[t!]
\includegraphics[scale=1]{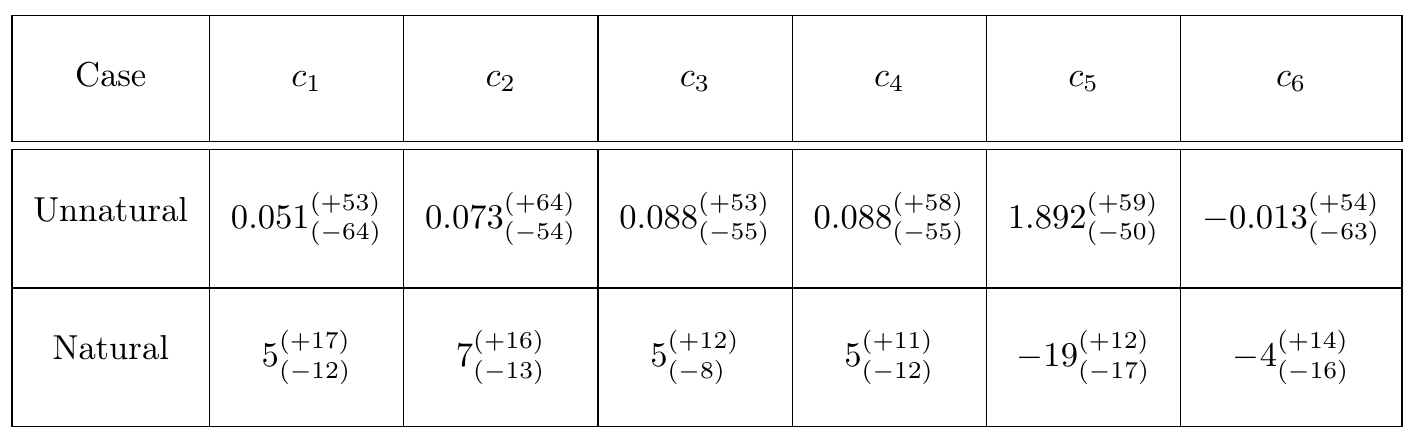}
\caption[.]{
Values of the coefficients of the LO $SU(3)$-symmetric interactions  
obtained by solving Eqs.~(\ref{eq:SW-combo-U-I})-(\ref{eq:SW-combo-U-VI}) for the unnatural case with $\mu=m_\pi$, and for the natural case with $\mu = 0$. The coefficients are expressed in units of $[\frac{2\pi}{M_B}]$, with $M_B$ being the baryon mass in this calculation, expressed in lattice units.}
\label{tab:SW-values}
\end{table}

\begin{figure}[h!]
\includegraphics[scale=0.575]{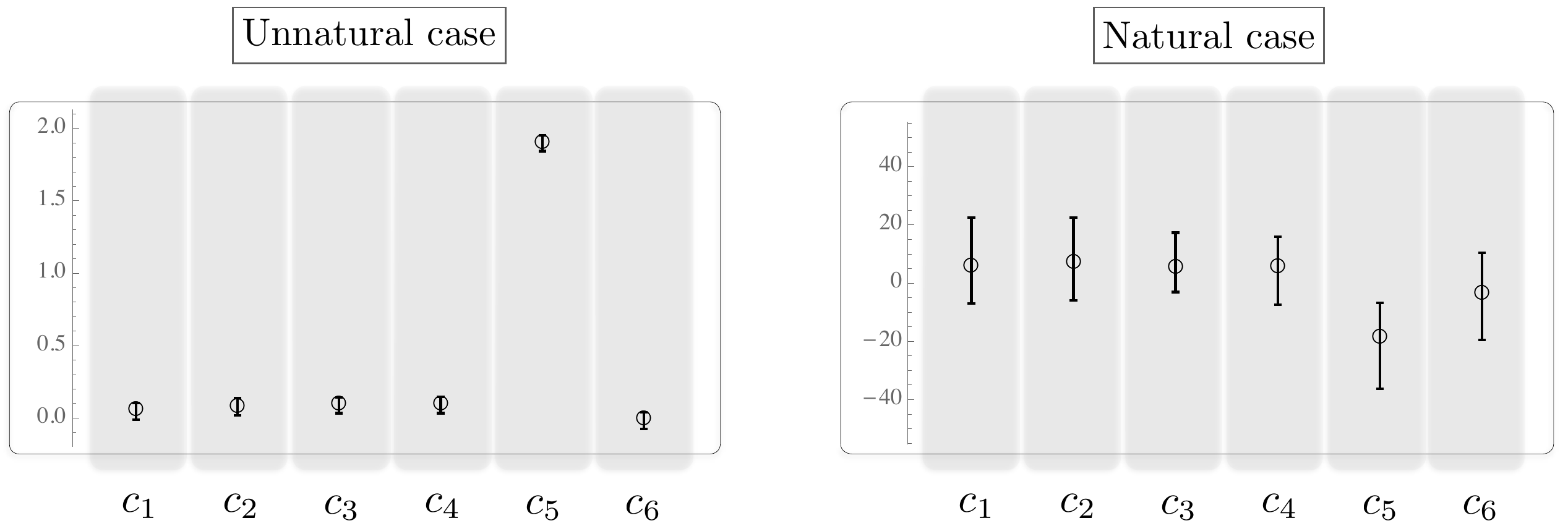}
\caption[.]{A comparison of the coefficients of the LO $SU(3)$-symmetric interactions. The left panel corresponds to the unnatural case with $\mu=m_\pi$, while the right panel represents the natural case with $\mu = 0$, corresponding to a tree-level expansion of the scattering amplitudes. The coefficients are expressed in units of $[\frac{2\pi}{M_B}]$, with $M_B$ being the baryon mass in this calculation, expressed in lattice units.}
\label{fig:sw-coeffs}
\end{figure}

Eqs.~(\ref{eq:SW-combo-U-I})-(\ref{eq:SW-combo-U-VI}) are solved to determine all six SW coefficients for unnatural interactions within the KSW-vK power counting at a renormalization scale of $\mu=m_{\pi}$, and for natural interactions through a tree-level expansion of the scattering amplitude, see Table \ref{tab:SW-values}. As is evident from these values, shown in Fig.~\ref{fig:sw-coeffs}, the unnatural scenario provides the most stringent constraints on the coefficients. In this case, the value of all SW coefficients except for $c_5$  are consistent with zero, a manifestation of the $SU(16)$  spin-flavor symmetry in the LO $SU(3)$ interactions, i.e., the $a \gg b/3$ hierarchy in the $SU(6)$ spin-flavor symmetric interactions. With these results, and the binding energies of light hypernuclei~\cite{Beane:2012vq}, ongoing \emph{ab initio} many-body calculations using the LQCD input at this value of the quark masses~\cite{Barnea:2013uqa, Kirscher:2015yda, Kirscher:2017fqc,Contessi:2017rww} can be extended to systems containing hyperons. Appendix~\ref{app:LO-amplitude} is devoted to summarizing the constraints obtained for the LO scattering amplitudes in flavor space.

%
%%%%%%%%%
%

%\newpage
%%%%%%%%%%%%%%%%%%%%%%%%%%%%%%%%%%%%%%%%%%%%
%%%%%%%%%%%%%%%%%%%%%%%%%%%%%%%%%%%%%%%%%%%%
%%%%%%%%%%%%%%%%%%%%%%%%%%%%%%%%%%%%%%%%%%%%
\section{Summary and conclusion
\label{sec:conclusion} 
}
\noindent
This paper presents the results of a Lattice QCD study of low-energy  $S$-wave scattering amplitudes of two octet baryons at an 
$SU(3)$ flavor-symmetric point, with a focus on underlying symmetry structures that are expected to emerge in the large-$N_c$ limit of QCD. At a pion mass of $\approx 806~\tt{MeV}$, 
$S$-wave interactions  between two baryons in the $27$, $\overline{10}$ and $8_A$ irreps 
(e.g., $NN~({^1}S_0)$, $NN~({^3}S_1)$ and $\frac{1}{\sqrt{2}}(\Xi^0n+\Xi^-p)~({^3}S_1)$, respectively) 
are found to induce bound states with binding energies: 
$20.6{}_{(-2.4)}^{(+1.8)}{}_{(-1.6)}^{(+2.8)}~\tt{MeV}$, 
$27.9{}_{(-2.3)}^{(+3.1)}{}_{(-1.4)}^{(+2.2)}~\tt{MeV}$ 
and $40.7{}_{(-3.2)}^{(+2.1)}{}_{(-1.4)}^{(+2.4)}~\tt{MeV}$, respectively, which  
are consistent with our previous analyses~\cite{Beane:2012vq} of the same correlation functions. 
The presence of a bound state in the $10$ irrep is not statistically significant, 
with a binding energy: $6.7{}_{(-1.9)}^{(+3.3)}{}_{(-6.2)}^{(+1.8)}~\tt{MeV}$. 
The scattering lengths and effective ranges in the four channels have been extracted, and suggest that all of these systems 
have unnaturally large scattering lengths, with $|r/a| \sim 0.5$.
If this feature is found to persist in the hyperon channels at the physical values of the quark masses,
its phenomenological consequences would be interesting to explore.

Utilizing KSW-vK power counting, that is appropriate in describing unnatural systems, and with three degenerate quark flavors, the values of the scattering parameters calculated in the two-baryon channels are found to be consistent with the $SU(6)$ spin-flavor symmetry in the nuclear and hypernuclear forces that is predicted in the large-$N_c$ limit of QCD~\cite{Kaplan:1995yg}. 
In addition, a suppressed contribution from one of the two large-$N_c$ low-energy constants is observed, which is consistent with an approximate accidental $SU(16)$ symmetry emerging from the underlying $SU(6)$ symmetry in the large-$N_c$ limit.
Therefore, to a good approximation, one universal coefficient determines low-energy $S$-wave  baryon-baryon scattering in all $SU(3)$ channels. Although the $S$-wave  scattering lengths in the  $8_S$ and $1$ irreps were not determined directly, 
$SU(6)$ symmetry relates them to those in the other channels. Quite precise results are found for the six natural-sized coefficients in the LO $SU(3)$-symmetric effective field theory describing low-energy baryon-baryon interactions. It will be interesting to see how the remnants of the $SU(6)$ and accidental $SU(16)$ symmetries, that are observed to be well satisfied at $N_c=3$ in the limit of $SU(3)$ flavor symmetry, are reflected in the hyperon-nucleon and hyperon-hyperon interactions at the physical  values of the quark masses.

This work demonstrates the role of LQCD in elucidating properties of systems involving hyperons, extending previous determinations of the nucleon-nucleon scattering parameters in Ref.~\cite{Beane:2013br} to hyperon-nucleon and hyperon-hyperon channels. Studies of such systems at lighter values of the quark masses already 
exist~\cite{Beane:2006gf, Nemura:2008sp, Beane:2009py, Beane:2010hg, Beane:2011zpa, Beane:2011iw, Inoue:2010es, Inoue:2011ai, Beane:2012ey, Yamada:2015cra, Nemura:2017bbw, Doi:2017cfx, Ishii:2017xud, Sasaki:2017ysy}, but higher precision and more comprehensive investigations are needed to be able to make reliable predictions for systems in nature. For future calculations closer to the physical values of the quark masses and at larger volumes, the signal-to-noise problem and the increasingly closely-spaced spectra at large volumes will pose challenges that were 
not prominent in the present study. It is expected that these challenges can be tackled with increased computational resources and algorithmic developments, such as signal-to-noise optimization~\cite{Detmold:2014hla} and phase-reweighting methods~\cite{Wagman:2017xfh, Wagman:2016bam}, promising significant progress in the near future.

%%%%%%%%%%%%%%%%%%%%%%%%%%%%%%%%%%%%%%%%%%%
\noindent
\subsection*{Acknowledgments}
We would like to thank Silas Beane and Assumpta Parre\~no for important discussions and significant contributions related to this project, and Marc Illa for valuable comments on the manuscript. We would also like to thank Francesco Pederiva for interesting discussions related to nuclear many-body calculations of light nuclei and hypernuclei. This research was supported in part by the National Science Foundation under grant number NSF PHY11-25915 and we acknowledge the Kavli Institute for Theoretical Physics for hospitality during preliminary stages of this work. 
Calculations were performed using computational resources provided by NERSC (supported by U.S. Department of Energy grant number DE-AC02-2705CH11231), and by the USQCD collaboration with support from the LQCD Ext-II Computing Project. 
This research used resources of the Oak Ridge Leadership Computing Facility at the Oak Ridge National Laboratory, which is supported by the Office of Science of the U.S. Department of Energy under Contract number DE-AC05-00OR22725. 
The PRACE Research Infrastructure resources at the Tr`es Grand Centre de Calcul and Barcelona Supercomputing Center were also used. 
Parts of the calculations used the {\tt chroma} software suite~\cite{Edwards:2004sx} and the {\tt quda} library~\cite{Clark:2009wm,Babich:2011np}. 
ZD, WD and PES were partly supported by U.S. Department of Energy Early Career Research Award DE-SC0010495 and grant number DE-SC0011090. 
The work of WD is supported in part by the U.S. Department of Energy, Office of Science, Office of Nuclear Physics, within the framework of the TMD Topical Collaboration. 
KO was partially supported by the U.S. Department of Energy through grant number DE-FG02-04ER41302 and through contract number DE-AC05-06OR23177 under which JSA operates the Thomas Jefferson National Accelerator Facility. 
MJS was supported by DOE grant number DE-FG02-00ER41132, and in part by the USQCD SciDAC project, the U.S. Department of Energy through grant number DE- SC00-10337. 
MLW was supported in part by DOE grant number DE-FG02-00ER41132. FW was partially supported through the USQCD Scientific Discovery through Advanced Computing (SciDAC) project funded by U.S. Department of Energy, Office of Science, Offices of Advanced Scientific Computing Research, Nuclear Physics and High Energy Physics and by the U.S. Department of Energy, Office of Science, Office of Nuclear Physics under contract DE-AC05-06OR23177. 

\bibliography{bibi}

\clearpage

\newpage

%%%%%%%%%%%%%%%%%%%%%%%%%%%%%%%%%%%%%%%%%%%%
%%%%%%%%%%%%%%%%%%%%%%%%%%%%%%%%%%%%%%%%%%%%
%%%%%%%%%%%%%%%%%%%%%%%%%%%%%%%%%%%%%%%%%%%%
\appendix
\section{TWO-BARYON STATES AT THE $SU(3)$ FLAVOR-SYMMETRIC POINT
\label{app:SU(3)}
}
%
%%%%%%%%%
\begin{figure}[b!]
\includegraphics[scale=0.771]{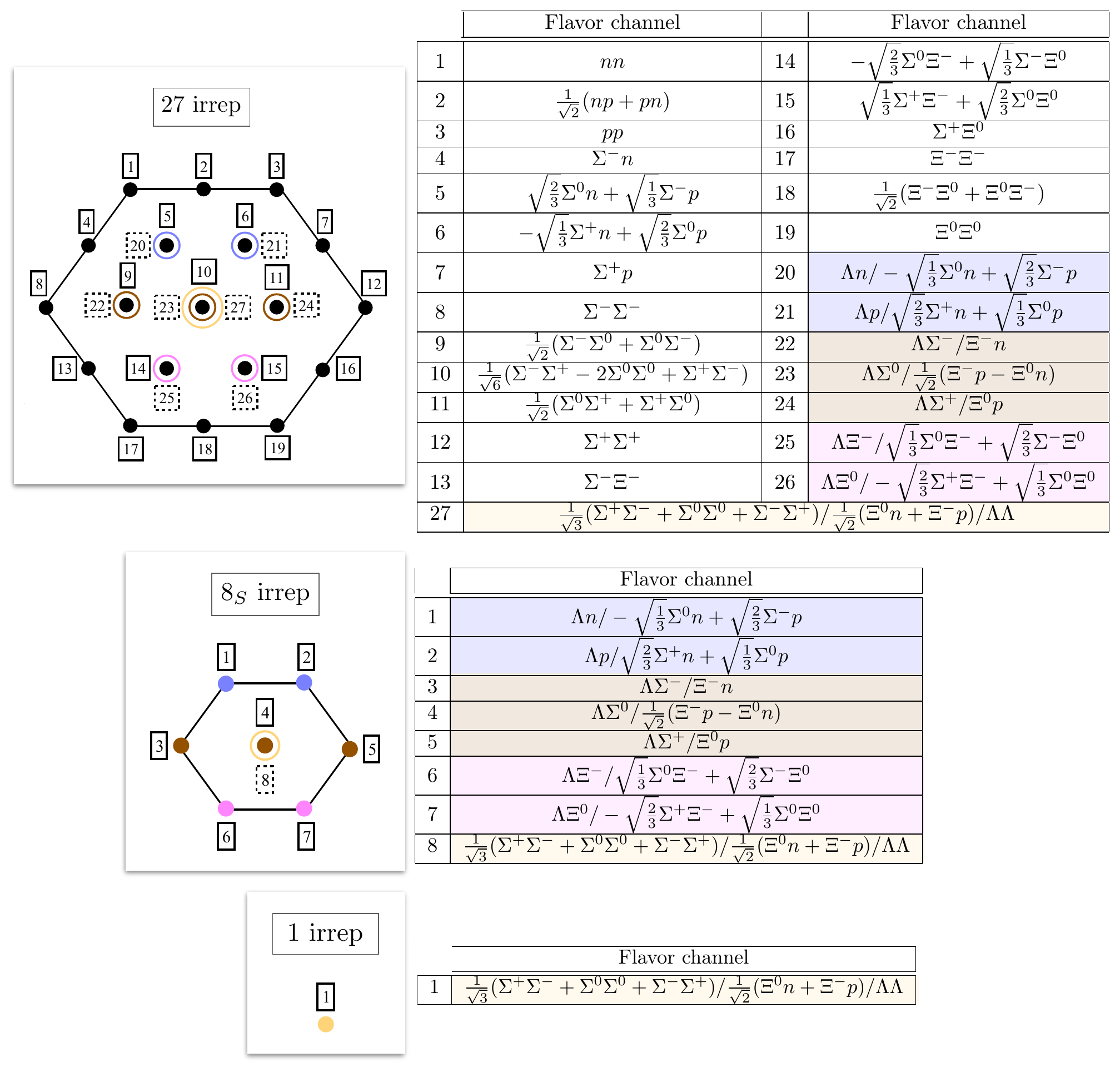}
\caption[.]{
Diagrammatic representation of the $27$, $8_S$ and $1$ irreps resulting from the $SU(3)$ decomposition of the product of two octet baryons, 
along with the corresponding two-baryon states with $J=0$. 
Strangeness decreases from top to bottom in the diagrams, 
while the third component of isospin increases from left to right. 
Mixed states that are colored alike have the same total isospin and strangeness quantum numbers.
}
\label{fig:J0-irreps}
\end{figure}
%
%%%%%%%%%
\begin{figure}[b!]
\includegraphics[scale=0.76]{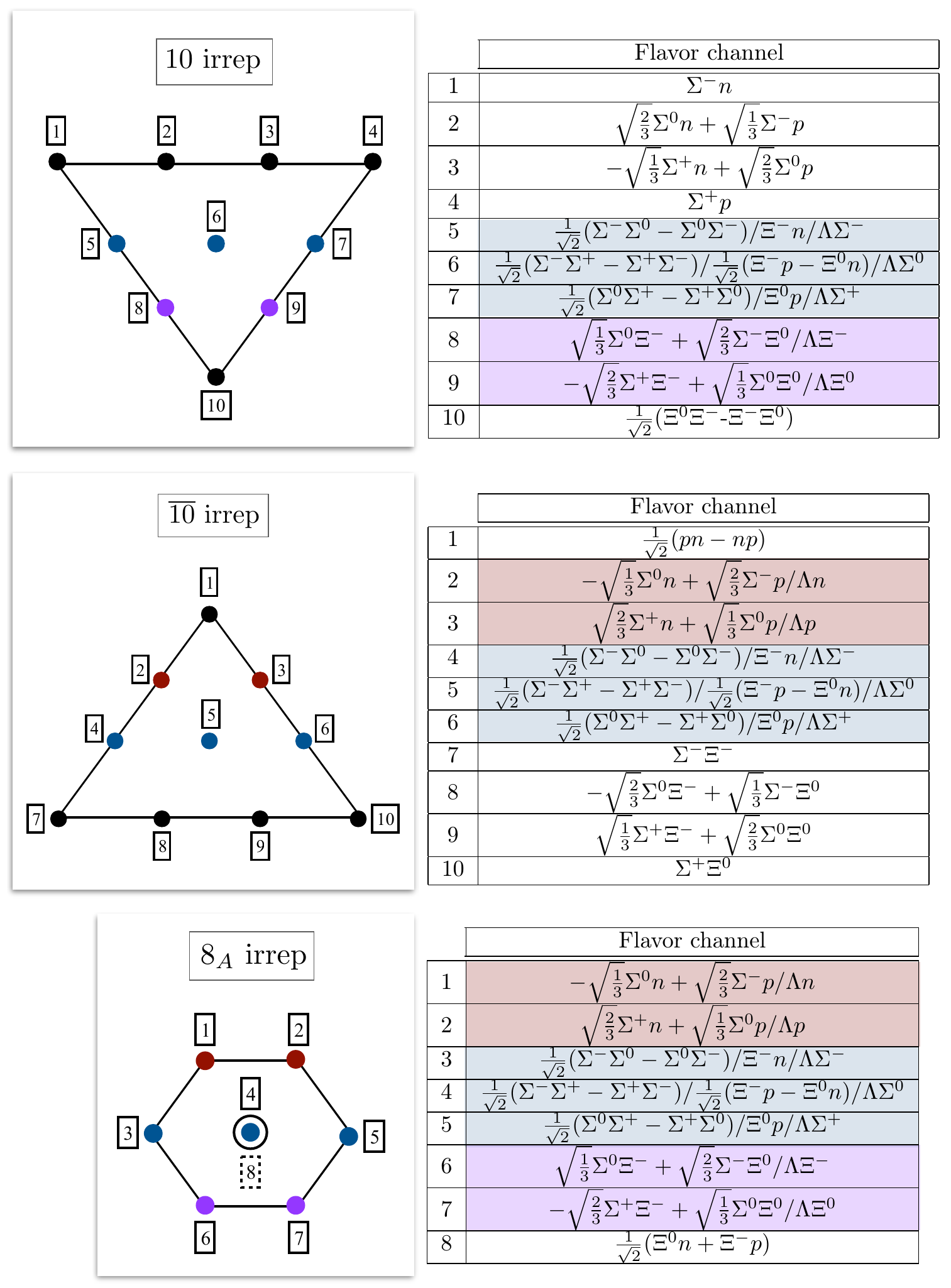}
\caption[.]{
Diagrammatic representation of the $10$, $\overline{10}$ and $8_A$ irreps resulting from the $SU(3)$ decomposition of the product of two octet baryons, 
along with the corresponding two-baryon states with $J=1$. 
Strangeness decreases from top to bottom in the diagrams, 
while the third component of isospin increases from left to right. 
Mixed states that are colored alike have the same total isospin and strangeness quantum numbers.
}
\label{fig:J1-irreps}
\end{figure}
\noindent
Two octet baryons (combined in a positive-parity state) can be arranged in 64 distinct flavor states when the up, down and strange quark masses are different. With $SU(3)$ flavor symmetry, these divide among 6 irreps of the $SU(3)$ decomposition of the product of two octet baryons, $27 \oplus 10 \oplus \overline{10} \oplus 8_S \oplus 8_A \oplus 1$. Besides parity and baryon number,  states are also classified according to the total angular momentum, i.e., either $0$ or $1$ for two baryons in an $S$ wave. Since the interpolating operators used in this work are constructed in the flavor basis, with only the isospin and strangeness quantum numbers governing the classification of states, it is useful to tabulate these flavor channels and their relation to the $SU(3)$ classifications. These are presented in Fig.~\ref{fig:J0-irreps} for the irreps with $J=0$, and in Fig.~\ref{fig:J1-irreps} for the irreps with $J=1$. The phase convention used in constructing the states in these tables is that dictated by the special embedding of the octet baryon fields in the octet baryon matrix in Eq.~(\ref{eq:octet-matrix}). As is seen in the tables, there occurs mixings among flavor states. Flavor states that mix with one another necessarily have the same electric charge, strangeness and total angular momentum. 
 %A coloring scheme is used to indicate the set of isospin partners that occur in multiple irreps. 
 %For example, there are two light pink entries in each of the tables corresponding to the $27$ and $8_S$ irreps.
 %These linear combinations can be uniquely found by a basis transformation from the flavor (isospin) to the $SU(3)$ basis.
 %Under the isospin subgroup of $SU(3)$
 As an example, consider the ``25'' entry of the $27$ irrep and the ``6'' entry of the $8_S$ irrep, denoted as $\Lambda\Xi^-/\sqrt{\frac{1}{3}} \Sigma^0 \Xi^-+\sqrt{\frac{2}{3}} \Xi^- \Sigma^0$. With an exact $SU(3)$ symmetry, two linear combinations of these degenerate flavor states can be formed such that each transforms in either the $27$ irrep or the $8_S$ irrep. In the absence of $SU(3)$ symmetry, the two flavor states are no longer degenerate; further, their mixing can no longer be uniquely determined via a straightforward basis transformation.
 %Similar discussions hold for the $I_3=+\frac{1}{2}$ partner, with mixing patterns identical to that in the $I_3=-\frac{1}{2}$ states when  the isospin symmetry is exact. 

\section{LO SCATTERING AMPLITUDES IN THE MIXED FLAVOR CHANNELS
\label{app:LO-amplitude}
}
\noindent
Away from the $SU(3)$-symmetric point, the flavor basis provides an appropriate classification of scattering states with the quantum numbers of two octet baryons, and it is useful to express the scattering amplitudes at the $SU(3)$-symmetric point in the flavor basis. When there is no mixing among flavor states (states with no colored background in the tables in Figs.~\ref{fig:J0-irreps}-\ref{fig:J1-irreps}), the scattering amplitudes are the same as those in the corresponding $SU(3)$ irreps, with constraints on the $S$-wave scattering amplitudes already obtained in Sec.~\ref{subsec:results}. For the coupled channels (states with colored background in the tables in Figs.~\ref{fig:J0-irreps}-\ref{fig:J1-irreps}), both the diagonal and off-diagonal elements of the scattering amplitude matrix can be constrained given the values of the SW coefficients obtained in Table~\ref{tab:SW-values}. Tables~\ref{tab:ME-flavor-basis-J0}-\ref{tab:ME-flavor-basis-J1} present the elements of the LO scattering amplitude matrix in each mixed-flavor channel in terms of the SW coefficients, along with their numerical values assuming unnatural interactions, see Table~\ref{tab:SW-values}. As is evident from these values, the off-diagonal elements are suppressed compared with the diagonal elements, indicating a small mixing among flavor channels. This is a consequence of the approximate accidental $SU(16)$ symmetry in the interactions. 
%%%%%%%%%%
\begin{table}[h!]
\includegraphics[scale=1.1]{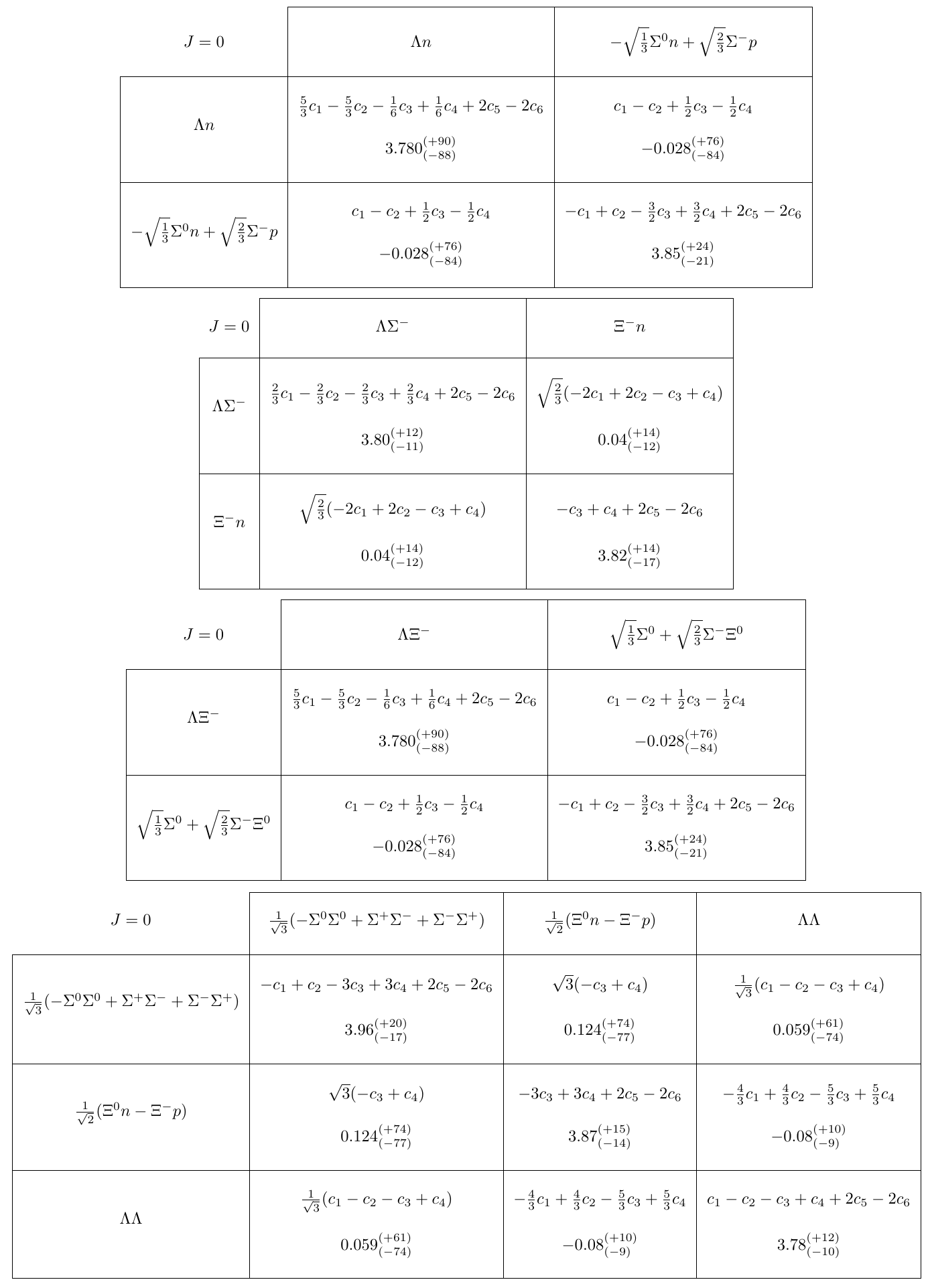}
\caption[.]{
The elements of the LO scattering amplitude matrix in the mixed flavor channels with J = 0. Using isospin symmetry, the scattering amplitudes in other mixed channels can be obtained from these results. The numerical values are obtained from the values of $c_i$ coefficients in the unnatural case with $\mu=m_{\pi}$ (see Table~\ref{tab:SW-values}), expressed in units of $[\frac{2\pi}{M_B}]$, where $m_{\pi}$ and $M_B$ are the pion mass and the baryon mass in this calculation in lattice units.}
\label{tab:ME-flavor-basis-J0}
\end{table}
%
%
%%%%%%%%%
\begin{table}[h!]
\includegraphics[scale=1.1]{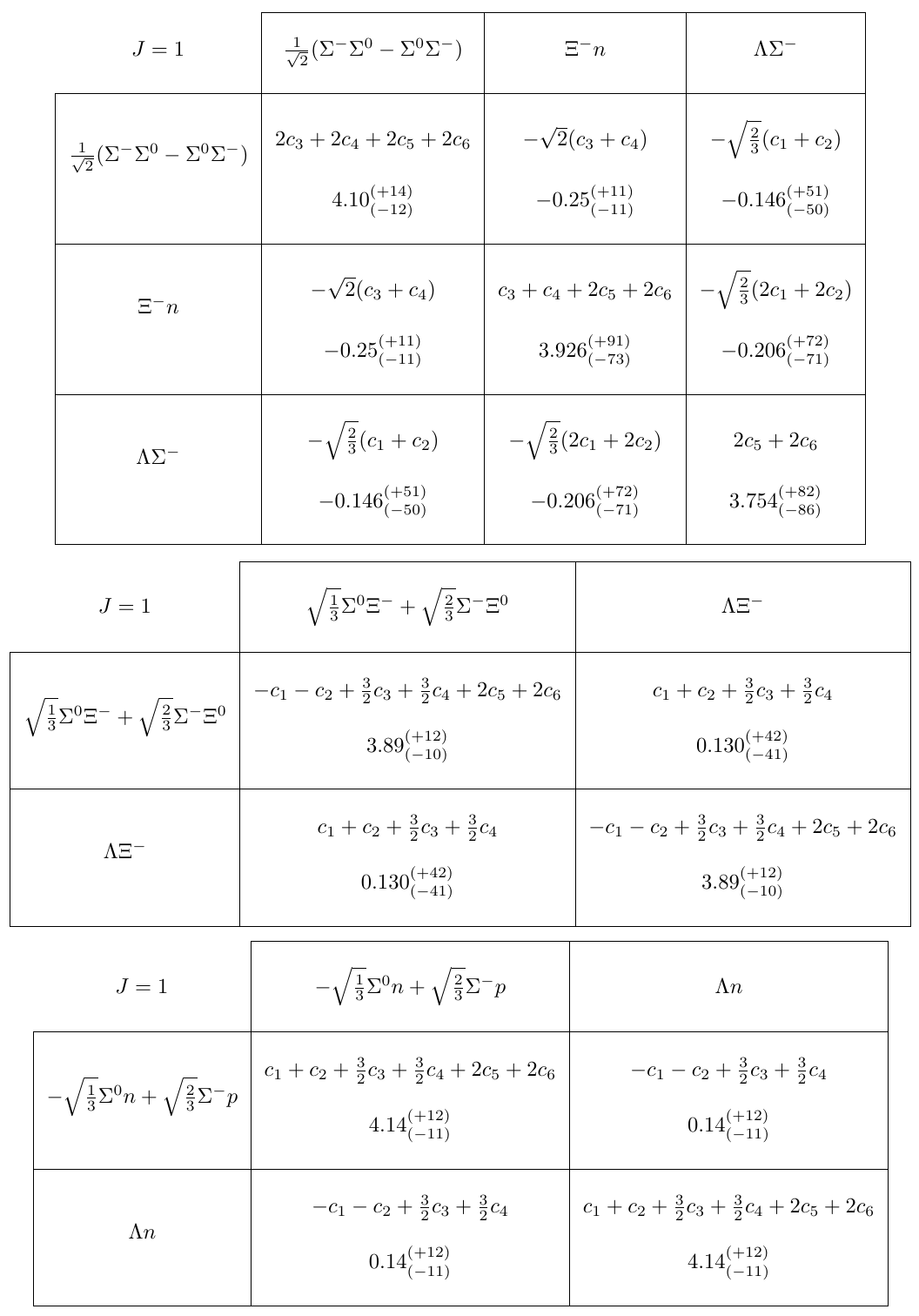}
\caption[.]{The elements of the LO scattering amplitude matrix in the mixed flavor channels with J = 1. Using isospin symmetry, the scattering amplitudes in other mixed channels can be obtained from these results. The numerical values are obtained from the values of $c_i$ coefficients in the unnatural case with $\mu=m_{\pi}$ (see Table~\ref{tab:SW-values}), expressed in units of $[\frac{2\pi}{M_B}]$, where $m_{\pi}$ and $M_B$ are the pion mass and the baryon mass in this calculation in lattice units.}
\label{tab:ME-flavor-basis-J1}
\end{table}
%

%%%%%%%%%%%%%%%%%%%%%%%%
\section{TABLES OF THE RESULTS
\label{app:tables}
}
\noindent
This appendix contains all numerical results that were omitted from the main body of the paper for brevity. These include the mass of the octet baryon measured on the three ensembles of this work (Table~\ref{tab:MB}), the shift in the energy of two baryons  in each irrep from two non-interacting baryons at rest, the corresponding CM momentum squared, ${k^*}^2$, and the value of $k^* \cot \delta$ obtained at these CM momenta (Tables~\ref{tab:27}-\ref{tab:8A}), and finally the necessary information to construct the confidence ellipses of the scatterings length and effective ranges obtained from a two-parameter ERE to $k^* \cot \delta$ in each irrep (Table~\ref{tab:ellipse}). All quantities in these tables are expressed in lattice units (l.u.). To convert to physical units, quantities must be multiplied by appropriate powers of the lattice spacing, $b=0.145(2)~{\tt{fm}}$.
%
%%%%%%%%%
\begin{table}[h!]
\includegraphics[scale=1]{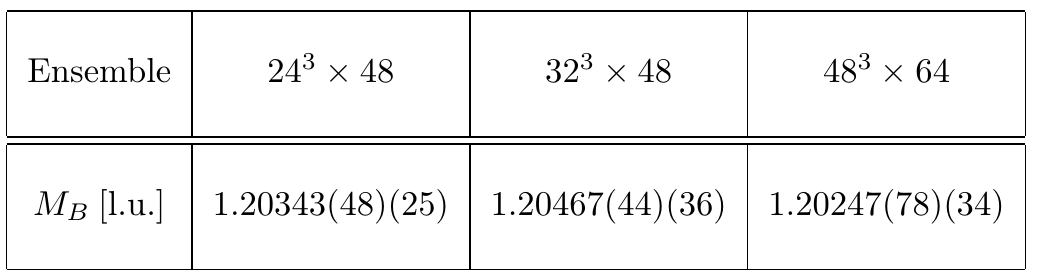}
\caption[.]{The baryon mass in lattice units (l.u.). The first uncertainty is statistical while the second uncertainty is the systematic associated with fitting.}
\label{tab:MB}
\end{table}
%
%
%%%%%%%%%
\begin{table}[b!]
\includegraphics[scale=1]{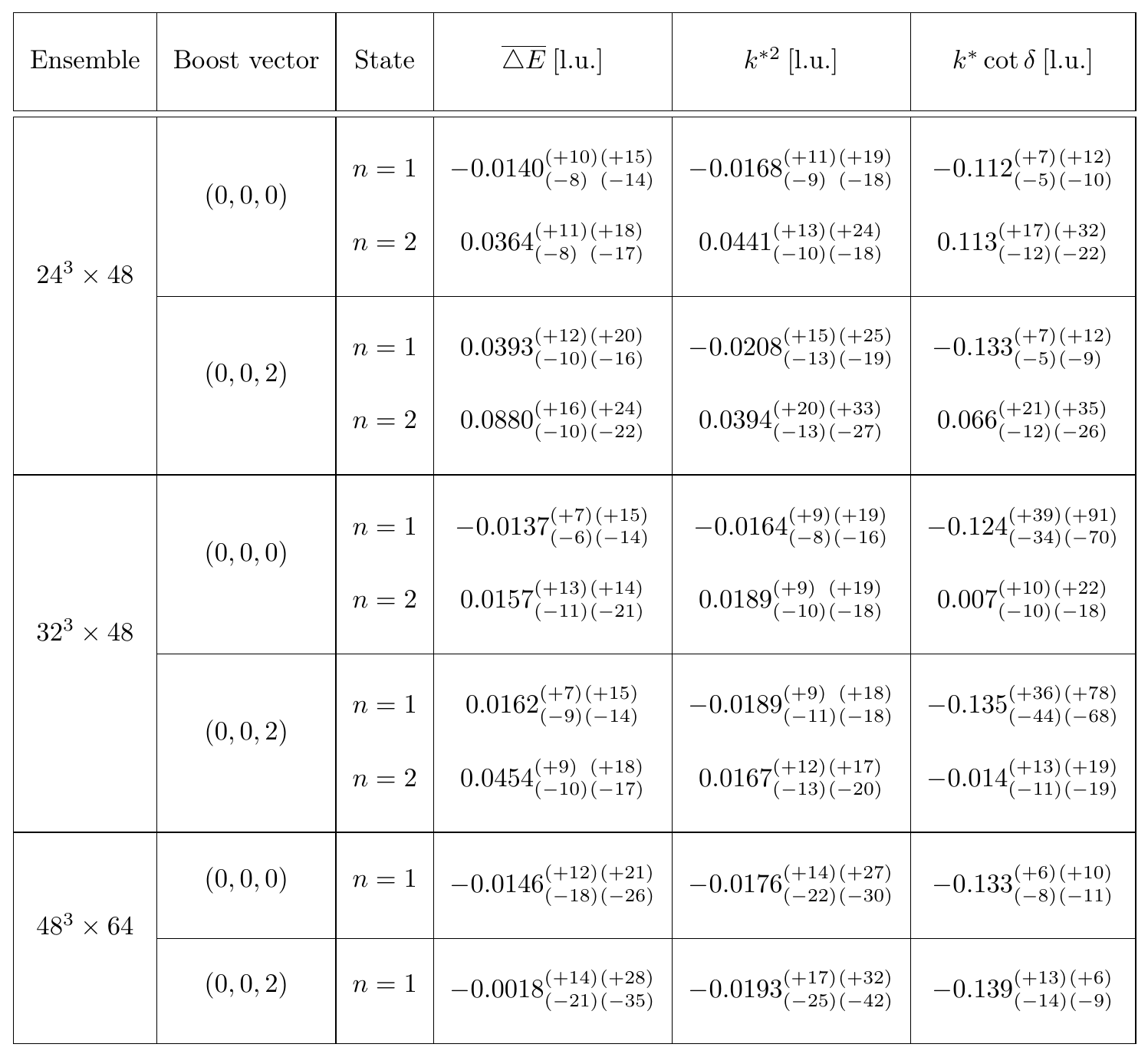}
\caption[.]{The values of energy shifts in the two-baryon system relative to two non-interacting baryons at rest, $\overline{\Delta E}$, the square of the CM momentum of the two baryons, ${k^*}^2$, and the corresponding value of $k^*\cot \delta$, in channels belonging to the 27 irrep. Energies correspond to the ground state ($n=1$) and the first excited state ($n=2$) of the system in a finite volume.  The first uncertainty is statistical while the second uncertainty is the systematic associated with fitting and the multiple analyses that are performed. All quantities are expressed in lattice units (l.u.).}
\label{tab:27}
\end{table}
%

%
%%%%%%%%%
\begin{table}
\includegraphics[scale=1]{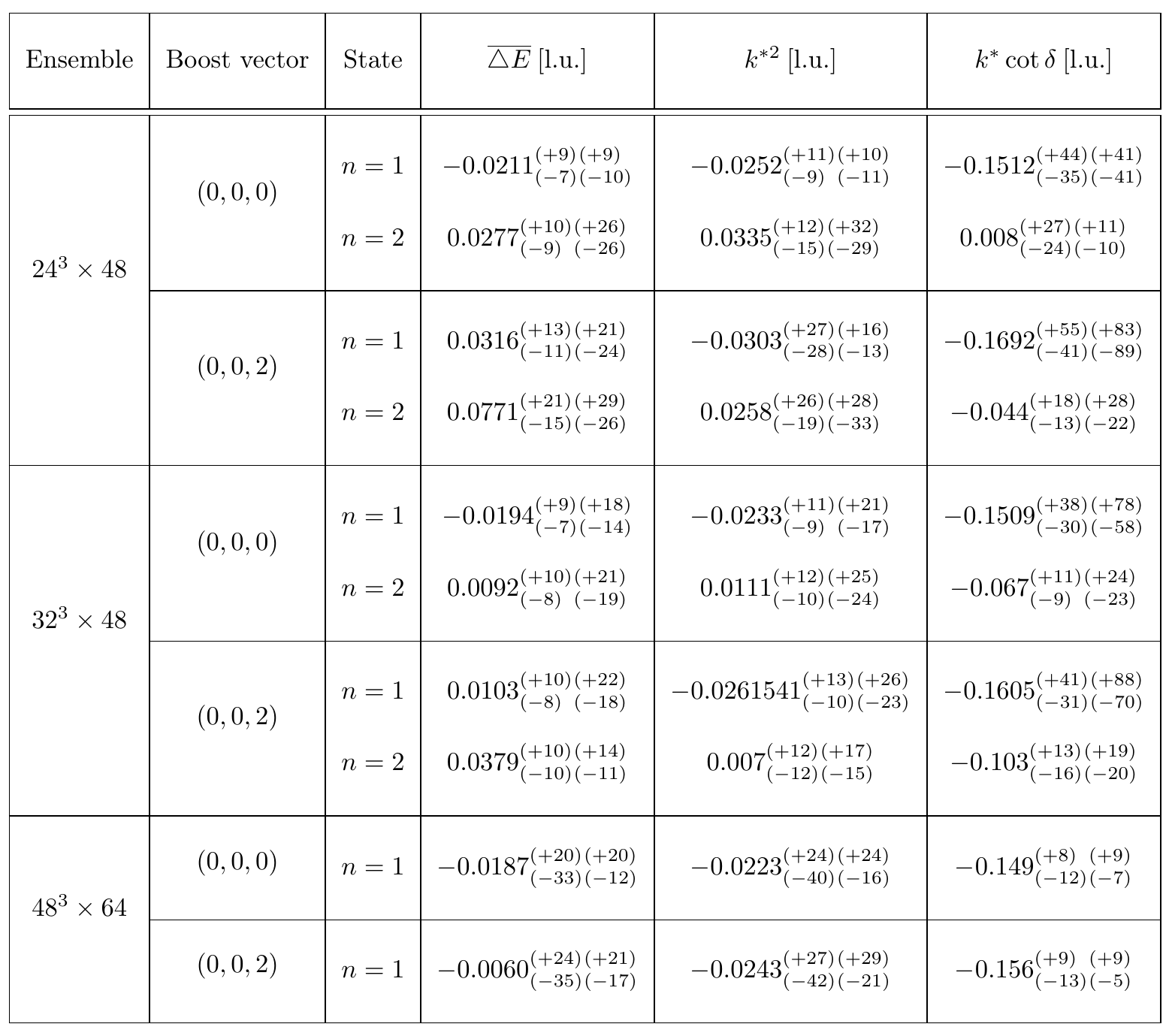}
\caption[.]{The values of energy shifts in the two-baryon system relative to two non-interacting baryons at rest, $\overline{\Delta E}$, the square of the CM momentum of the two baryons, ${k^*}^2$, and the corresponding value of $k^*\cot \delta$, in channels belonging to the $\overline{10}$ irrep. Energies correspond to the ground state ($n=1$) and the first excited state ($n=2$) of the system in a finite volume.  The first uncertainty is statistical while the second uncertainty is the systematic associated with fitting and the multiple analyses that are performed. All quantities are expressed in lattice units (l.u.).}
\label{tab:10-bar}
\end{table}
%

%
%%%%%%%%%
\begin{table}
\includegraphics[scale=1]{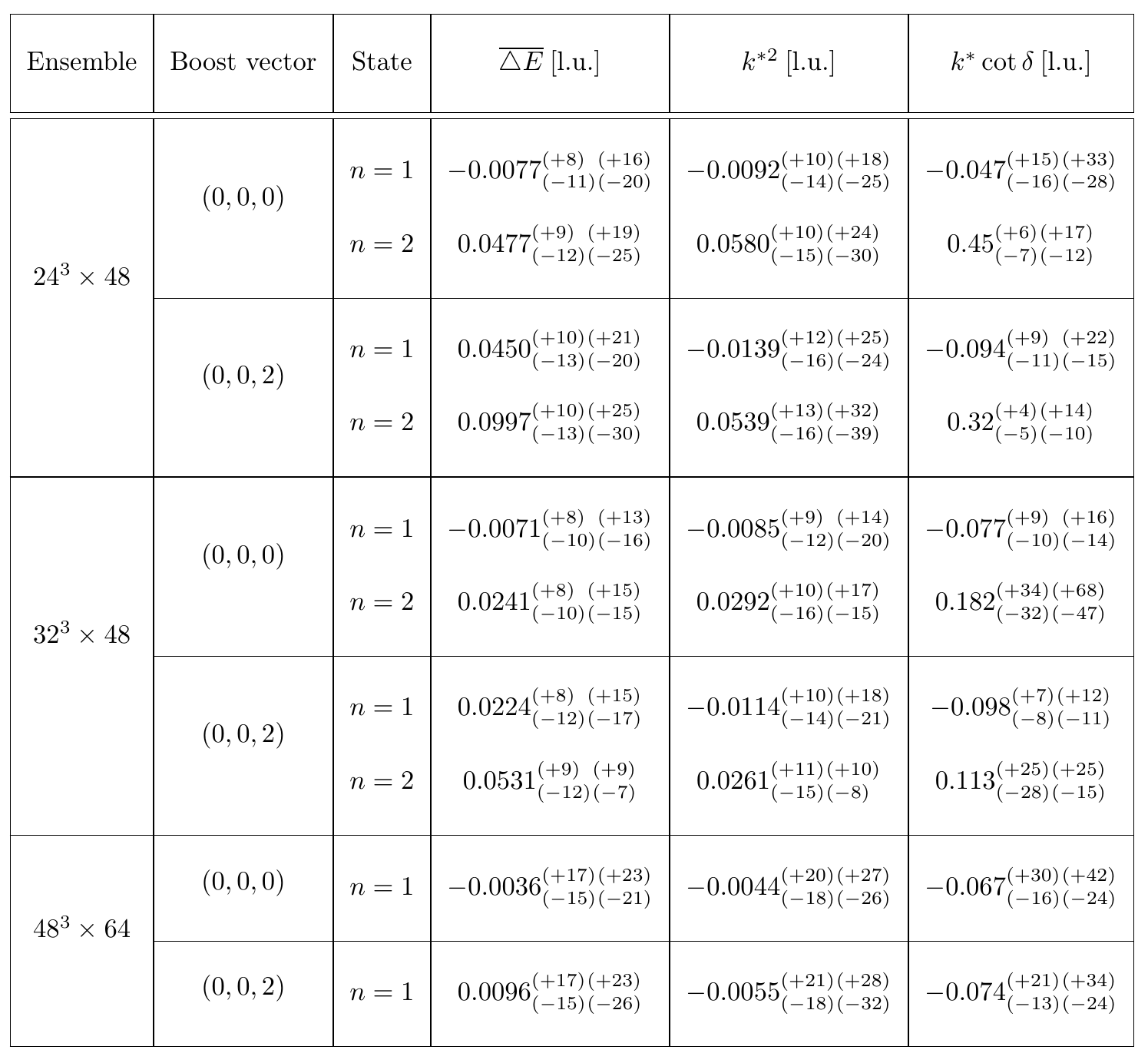}
\caption[.]{The values of energy shifts in the two-baryon system relative to two non-interacting baryons at rest, $\overline{\Delta E}$, the square of the CM momentum of the two baryons, ${k^*}^2$, and the corresponding value of $k^*\cot \delta$, in channels belonging to the 10 irrep. Energies correspond to the ground state ($n=1$) and the first excited state ($n=2$) of the system in a finite volume.  The first uncertainty is statistical while the second uncertainty is the systematic associated with fitting and the multiple analyses that are performed. All quantities are expressed in lattice units (l.u.).}
\label{tab:10}
\end{table}
%

%
%%%%%%%%%
\begin{table}[t!]
\includegraphics[scale=1]{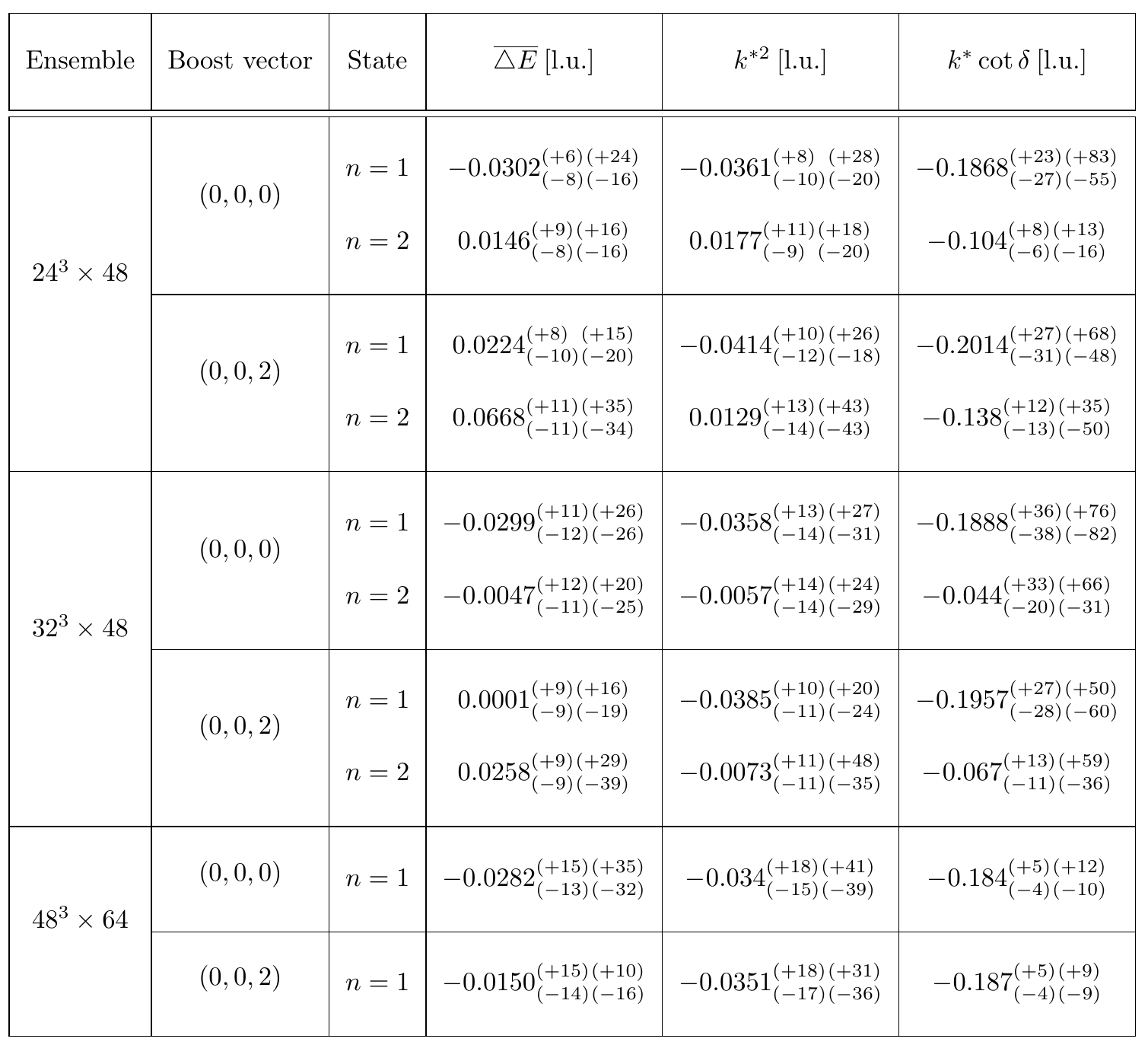}
\caption[.]{The values of energy shifts in the two-baryon system relative to two non-interacting baryons at rest, $\overline{\Delta E}$, the square of the CM momentum of the two baryons, ${k^*}^2$, and the corresponding value of $k^*\cot \delta$, in channels belonging to the $8_A$ irrep. Energies correspond to the ground state ($n=1$) and the first excited state ($n=2$) of the system in a finite volume. The first uncertainty is statistical while the second uncertainty is the systematic associated with fitting and the multiple analyses that are performed. All quantities are expressed in lattice units (l.u.).}
\label{tab:8A}
\end{table}

\clearpage

\newpage

\section{THE FOUR SO-CALLED ``SANITY CHECKS'' OF Ref.~\cite{Iritani:2017rlk} ARE PASSED
\label{app:checks}
}
%
%%%%%%%%%
\begin{table}[t!]
\includegraphics[scale=1]{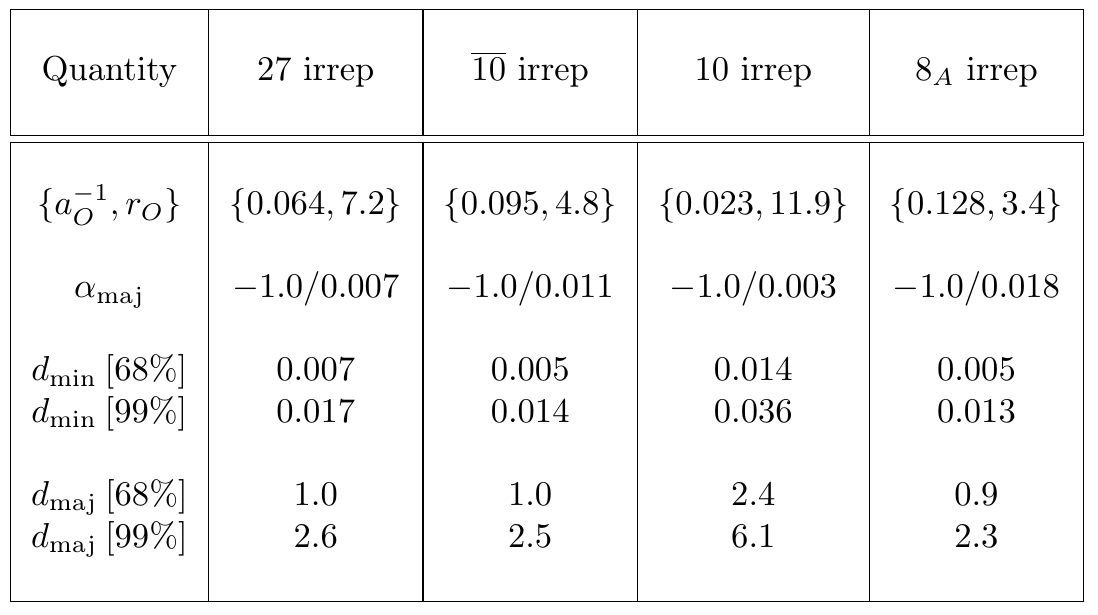}
\caption[.]{
Specifications of the $68\%$ and $99\%$ confidence ellipses arising from the correlation between the scattering length and effective range in a two-parameter ERE fit to $k^*\cot \delta$ in each channel. These can be used to reconstruct the corresponding ellipses in Fig.~\ref{fig:ar-confint}. $\{a^{-1}_O,r_O \}$ denotes the coordinate of the center of ellipse, $\alpha_{\text{maj}}$ is the slope of the semi-major axis, $d_{\text{min}}~[68\%]$ ($d_{\text{min}}~[99\%]$) is the semi-minor axis and $d_{\text{maj}}~[68\%]$ ($d_{\text{maj}}~[99\%]$) is the semi-major axis of the $68\%$ ($99\%$) confidence ellipse. All quantities are expressed in lattice units (l.u.).}
\label{tab:ellipse}
\end{table}
\noindent
In Ref~.\cite{Iritani:2017rlk}, Iritani,~et~al. suggest a set of tests that the two-nucleon scattering amplitudes must pass. They state that if the amplitudes obtained using L\"uscher's method fail these checks, the energy levels determined from the late-time behavior of two-nucleon correlation functions do not correspond to the correct energy eigenvalues of the system, and according to these authors, all calculations of baryon-baryon interactions performed by researchers other than themselves have been misled by fake intermediate plateaus in EMPs at early times. Such consistency checks are important but must be carried out carefully. Among other works, Iritani,~et~al. study the work by the NPLQCD collaboration on the ensembles with $m_{\pi} \approx 806~\tt{MeV}$~\cite{Beane:2012vq, Beane:2013br} that are used in the present work, and conclude that at least two of the ``sanity checks'' are not passed for the results presented in these references. Consequently, they conclude that there is no bound states present in both the isosinglet and isotriplet two-nucleon channels, consistent with their previous studies of these channels at similar quark masses using HALQCD's ``potential method''~\cite{Aoki:2008hh, Aoki:2009ji, Aoki:2011ep, Inoue:2011ai}.\footnote{The HALQCD method is subject to several unquantified systematic uncertainties, as have been previously pointed out in literature, see e.g., Refs.~\cite{Beane:2010em, Detmold:2007wk, Detmold:2015jda, Walker-Loud:2014iea, Yamazaki:2015nka, Savage:2016egr}.} Arguments against the claims of ``mirage plateaus'' in the two-baryon calculations of this work are already presented in Sec.~\ref{subsec:EMP}. Our results are tested against the four so-called ``sanity checks'' of Ref.~\cite{Iritani:2017rlk}, and unambiguously pass these checks. This conclusion applies to our previous results in Ref.~\cite{Beane:2012vq, Beane:2013br}.
\begin{itemize}
\item[--]{\emph{``Sanity check'' (0) \textbf{passed}:} 
EMPs corresponding to correlation functions with different interpolating operators, 
but with the same quantum numbers, must agree at large times, and the energies extracted from these correlation functions should be consistent with each other within the uncertainties of each calculation. Although  calculations with $\mathbf{d}=(0,0,0)$ and $(0,0,2)$ correspond to different Fourier transforms of the same interpolating operator structure, these are noted as different sources by Iritani,~et~al., and their consistency has been examined by these authors. Under the assumption that these represent independent measurements (up to the common gauge configurations used), we have examined the consistency of the results obtained from the $\mathbf{d}=(0,0,0)$ and $(0,0,2)$ correlation functions. Note that these transform similarly in the CM frame under the cubic group -- a statement that holds up to relativistic corrections which are at sub-percent level in this calculation. As a result, the CM energies (momenta) obtained from correlation functions with $\mathbf{d}=(0,0,0)$ and $(0,0,2)$ should be approximately identical. Indeed, as is evident from the ${k^*}^2$ values in Tables~\ref{tab:27}-\ref{tab:8A}, such ``source independence'' is a feature of the calculations performed, in contradiction with the claim of Ref.~\cite{Iritani:2017rlk}. For clarity, the ${k^*}^2$ values in each ensemble are plotted against  each other in Fig.~\ref{fig:ksq-comparisons}, demonstrating the consistency between the $\mathbf{d}=(0,0,0)$ and $(0,0,2)$ cases, up to minor statistical deviations. Additionally, the values of the binding momenta of the bound states obtained from volume extrapolations in each of the $\mathbf{d}=(0,0,0)$ and $(0,0,2)$ cases are perfectly consistent with each other, as was noted in Table~\ref{tab:binding-momenta}. This feature existed also in our previous analysis of these correlation functions~\cite{Beane:2012vq}, with agreement at the level of two standard deviations or better~\cite{Beane:2017edf}.\footnote{It must be pointed out that Iritani~et~al.'s claim of source dependence of the calculations performed by Berkowitz,~et~al. in Ref.~\cite{Berkowitz:2015eaa} (using the NPLQCD ensembles) is flawed by the fact that the states obtained in Ref.~\cite{Berkowitz:2015eaa} using a displaced interpolating operator are noted as the first excited states by the authors, while they have been identified as ground states by Iritani~et~al. in both channels. Considering only the ground states, the NPLQCD results and those by Berkowitz, {\it et al.} are consistent, see Ref.~\cite{Beane:2017edf}.}} Further, the full agreement between the SS and SP combinations of the source and sink operators is another confirmation of the source and sink independence of the present results.\footnote{Our recent calculations of matrix elements of the axial current in light nuclei~\cite{Savage:2016kon} employ different source and sink smearing than in this work. The same binding energies are recovered in those calculations, further supporting source independence of the extracted two-baryon energies.}
%
%%%%%%%%%
\begin{figure}[t!]
\includegraphics[scale=0.635]{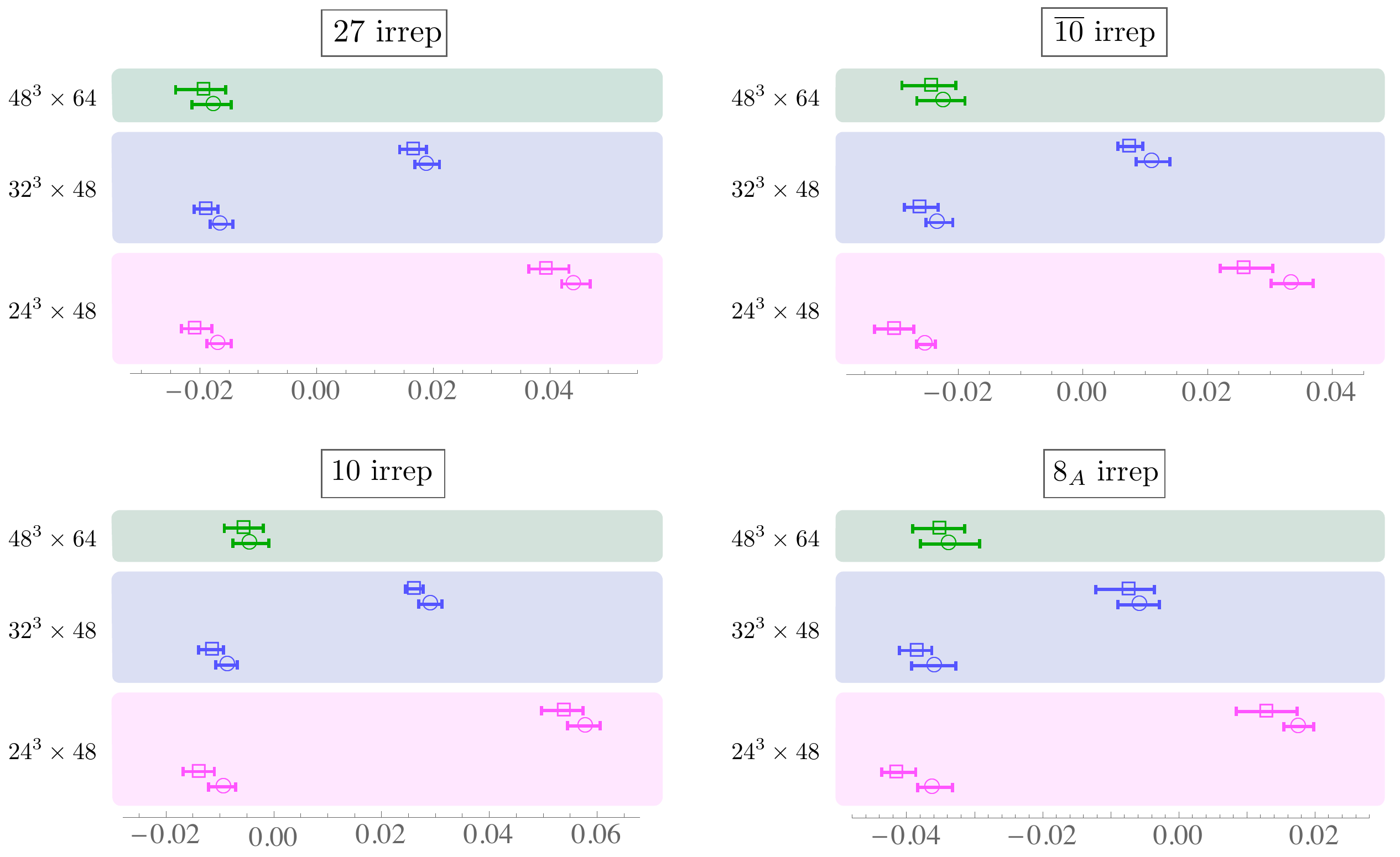}
\caption[.]{The values of ${k^*}^2$ obtained in this work for two-baryon systems 
in the $27$, $\overline{10}$, $10$ and $8_A$ irreps, expressed in lattice units (l.u.). 
The circle and square symbols correspond to $\mathbf{d}=(0,0,0)$ and $(0,0,2)$, respectively, and the agreement within uncertainties in each pair of data points demonstrates the source independence of this calculation.}
\label{fig:ksq-comparisons}
\end{figure}

\item[--]{\emph{``Sanity check'' (1) \textbf{passed}:} If the ERE is a valid parametrization of the scattering amplitude at low energies, the analyticity of the amplitude as a function of the CM energy implies that the  ERE obtained from states with positively-shifted energies (${k^*}^2>0$) must be consistent with that obtained from states with negatively-shifted energies (${k^*}^2<0$). Iritani~et~al. find that the NPQCD results pass this test, and for completeness, we demonstrate this consistency in Fig.~\ref{fig:ERE-n1-n2}. Fits to EREs using both the ground states ($n=1$) and the first excited states ($n=2$) (color-filled bands) are overlaid on fits to EREs using only the ground states (hashed bands). The two sets of bands are consistent with each other, showing that this check is unambiguously passed. The same feature is seen for the three-parameter ERE fits, with significantly larger uncertainties, particularly in the case of fits to only the ground states. It is important to note that \emph{a priori} the radius of convergence of the ERE is unknown, so an inconsistency between ERE fits to ${k^*}^2<0$ and ${k^*}^2>0$ regions can imply that either higher-order terms in the ERE are required or that the ERE does not apply. As a result, this ``sanity check'' is not a rigorous diagnostic of the validity of energy extractions. 
}
%
%%%%%%%%%
\begin{figure}[h!]
\includegraphics[scale=0.701]{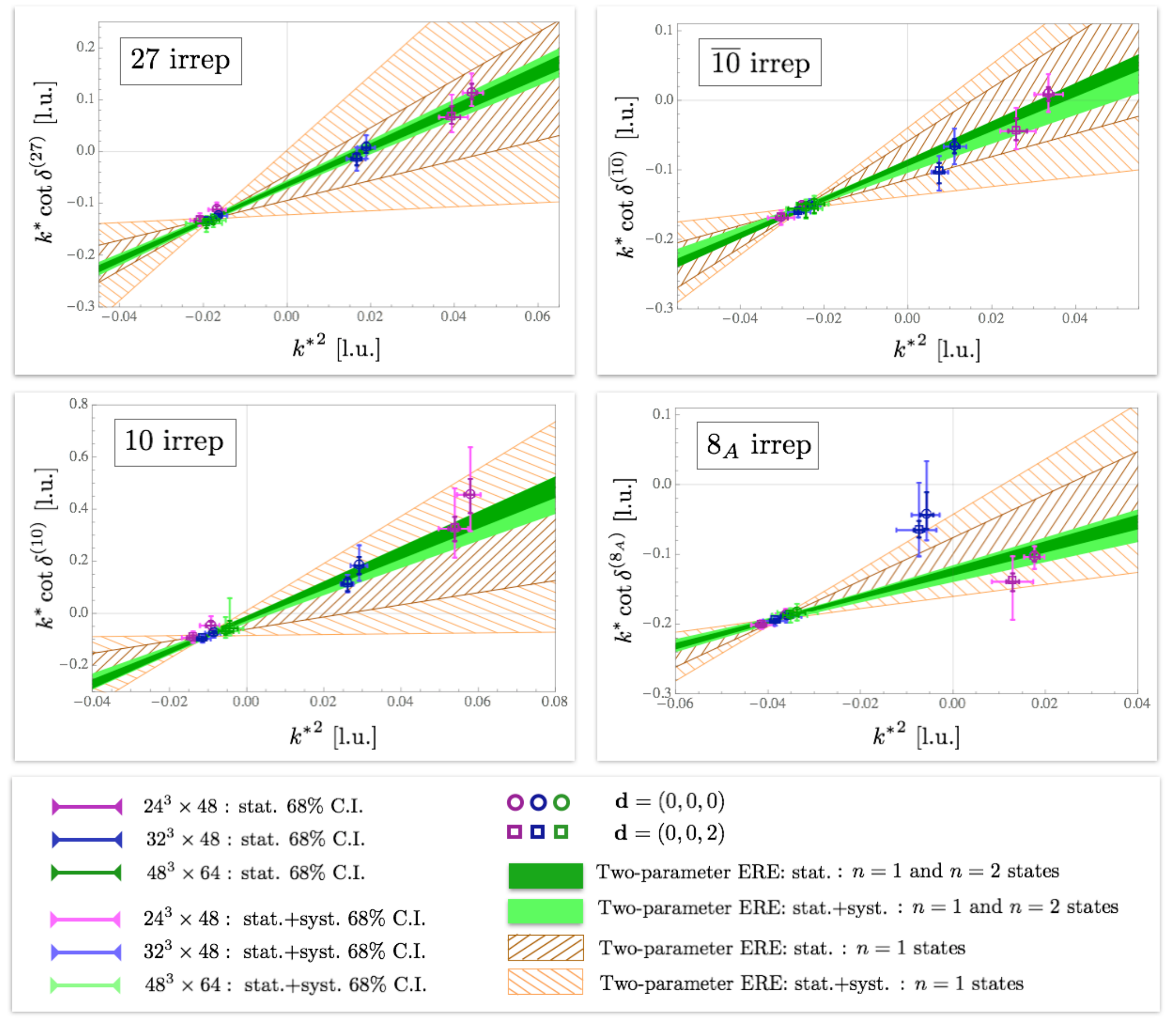}
\caption[.]{$k^*\cot \delta$ versus the square of the CM momentum of the two baryons, ${k^*}^2$, along with the bands representing fits to two-parameter EREs obtained from i) only the ground states ($n=1$) and ii) from both the ground states ($n=1$) and the first excited states ($n=2$). The plots demonstrate the consistency of the EREs between negative and positive ${k^*}^2$ regions in all channels. Quantities are expressed in lattice units (l.u.).}
\label{fig:ERE-n1-n2}
\end{figure}
\item[--]{\emph{``Sanity check'' (2) \textbf{passed}:} This check states that the value of all scattering parameters must be non-singular. This check is immediately passed, as also noted by Iritani,~et~al., for the results presented in Refs.~\cite{Beane:2012vq, Beane:2013br}, as well as those presented in this work for all two-baryon channels. The values of the scattering parameters obtained in this work are tabulated in Table~\ref{fig:scatt-param-table}. None of the parameters $a^{-1}$, $r$ and $P$ resulting from the two and three-parameter ERE fits are singular. 
}

\item[--]{\emph{``Sanity check'' (3) \textbf{passed}:} The sign of the residue of the S-matrix at the bound-state pole is fixed. This requirement leads to the following condition on the $k^*\cot \delta$ function:
\begin{eqnarray}
\left. \frac{d}{d{k^*}^2}(k^*\cot \delta+\sqrt{-{k^*}^2}) \right |_{{k^*}^2=-{\kappa^{(\infty)}}^2} < 0,
\label{eq:slope}
\end{eqnarray}
where $\kappa^{(\infty)}$ is the binding momentum. Despite the claim of Ref.~\cite{Iritani:2017rlk}, the results presented here and in Refs.~\cite{Beane:2012vq, Beane:2013br} pass this check as well. As is seen from Fig.~\ref{fig:ERE-tangent}, at the level of one standard deviation, the slope of the two-parameter ERE fit to the $k^*\cot \delta$ function (color-filled bands) in all channels is never greater than the slope of the ($-\sqrt{-{k^*}^2}$) function (gray bands) at the corresponding bound-state pole. The uncertainty in the tangent line to ($-\sqrt{-{k^*}^2}$) at ${k^*}^2=-{\kappa^{(\infty)}}^2$ arises from the uncertainty in the values of $\kappa^{(\infty)}$ given in Table~\ref{tab:binding-momenta}. A similar conclusion can be made from the three-parameter ERE fits.}

\end{itemize}
% 

%
%%%%%%%%%
\begin{figure}[h!]
\includegraphics[scale=0.701]{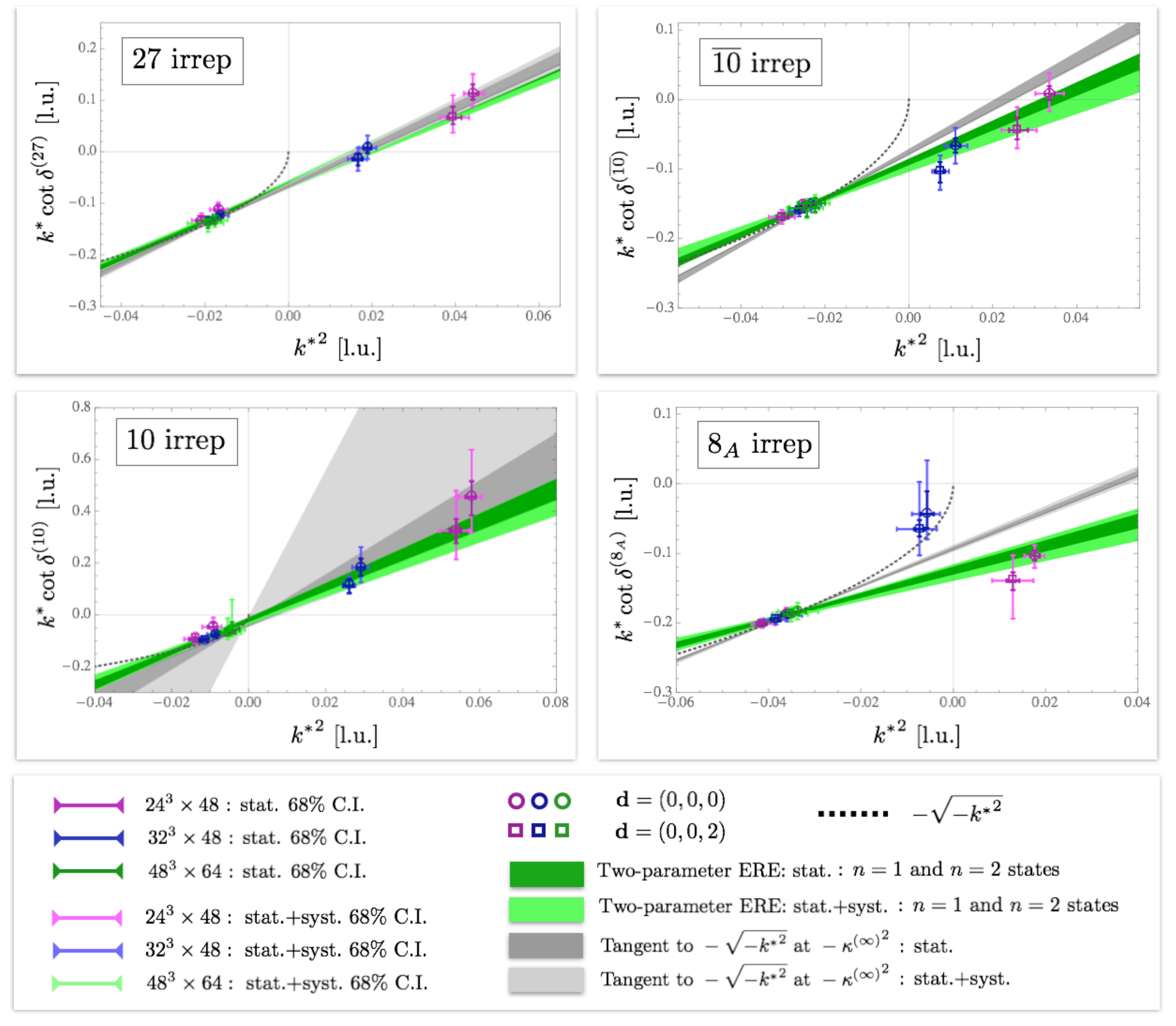}
\caption[.]{The two-parameter ERE is compared with the tangents to the ($-\sqrt{-{k^*}^2}$) curve at values of ${k^*}^2=-{\kappa^{(\infty)}}^2$, with $\kappa^{(\infty)}$ values given in Table~\ref{tab:binding-momenta}. The plots verify that all the identified bound states in this work are consistent with the criterion in Eq.~(\ref{eq:slope}) within uncertainties. Quantities are expressed in lattice units (l.u.).}
\label{fig:ERE-tangent}
\end{figure}

\end{document}